\documentclass[iop]{emulateapj}

\shorttitle{RESOLVED WHITE DWARF -- RED DWARF SYSTEMS}
\shortauthors{Farihi, Hoard, \& Wachter}

\slugcomment{Accepted to ApJ Supplement Series}

\begin{document}

\title{WHITE DWARF -- RED DWARF SYSTEMS RESOLVED
		WITH THE {\em HUBBLE SPACE TELESCOPE}.\\ II.
		FULL SNAPSHOT SURVEY RESULTS}
		
\author{J. Farihi\altaffilmark{1},
	D. W. Hoard\altaffilmark{2},
	\& S. Wachter\altaffilmark{2}}
	
\altaffiltext{1}{Department of Physics \& Astronomy,
			University of Leicester,
			Leicester LE1 7RH, 
			UK; jf123@star.le.ac.uk}
			
\altaffiltext{2}{Spitzer Science Center,
			California Institute of Technology,
			MS 220-6,
			Pasadena, CA 91125}

\begin{abstract}

Results are presented for a {\em Hubble Space Telescope} Advanced Camera for Surveys 
high-resolution imaging campaign of 90 white dwarfs with known or suspected low mass 
stellar and substellar companions.  Of the 72 targets which remain candidate and confirmed 
white dwarfs with near-infrared excess, 43 are spatially resolved into two or more components, 
and a total of 12 systems are potentially triples.  For 68 systems where a comparison is possible, 
50\% have significant photometric distance mismatches between their white dwarf and M dwarf 
components, suggesting white dwarf parameters derived spectroscopically are often biased due 
to the cool companion.  Interestingly, nine of 30 binaries known to have emission lines are found
to be visual pairs and hence widely separated, indicating an intrinsically active cool star and not 
irradiation from the white dwarf.  There is a possible, slight deficit of earlier spectral types (bluer
colors) among the spatially unresolved companions, exactly the opposite of expectations if 
significant mass is transferred to the companion during the common envelope phase.

Using the best available distance estimates, the low mass companions to white dwarfs exhibit 
a bimodal distribution in projected separation.  This result supports the hypothesis that during 
the giant phases of the white dwarf progenitor, any unevolved companions either migrate 
inward to short periods of hours to days, or outward to periods of hundreds to thousands of 
years.  No intermediate projected separations of a few to several AU are found among these 
pairs.  However, a few double M dwarfs (within triples) are spatially resolved in this range, 
empirically demonstrating that such separations were readily detectable among the binaries 
with white dwarfs.  A straightforward and testable prediction emerges: all spatially unresolved, 
low mass stellar and substellar companions to white dwarfs should be in short period orbits.  
This result has implications for substellar companion and planetary orbital evolution during 
the post-main sequence lifetime of their stellar hosts.

\end{abstract}
	
	
\section{INTRODUCTION}

Unlike main-sequence stars, white dwarfs allow one to see more deeply into their local 
spatial environments due to low (and blue-peaked) luminosities; quite possibly providing 
an excellent window to directly detect extrasolar planets \citep{hog09,bur02}.  While surviving 
planets orbiting white dwarfs probably exist in modest abundance \citep{far09a}, and their 
post-main sequence evolution is of obvious interest, they are currently difficult to detect.  
Substellar companions to white dwarfs are the best proxy for any giant planets that survive 
beyond the main sequence, but white dwarf -- brown dwarf binaries are rare \citep{far08b,
mul07,hoa07,far05a}.  Low mass stellar companions to white dwarfs, however, are relatively 
ubiquitous and produce a composite color signature that is readily detectable \citep{hoa07,
sil06,pro83}.  These binary systems can provide basic information on the long-term evolution 
of low mass objects in the presence of host giant stellar envelopes, and the initial mass function 
near the bottom of the main sequence \citep{sch96,zuc92}.

Understanding common envelope evolution is critical for additional astrophysical phenomena 
such as planetary nebulae, cataclysmic variables, X-ray binaries, and compact stellar mergers.
The latter are perhaps the best candidates for type Ia supernovae and short $\gamma$-ray 
bursts \citep{dem09,tap09,nel05}.  Here, low mass stellar companions to white dwarfs are used 
to constrain the effects of common envelope evolution by directly measuring the distribution of 
their projected separations and secondary star colors; these quantities are excellent indicators 
of orbital semimajor axes and masses.  This is a fundamental astrophysical question: how does 
post-main sequence evolution influence the distribution of binary parameters?

This paper presents astrometric and photometric data for 90 white dwarfs with suspected
near-infrared excess owing to a low mass stellar or substellar companion.  The program,
data acquisition, and analysis is described in \S2, and the results are discussed in \S3.  
The first half of this dataset was presented in Farihi, Hoard, \& Wachter (2006; Paper I), 
but the results given here, which were produced using a uniform application of a more
recent processing pipeline, supersede those in Paper I.

\section{THE PROGRAM AND OBSERVATIONAL DATA}

\begin{figure*}
\epsscale{1.2}
\figurenum{1a}
\plotone{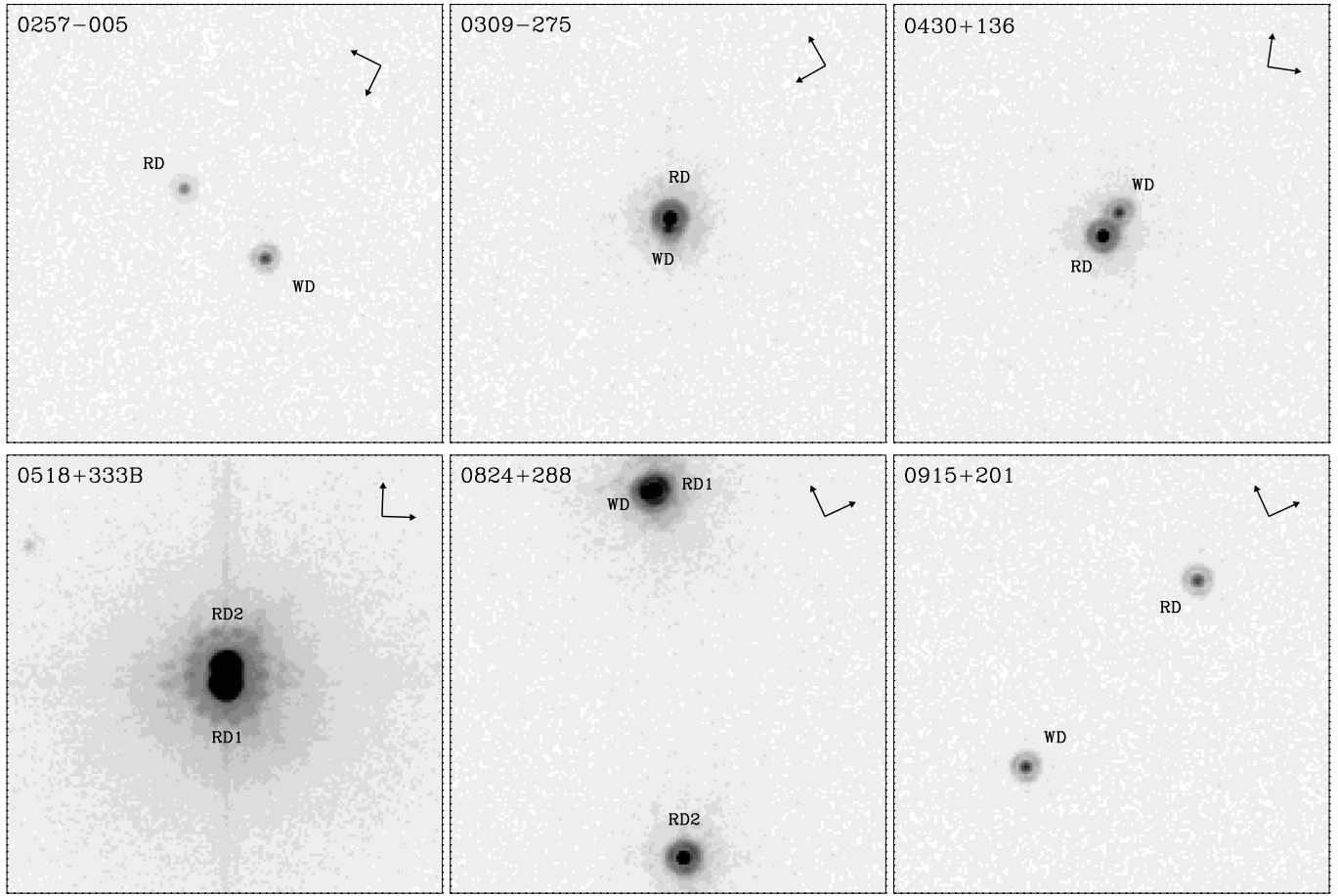}
\caption{Multidrizzled, ACS HRC images of all 17 newly detected, totally or partially resolved, multiple 
systems.  All data were taken in the F814W filter and are displayed at a single color scale and stretch.
The panels are $4\farcs0\times4\farcs0$ and displayed in the native spacecraft orientation with arrows
indicating the directions of North and East.
\label{fig1a}}\
\epsscale{1.0}
\end{figure*}

\begin{figure*}
\epsscale{1.2}
\figurenum{1b}
\plotone{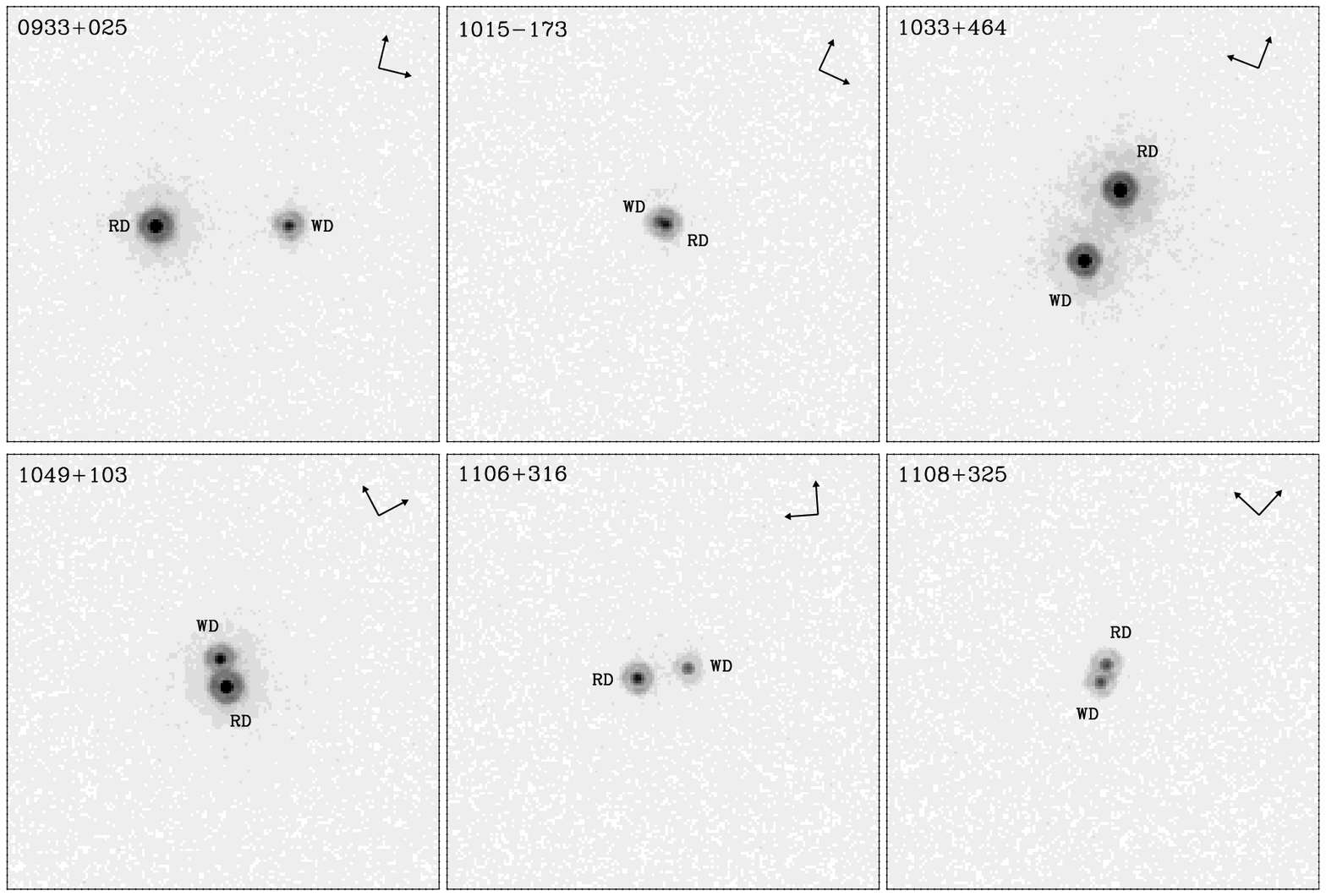}
\caption{{\em Continued}
\label{fig1b}}
\epsscale{1.0}
\end{figure*}

\begin{figure*}
\epsscale{1.2}
\figurenum{1c}
\plotone{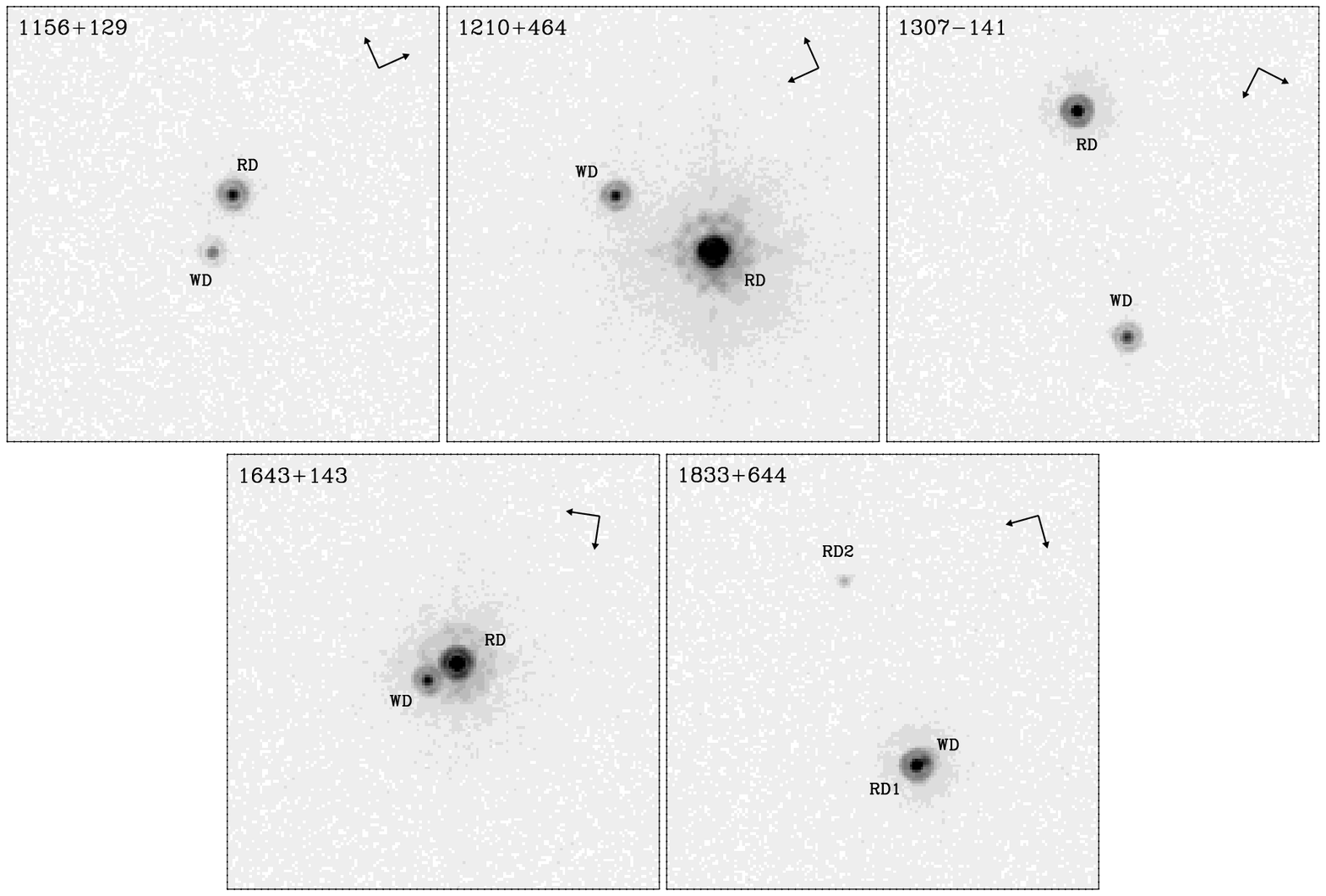}
\caption{{\em Continued}
\label{fig1c}}
\epsscale{1.0}
\end{figure*}

The observational goal of the program was to image white dwarfs with low mass, cool 
companions at sufficient spatial resolution to directly detect any secondaries in the few to 
several AU range; i.e.\ orbital periods of one to tens of years which are difficult to probe from 
the ground.  The targets were chosen from ground-based datasets as confirmed or candidate, 
spatially unresolved binaries consisting of a known or suspected white dwarf with near-infrared 
excess emission \citep{hoa07,far05a,wac03,mcc99}.

In Cycle 13, {\em Hubble Space Telescope} ({\em HST\,}) Snapshot Program 10255 was 
awarded 100 targets for single bandpass imaging with the Advanced Camera for Surveys 
(ACS; \citealt{for98}) High-Resolution Camera (HRC).  The F814W filter was chosen as the 
most likely wavelength range in which both the white dwarf and low mass components would 
be relatively bright, and would still be sensitive to the latest M dwarf companions.  The images 
were acquired in a four-point dither pattern with individual 10\,s exposures for all but two stars 
that had suspected or known brown dwarf companions; for those targets the exposures were 
270\,s each.  The data for each target were processed into a single, multi-drizzled image with 
the ACS calibration software and pipeline CALACS 4.5.6 and OPUS 15.7; this represents a 
more recent pipeline than that used for the subset of observations analyzed for Paper I and 
may account for some of the minor numerical differences (see descriptions of individual 
targets in the Appendix).

Figures \ref{fig1a} -- \ref{fig1c} display the ACS images of 17 newly detected multiple systems 
with spatially-resolved components.  Images of the remaining 28 systems with components
that were spatially resolved in the ACS program are shown in Paper I.  All other science
target images were found to be consistent with point sources.

\subsection{Photometry and Astrometry}

All flux measurements were carried out using an aperture radius of $r=0\farcs20$ (8\,pixels; 
includes first airy maximum and 85\% of energy), with standard sky annulus, aperture 
corrections, photometric transformations following \citet{sir05}, and Vega zeropoints 
from \citet{boh07}.  Photometry was performed with the IRAF task {\sf apphot} for all point 
sources isolated by a radius of at least $0\farcs4$.  For pairs more closely separated than 
this, {\sf daophot} was used to simultaneously photometrically deconvolve the sources.  A 
total of 12 point spread function (PSF) stars were chosen among the highest signal-to-noise 
(S/N), isolated point sources; four of these were white dwarfs, four were M dwarfs, and four 
were spatially-unresolved white dwarf -- M dwarf pairs.  For the 22 close visual doubles and 
triples, the components were fitted with each of the 12 PSF stars, and photometry was 
determined by averaging across values within the appropriate subtype.  In a few cases, 
only one or two PSF stars produced consistent results with acceptable errors, and other fits 
were rejected.  Photometric errors are the quadrature sum of the measurement errors via 
{\sf apphot} or {\sf daophot} and a 2\% calibration error; the Cousins $I$-band magnitude 
errors include an additional uncertainty from the transformation equations \citep{sir05}.

Astrometry was performed with IDP3\footnote{http://nicmos.as.arizona.edu/software} radial 
profile fitting at $r=3$\,pixels.  This radius is an ideal choice for image centroiding and PSF 
characterization as it occurs near the first Airy pattern minimum, and interior to this radius 
the PSF is well-reproduced by a Gaussian radial profile. For isolated sources, Gaussian fit 
centroid errors were typically less than 0.03\,pixels along both the x- and y-axes.  Close 
visual pairs had centroid errors somewhat larger and determined by the scatter in positions 
from multiple PSF fits and/or images in which the neighboring source was fitted and subtracted; 
where multiple methods were used, a weighted average was taken.  The errors in binary angular 
separation and position angle were calculated by direct propagation of measured x- and y-axis 
positional errors through the formulae.  Full widths at half maximum (FWHMs) were measured 
using Gaussian fits of the radial profile out to $r=3$\,pixels, corresponding to a diameter of 2 
typical FWHMs.

The size of the F814W aperture correction and subsequent photometric transformation to 
Cousins $I$-band are both color-dependent, and were determined using Tables 7, 8, and 
23 of \citet{sir05}.  First, the published or best estimate of the $V-I$ (for white dwarfs) or $I-K$ 
(for M dwarfs) color of a given star was used to look up a color-dependent aperture correction,
utilizing any available colors or spectral type estimate.  Second, a Cousins $I$ magnitude was 
calculated on the basis of this estimated color and the aperture-corrected F814W magnitude.
Third, this $I$-band magnitude was used to re-determine $V-I$ or $I-K$, and the two previous
steps were repeated until consistent results were achieved.  This process required 3 or fewer
iterations, and a typical error is 0.02\,mag or less.  For spatially unresolved binaries, the above 
process was repeated for each component until the flux ratio and the composite color were 
consistent (see \S2.2).

All astrometric and photometric data are listed in Table \ref{tbl1}.  The first and second columns 
list the white dwarf designations \citep{mcc99} and common names of all 90 targets.  The third 
column indicates if the target is spatially resolved into multiple components.  The fourth and fifth 
columns list the FWHM of the stellar PSF and the ratio of the semimajor to semiminor axis ($a/b$) 
of the radial profile fit.  For close visual pairs, the sixth and seventh column list the  angular 
separation and position angle of the secondary star.  The eighth and ninth columns give the 
F814W and Cousins $I$-band photometry for all single and composite point sources.  The 
final column contains information on the number and nature of the imaged sources.  Table 
\ref{tbl2} lists the coordinates of all primary (white dwarf) targets within the ACS images, 
some of which, as noted in the Table, have erroneous or inaccurate values in SIMBAD 
and the literature.

\subsection{Binary and Multiple Component Analysis}

\begin{figure}
\epsscale{1.2}
\figurenum{2}
\plotone{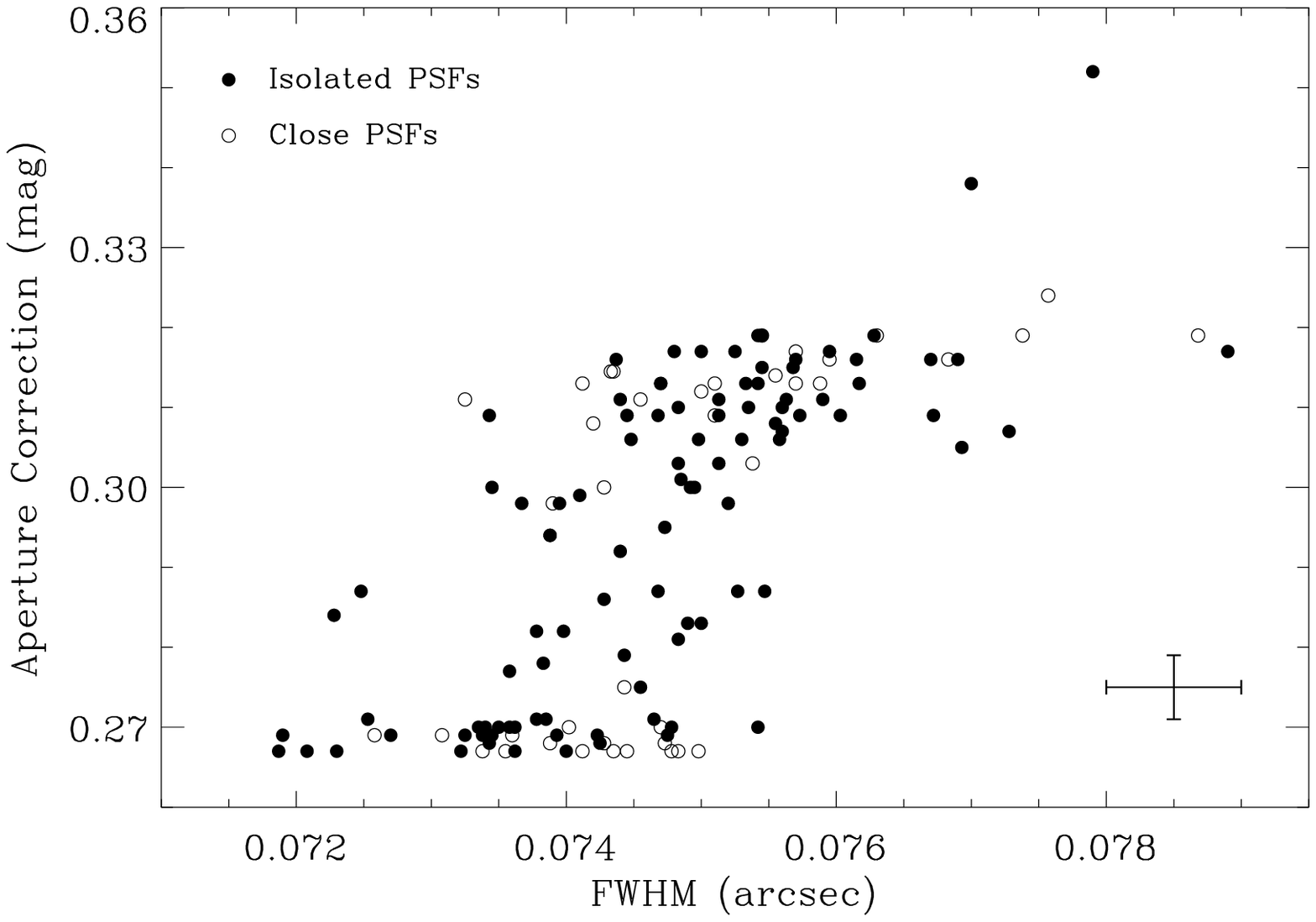}
\caption{The adopted, color-dependent aperture corrections in F814W Vega magnitudes for 
all 141 directly imaged white dwarf and companion point sources, plotted versus the measured
FWHM.  This correction is a reflection of the best available estimate of the color (spectral type) of 
each star, and is sometimes uncertain where little independent data exist.  The error bars at the
lower right display a $\pm0.02$\,pixel uncertainty in FWHM measurements, and an uncertainty 
in aperture correction corresponding to $\pm1$ spectral type at M4, with bluer objects having
smaller errors than this benchmark.  The filled circles represent stellar images that are well 
isolated from any companions, while the open circles are plotted for sources with companions 
within $0\farcs4$ (i.e.\ the two photometric apertures overlap).
\label{fig2}}
\epsscale{1.0}
\end{figure}

Component identification for spatially-resolved multiples imaged in a single filter is difficult, but 
fortunately the ACS F814W PSF contains color information \citep{sir05}.  Figure \ref{fig2} plots 
the applied, color-dependent, aperture correction (i.e.\ the expected $V-I$ color based on the 
best spectral type or temperature estimate) versus the measured FWHM.  The isolated white 
dwarfs occupy the region blueward of 0.28\,mag, and the isolated low mass stars all sit redward 
of 0.30\,mag, while the region between represents composite systems.  While the FWHMs for
different stellar types somewhat overlap, there is a clear trend as well as a number of outliers, 
some of which are certain or suspected binaries.

Combining the well-understood ACS PSF color trend \citep{sir05} with the expected $I$-band 
brightness for the components (based on photometric catalogs and the literature) yields an
unambiguous identification of the white dwarf and cool companion within all the ACS imaged, 
spatially-resolved binaries.  First, published ultraviolet-blue and near-infrared data for each pair 
provide a broad but good indication of the expected brightness for both the relatively hot and 
cool stars in the F814W filter.  Only in a single case were the brightnesses sufficiently close for 
identification to remain uncertain.  Second, for all spatially well-resolved binaries where the 
recognition of the white dwarf and M dwarf components is solid, the FWHM of the M dwarf is 
larger than that of the white dwarf by an average of 0.1\,pixel (a single exception is a suspected 
unresolved binary).

For each white dwarf, the literature was scoured for available optical photometry \citep{aba09,
cmc06,den05}, effective temperature, and surface gravity determinations \citep{mcc08,mcc99}.
Near-infrared photometry was taken from the Two Micron All Sky Survey (2MASS; \citealt{skr06}), 
while far- and near-ultraviolet fluxes were taken from archival {\em Galaxy Evolution Explorer} 
({\em GALEX}; \citealt{mar05}) data where available.  The {\em GALEX} fluxes were used to 
constrain the effective temperature in systems lacking a reliable estimate, via comparison with 
ultraviolet-optical flux ratios for stars of known temperature.  In ideal cases, SDSS $ug$-band 
or Johnson $UB$-band photometry could be combined with models to predict the white dwarf 
flux at longer wavelengths.  Generally, any photometric data unlikely to be contaminated by 
the cool companion was utilized to predict the white dwarf brightness at $VIJHK$.  For those 
systems in which the white dwarf was spatially resolved from its secondary, the model and 
measured $I$-band magnitude could be directly compared.  

Table \ref{tbl3} lists the adopted white dwarf parameters.  All model effective temperatures, 
surface gravities, masses, absolute magnitudes, and colors were taken from \citet{hol06,
fon01}.  Where a reliable surface gravity determination is lacking, 0.6\,$M_{\odot}$ was 
assumed, equivalent to $\log\,[\,g\,({\rm cm\,s}^{-2})\,]\approx8$ and represented by single 
decimal place precision in the third column.  The $V$-band magnitudes in the fourth column 
are values predicted for the white dwarf alone, based on model colors and photometry in 
bandpasses that are certainly or unlikely to be contaminated by the M dwarf. The photometric 
distances to the white dwarfs were derived from $V-M_V$.  

For spatially unresolved white dwarf -- M dwarf pairs, the $IJHK$ magnitudes of the M dwarf 
component were calculated by subtracting the expected white dwarf flux in those bandpasses
from the composite photometry.  If the M dwarf was spatially resolved from the white dwarf, its 
$I$-band magnitude was measured directly.  Tables \ref{tbl4} and \ref{tbl5} list the parameters
derived for all low mass companions.  Spectral type estimates were assigned based on $I-K$ 
color, with empirical absolute magnitude, color, and spectral type relations for M dwarfs taken 
from \citep{dah02,kir94,leg92}.  The photometric distances for these stars were calculated from 
the average of $I-M_I$ and $K-M_K$.

\begin{figure}
\epsscale{1.2}
\figurenum{3}
\plotone{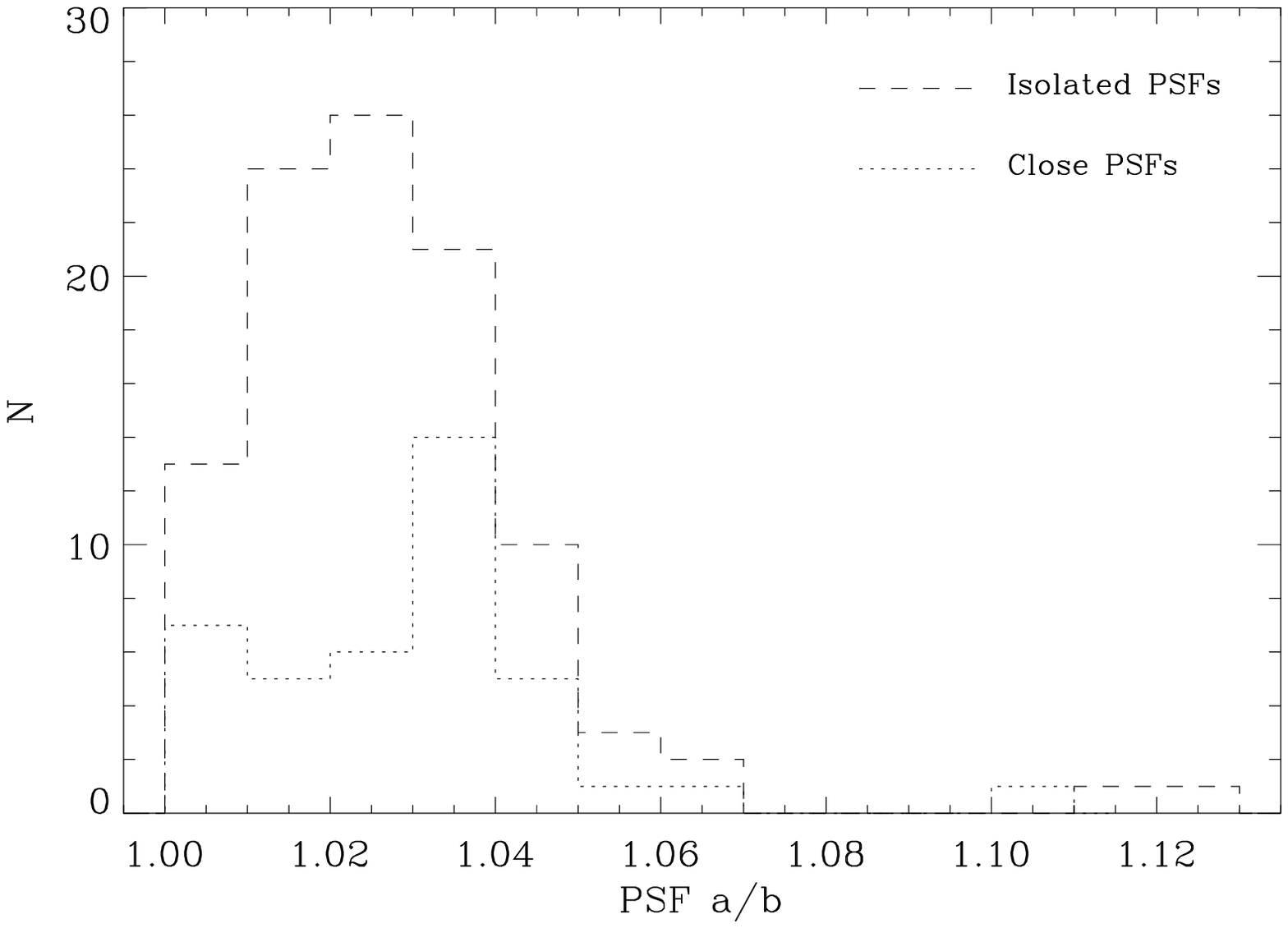}
\caption{Histogram of the PSF semimajor to semiminor axis.  The dashed line is for stellar 
images that are well isolated from any companions, while the dotted line is for sources with
companions within $0\farcs4$.
\label{fig3}}
\epsscale{1.0}
\end{figure}

\section{RESULTS}

Of the 72 certain or highly probable white dwarf binary systems, the ACS images spatially 
resolve 43 systems into two or more components, a fraction of 60\%.  Given that the spatial 
resolution of 2MASS is a few arcseconds, and the targets were all selected for near-infrared
excess in their 2MASS photometry, the wide binary fraction among white dwarfs with low 
mass companions is at least 60\%.  These results can be compared with the extensive 
ground-based study of \citet{far04a}, where it was found that white dwarf -- M dwarf binaries 
with angular separations smaller than $4''$ represent 63\% of all such systems.  Because the 
ACS program targeted stars in this latter group, one can estimate (all else assumed equal) 
that 75\% ($0.37+(0.60\times0.63$)) of all low mass companions to white dwarfs are widely 
separated and never experienced a common envelope phase.  It then follows that around 
25\% of all such binaries are post-common envelope systems, consistent with the findings 
of \citet{reb07} based on a similar number of SDSS spectra.

\subsection{Three Very Close Visual Doubles}

Figures \ref{fig2} and \ref{fig3} demonstrate there are a few sources whose FWHM or 
$a/b$ values are clear outliers.   The three objects with $a/b>1.1$ in Figure \ref{fig3} are 
particularly interesting and are just discernibly (by human eye) elongated with respect to 
their visual companions.  Each of these three sources is a visual M dwarf companion to a 
white dwarf, whose PSF is point-like and narrow in comparison.  With little difficulty, {\sf 
daophot} was able to fit two sources to each of these elongated PSFs, confirming they are 
all barely resolved binaries at angular separations around $0\farcs03$ (the separations of 
these three close doubles were underestimated in Paper I).  Photometry and astrometry for 
these three binary M dwarfs is listed in Table \ref{tbl6}; these measurements demonstrate 
that binaries with $\Delta I \la1$\,mag and with separations as small as 1\,pixel were readily 
detectable in this survey.

\subsection{Photometric Distances:  A Need for Better Agreement}

In order to convert angular separation in arcseconds into a projected orbital separation in 
AU, a distance estimate is necessary; only four stars in this survey have reliable trigonometric 
parallaxes \citep{mcc08}.  Unfortunately, the white dwarf and M dwarf components within each 
binary often give discrepant distance estimates when obtained using other methods.  To 
measure this effect, define the quantity:

\begin{equation}
\Delta(m-M) = (m-M)_{\rm wd} - (m-M)_{\rm rd}
\end{equation}

\noindent
There is a significant mismatch in photometric distances when $|\Delta (m-M)| > 0.5$\,mag, 
corresponding to over 25\% disagreement between distance estimates.  In this view, exactly 
half of the binaries without parallaxes have significant discrepancies in the estimates obtained
from each component.

Figure \ref{fig4} plots a histogram of this distance modulus discrepancy, demonstrating that 
the most common type of mismatch is for $\Delta (m-M) > +0.5$\,mag; i.e.\ the white dwarf 
appears to be further from the Sun and overluminous relative to the M dwarf.  In comparing 
parameters derived from simultaneous spectral fitting of both white dwarf and M dwarf 
components in a composite spectrum versus fitting only the white dwarf (as generally done 
in the literature for the systems in this study), \citet{tap09} find this same bias.  They conclude 
that when Balmer lines are uncorrected for the M dwarf contribution, the resulting equivalent 
widths are underestimated, leading to an overestimated temperature and luminosity.  There 
is little doubt this effect is present here, and it underscores the need for reliable and robust 
white dwarf parameter determinations for these systems.

Another source of error in the photometric distances lies in the fact that up to one dozen 
white dwarfs in this study have little or no observational constraints on their effective temperatures 
or surface gravities \citep{mcc08}.  While there is intrinsic scatter in the absolute magnitudes of M 
dwarfs \citep{kir94,leg92} that depends primarily on metallicity (difficult to measure for cool stars), 
it is likely that many of the distances discrepancies would be improved with a standard optical 
spectral analysis of the white dwarfs.

There are four distinct types of photometric distance mismatches and each is worth brief mention:

\begin{figure}
\epsscale{1.2}
\figurenum{4}
\plotone{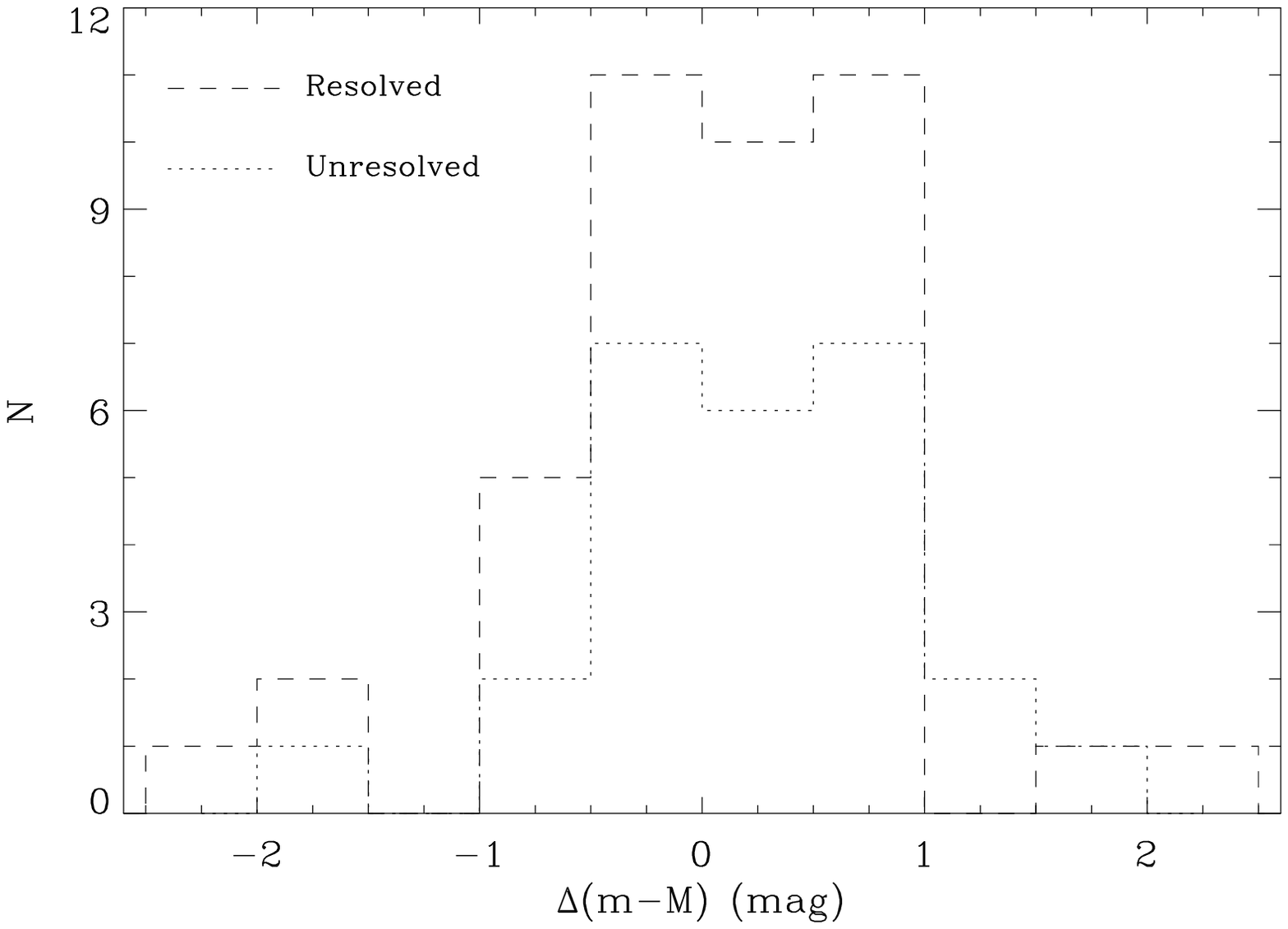}
\caption{Histogram of the difference in distance moduli between the white dwarf and M 
dwarf components of all such binaries.
\label{fig4}}
\epsscale{1.0}
\end{figure}

\begin{enumerate}

\item{
$\Delta (m-M) > +0.5$\,mag visual binaries; 14 systems.  These may be explained if the 
widely separated M dwarf is itself double or the white dwarf luminosity is overestimated 
(higher mass or lower temperature); the latter is plausible given the optical spectroscopic 
contamination by the companion.}

\item{
$\Delta (m-M) > +0.5$\,mag unresolved doubles; 12 systems.  Similar to type 1), but 
binarity in the M dwarf component itself (i.e.\ a spatially unresolved triple system) is 
unlikely.}

\item{
$\Delta (m-M) < -0.5$\,mag visual binaries; 6 systems.  Somewhat puzzling, these cases
suggest the white dwarf has an underestimated luminosity and may be binary or hotter 
than current estimates.  A somewhat underluminous M dwarf is also possible.}

\item{
$\Delta (m-M) < -0.5$\,mag unresolved doubles; 3 systems.  Similar to 3), a relatively 
low mass white dwarf can account for this type of mismatch, which is a realistic possibility 
for these post-common envelope systems.}

\end{enumerate}

\begin{figure}
\epsscale{1.2}
\figurenum{5}
\plotone{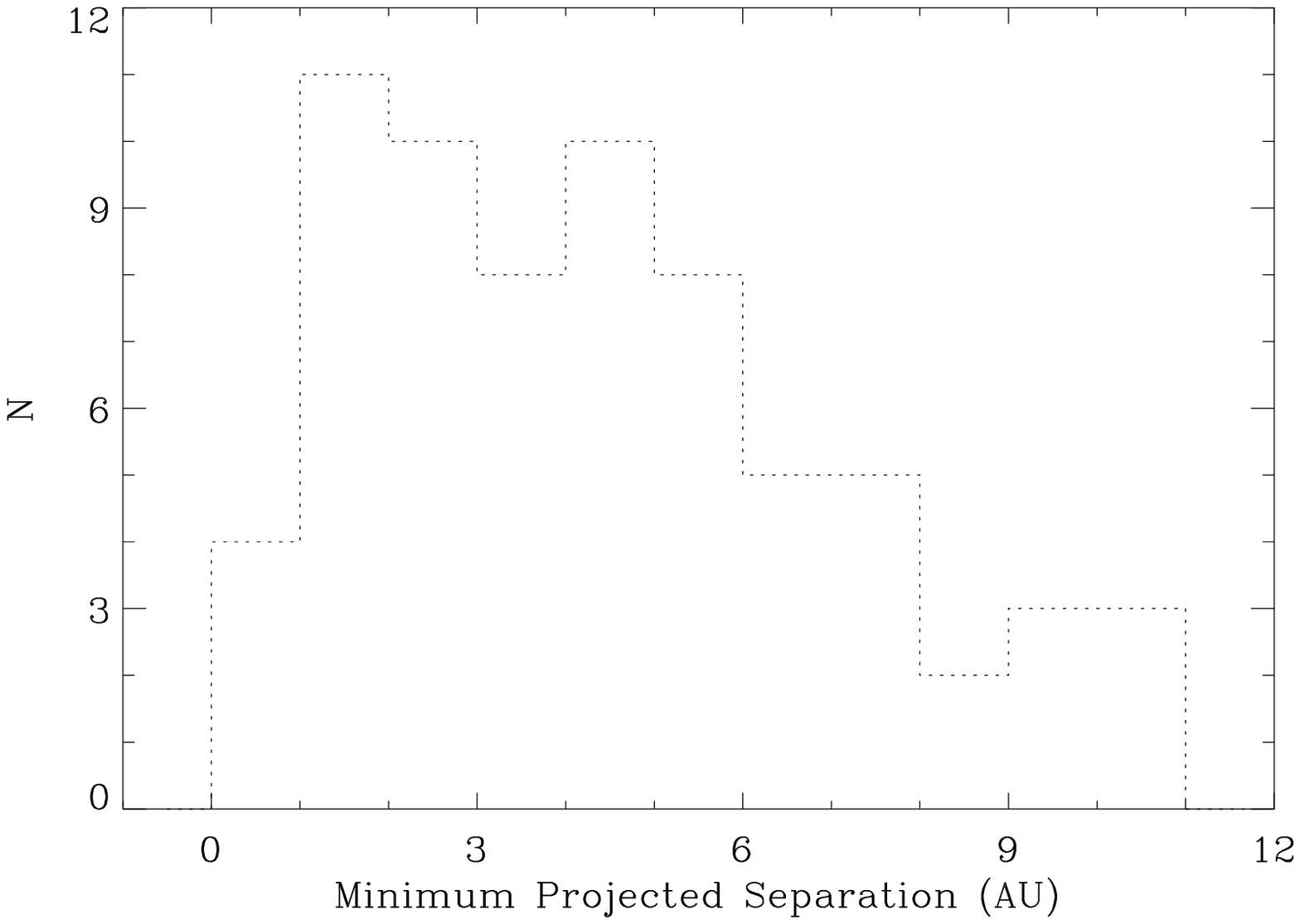}
\caption{Histogram of minimum projected separations to which this study was sensitive for all 
white dwarf -- M dwarf binaries, assuming an angular separation of a single ACS / HRC pixel 
of $0\farcs025$.
\label{fig5}}
\epsscale{1.0}
\end{figure}

\subsection{Distribution of Projected Orbital Separations}

All double star parameters are listed in Table \ref{tbl7}, triples are listed once for each 
unique pair.  The adopted distance to the system was taken to be the average of the white
dwarf and M dwarf photometric distances.  Column eight lists the projected separation in AU 
for all visual doubles, which is actually the minimum value for the binary semimajor axis (for
zero inclination).  Column nine lists an upper limit to the binary semimajor axis for all spatially
unresolved doubles, assuming an angular separation less than $0\farcs025$ or 1 ACS / HRC 
pixel; given the three detections in Table \ref{tbl6} (see also \S3.1), this is likely to be a safe 
assumption.

Figure \ref{fig5} plots a histogram of this limiting 1-pixel resolution multiplied with the adopted 
distance for all 72 white dwarfs with low mass companions.  From this plot and Table \ref{tbl7}, 
it is clear the observations were sufficiently sensitive to detect binaries with semimajor axes 
down to $1-2$\,AU; the Table \ref{tbl6} double M dwarfs have projected separations ranging 
from 2 to 9\,AU.

\begin{figure}
\epsscale{1.2}
\figurenum{6}
\plotone{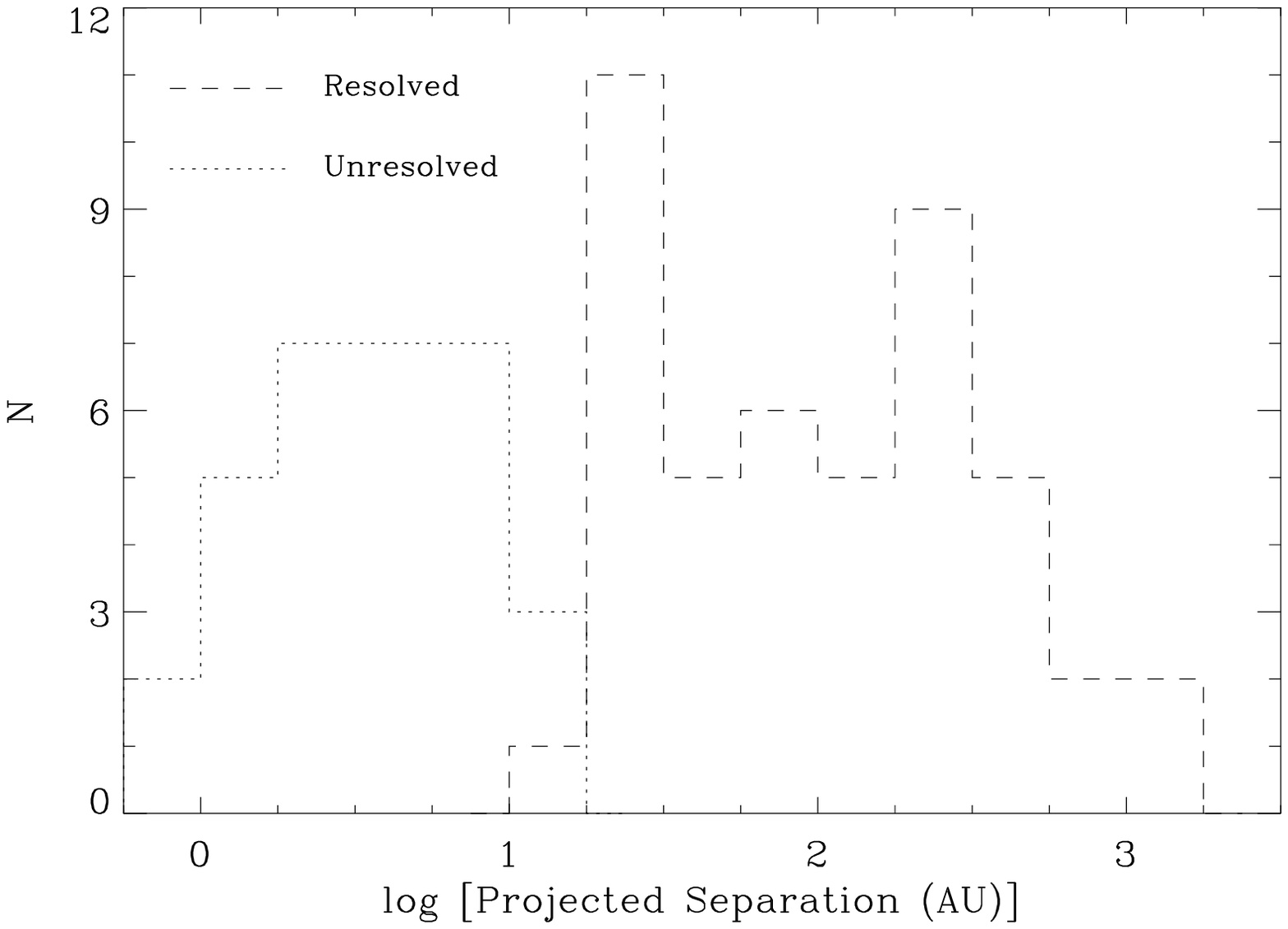}
\caption{Histogram of spatially-resolved (minimum, dashed line) and -unresolved (maximum, 
dotted line) orbital separations for all 72 white dwarf -- M dwarf binary and triple systems in the 
study.  The spatially-unresolved pairs were assumed to have separations less than a single ACS
/ HRC pixel of $0\farcs025$.
\label{fig6}}
\epsscale{1.0}
\end{figure}

Figure \ref{fig6} plots histograms of the minimum and maximum orbital separations for all 
double components within these 72 white dwarf -- red dwarf systems.  This is the main result 
of this work:  there is essentially no overlap between these two populations and the region 
between a few to one dozen AU is empty for low mass companions to white dwarfs.  While 
the displayed histograms overlap slightly due to the plot resolution, they represent minimum 
and maximum values for orbital separations, and hence any overlap does not indicate the 
two populations occupy the same region.  The actual values, for which the unresolved 
systems would shift to smaller separations and the resolved systems would shift to larger 
separations, almost certainly do not overlap by any significant amount.

\begin{figure}
\epsscale{1.2}
\figurenum{7}
\plotone{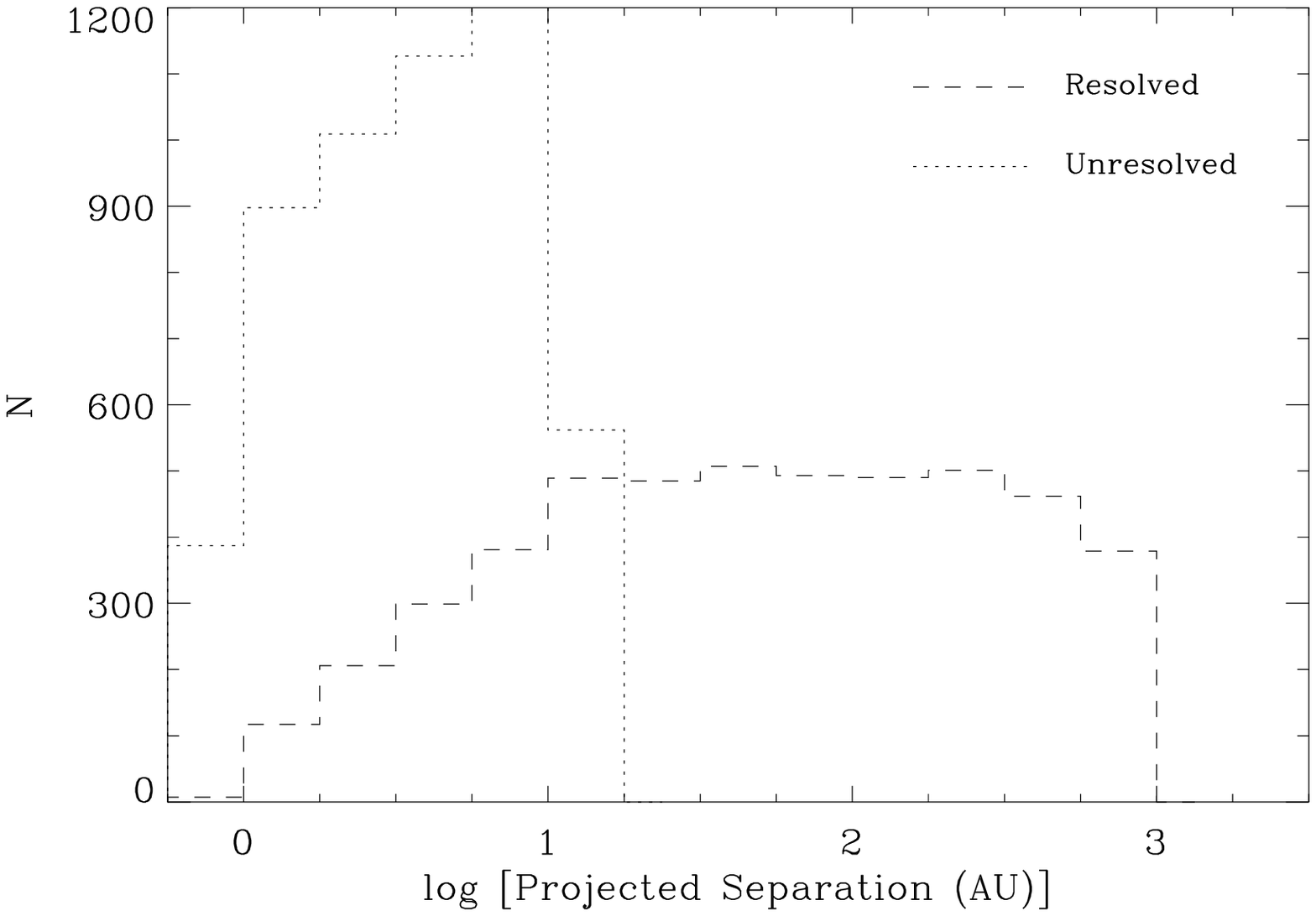}
\caption{Same as Figure \ref{fig6} but for a Monte Carlo-simulated population of 10,000 binaries 
with a {\em uniform} distribution of orbital separations (see \S3.7 for details).  For each simulated 
binary, if the projected angle on the sky is less than a single ACS / HRC pixel, then the system is 
counted as spatially-unresolved, and plotted with an upper limit to its actual separation.  If the 
angular separation is greater than or equal to $0\farcs025$, the binary is counted as spatially-$
$resolved and plotted with this lower limit to its actual semimajor axis.  The simulations clearly 
demonstrate that binaries in the few to several AU range were detectable in the ACS study.  A 
K-S test supports the conclusion that the actual data exhibit a bimodal distribution in semimajor 
axis at 99.95\% confidence.
\label{fig7}}
\epsscale{1.0}
\end{figure}

To further test this conclusion, Monte Carlo simulations were performed.  A sample of 
10,000 binaries were randomly selected from a population with the following characteristics:  
1) a logarithmically-uniform distribution of semimajor axes between 0.01 and 1000\,AU, 2) a 
logarithmically-uniform distribution of distances from the Sun between 30 and 500\,AU, 3) a 
uniform distribution of orbital inclinations with respect to the plane of the sky, and 4) a uniform 
distribution of orbital phases within the binary reference frame.  For simplicity, the orbits were
assumed to be circular.  Each randomly selected binary was projected onto the plane of the
sky and its angular separation was calculated.  If this angle was smaller than 1 ACS / HRC
pixel, the binary was deemed to be spatially-unresolved and an upper limit to the orbital 
separation was calculated as above.  If instead the angle on the sky was greater than or 
equal to $0\farcs025$, the binary was declared to be spatially-resolved and a projected 
separation was calculated.  The results of the simulations are plotted in Figure \ref{fig7}
in precisely the same manner as the actual data in \ref{fig6}.

The simulated data describe a population that is clearly distinct from the observed binary
population, despite representing the same (potential) range of semimajor axes, distances 
from the Sun, and angular detection criterion.  A Kolmogorov-Smirnov (K-S) test was performed 
on observed and simulated samples of spatially-resolved binaries, resulting in a probability of 
99.95\% that the two populations are not related.  These results demonstrate that the actual 
population of white dwarf plus low mass stellar companions does not have a uniform distribution 
of orbital semimajor axes, but is a population with a deficit of systems with semimajor axes below 
20\,AU.  From radial velocity studies sensitive to the other end of the distribution, the observed 
population also has a distinct lack of binaries with periods longer than 10\,d \citep{reb08,mor05}.  
Therefore, the {\em HST} / ACS observations complete an empirical picture of the bimodal 
semimajor axis distribution among white dwarfs with low mass, unevolved (stellar or substellar) 
companions.

\subsection{Companion Sensitivity}

Regarding the sensitivity to spatially-resolved companions as a function of angular separation
and contrast ratio, the study was designed to be intrinsically robust in the following manner. For
the 72 systems where a reliable near-infrared excess is associated with the white dwarf, in all 
but one instance there is a corresponding, measurable $I$-band excess.  In each case, this 
excess was predicted on the basis of the near-infrared photometry, and hence the choice of 
the F814W filter to exploit the expected favorable contrast.  These 71 systems were either 1) 
spatially resolved into two or more components and the dM companion was directly detected, 
or 2) measured to have a clear ($I_{\rm ACS} - I_{\rm WD} > 0.5$\,mag) photometric excess 
indicating the presence of the dM companion.  Thus, among these targets, all previously-$
$suspected companions have been directly or indirectly detected; none were missed.

The single exception is 0145$-$221 where the L dwarf companion is too cool to reveal itself 
via excess emission at F814W \citep{far04b}.  However, this system has recently been found 
to be a short-period binary and is therefore not expected to be directly imaged in the ACS 
observations (see Appendix for further details).  Apart from this one case, the remaining 71 
systems all have $\Delta I < 3.0$\,mag between the white dwarf and M dwarf components, 
with all but a few having contrasts milder than 10:1 in flux.  One of the companions to 1833$
+$464 is readily detected at $0\farcs08$ and $\Delta I = 1.8$\,mag, while two of the three close 
M dwarf pairs in Table \ref{tbl6} have $0\farcs03$ and $\Delta I \approx 1$\,mag.

\subsection{Distribution of Companion Spectral Types}

\begin{figure}
\epsscale{1.2}
\figurenum{8}
\plotone{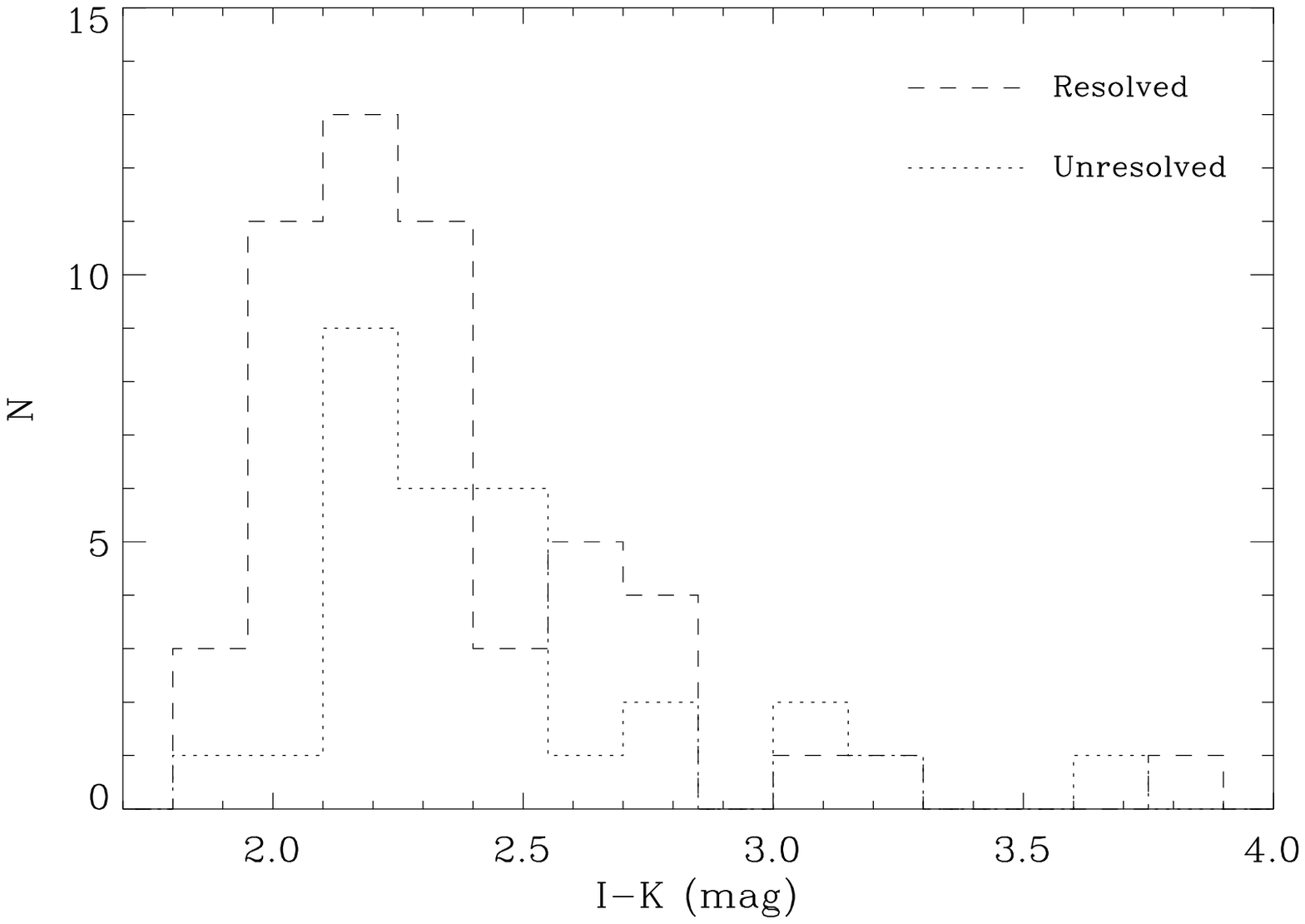}
\caption{Histogram of the $I-K$ color determinations for all M dwarf companions observed in
the program.  The populations are similar except for a relative dearth of the bluest colors among
the spatially-unresolved companions.
\label{fig8}}
\epsscale{1.0}
\end{figure}

\begin{figure}
\epsscale{1.2}
\figurenum{9}
\plotone{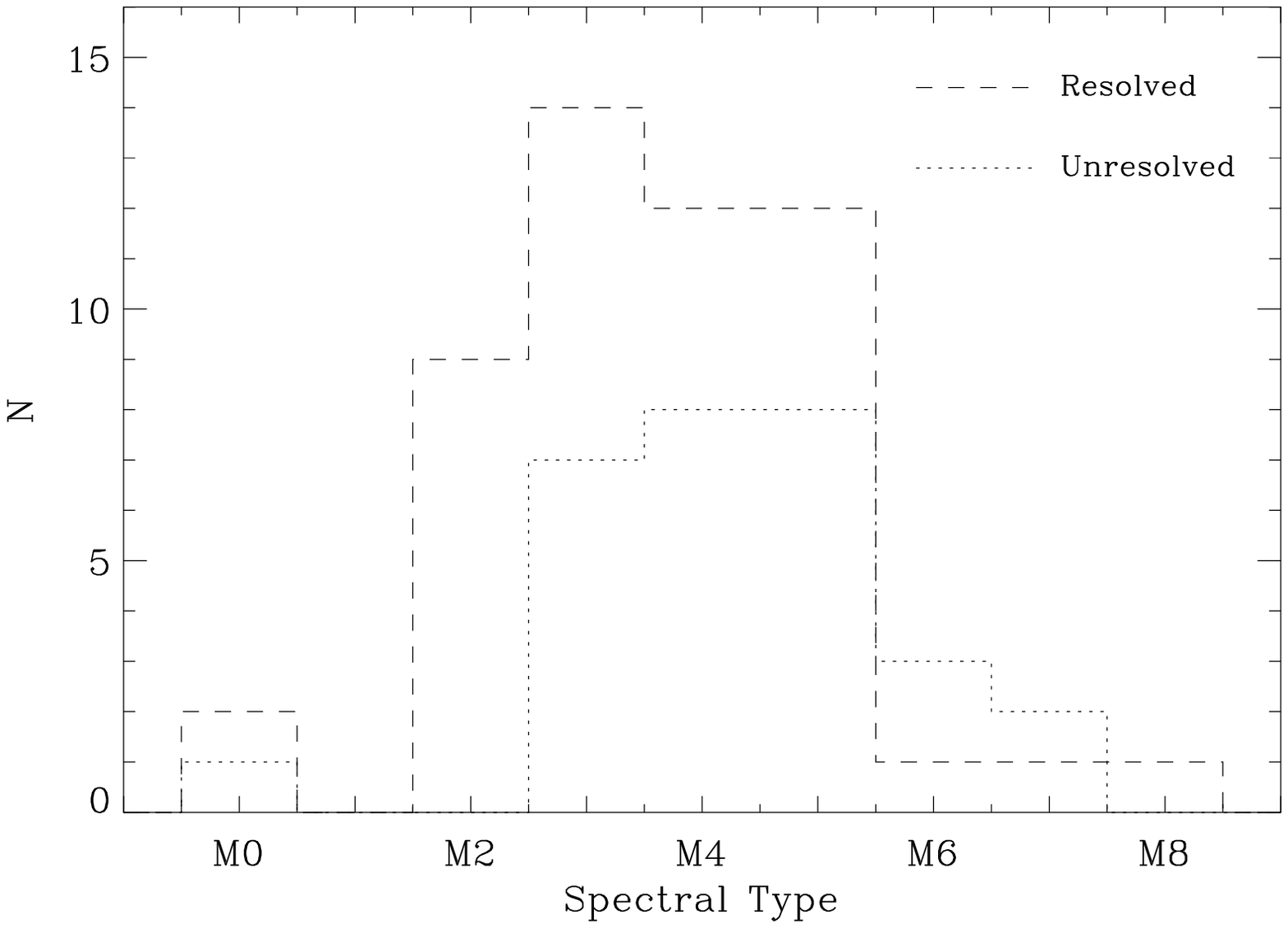}
\caption{Histogram of M dwarf companion spectral types inferred from $I-K$ colors.  As seen in
the previous figure, there are relatively fewer early M dwarfs among the spatially-unresolved
companions.
\label{fig9}}
\epsscale{1.0}
\end{figure}

Figures \ref{fig8} and \ref{fig9} plot the relative frequency of M dwarf companions as a 
function of their $I-K$ color and spectral type inferred from this index.  The two populations of 
spatially-resolved and -unresolved companions appear generally similar, and agree well with 
the distribution found by \citet{far05a} with a peak around M3.5.  Interestingly, there appears to 
be a relative dearth of early M dwarfs among the spatially-unresolved companions.  The white 
dwarf primaries in the sample are not sufficiently well-characterized to instill high confidence in 
their predicted $I$-band fluxes necessary to derive the $I-K$ colors of the spatially unresolved 
M dwarfs.  On the other hand, the distribution of spectral types based on these derivations 
otherwise matches expectations based on the distribution of field M dwarfs \citep{far05a,
cru03} and that of the spatially-resolved companions.

K-S tests were employed to compare the cumulative distributions of $I-K$ color and 
spectral type between the spatially-resolved and -unresolved systems.  The results yielded 
$P$ values of 0.12 in $I-K$ color and 0.37 for spectral type, implying the two datasets are 
most likely distinct.  The $P$ value for spectral type is understandably higher as the raw $I-K$ 
data have been smoothed during transformation into spectral type.  The data are not sufficient 
to conclude with certainty if the measured difference is the result of small number statistics, 
errors in deconvolving the $I$-band photometry for spatially-unresolved pairs, or real.  It is 
noteworthy that with a similar yet much larger sample of low mass, main-sequence companions 
to white dwarfs from the SDSS, \citet{sch10} find the distribution of spectral types for close (post-$
$common envelope) and wide secondaries to be distinct with high confidence.

Again, the primary deviation between the subsamples here is an apparent deficit of earlier 
M types among the spatially-unresolved secondaries.  While some caution is warranted, this 
putative trend is {\em exactly the opposite} of what would be expected if the secondary stars 
accreted a significant amount of mass during the common envelope phase.  This dataset is 
insufficient to rule out or uncover mass transfer at the several to tens of percent level, but it is 
clear that $0.1-0.2\,M_{\odot}$ stars do not emerge from the common envelope with $0.3-0.5
\,M_{\odot}$.  This result emphasizes the fact that mass transfer via Roche lobe overflow from 
a higher mass to lower mass star is {\em unstable} \citep{pac76}.

\subsection{Confirmed and Candidate Close Binaries}

From these results, and as noted in Table \ref{tbl7}, all 29 spatially-unresolved binaries 
are predicted to have short orbital periods; this can be easily tested with radial velocity or 
photometric monitoring.  Some of these systems have additional evidence that favors close 
binarity, such as variable Balmer line emission, yet none have established periods \citep{sil06,
lie05a,sil05,koe01,ste01,max00,put97,hut96,sch96,rei96}.  Since 2005 when this study was 
initiated, four of these 29 candidate short period binaries have been confirmed via independent 
studies: 0145$-$221, 1247$-$176, 1334$-$326, and 2009$+$622 (Burleigh et al.\ 2010, in
preparation; \citealt{tap09,saf93,mor05}).  The short-period binary 1026$+$002 \citep{saf93}
went unrecognized during target selection, and was included in the ACS imaging program.  
Table \ref{tbl7} lists only those systems with established periods as known close binaries, 
because variability due to activity is often seen in widely bound systems (nine of these are 
identified in this study and discussed in the Appendix).

\subsection{Non-Binaries and Misidentifications}

Table \ref{tbl8} lists 16 ACS targets that are no longer considered white dwarf -- M dwarf 
binary candidates.  These consist of: 1) high-risk targets, white dwarfs whose 2MASS data 
suggested only modest $H$- or $K_s$-band excess (which might have indicated a substellar
companion), 2) low mass stars incorrectly classified as white dwarfs, and 3) white dwarfs 
co-identified with nearby, bright 2MASS sources (including wide visual companions).  Two 
additional systems (1156$+$132, 2237$-$365) were not imaged because the white dwarf 
fell outside of the narrow ACS field of view due to inaccurate coordinates in SIMBAD and 
the literature.  Each of these 18 systems is briefly discussed in Paper I or the Appendix.

\subsection{Two Dwarf Carbon Companions}

Two white dwarfs in this study have low mass stellar companions which are carbon-enriched 
counterparts to M dwarfs; 0824$+$288 and 1517$+$502 \citep{lie94,heb93}.  The latter binary
is spatially unresolved and a strong candidate to have a short orbital period, consistent with 
the paradigm that dwarf carbon stars are the product of mass transfer from a carbon-rich giant 
\citep{dah77}.  The former, however, is a spatially-resolved triple system consisting of a wide 
M dwarf secondary and a close visual tertiary.  Assuming the tertiary in 0824$+$288 is the 
dwarf carbon star, the lower limit for its semimajor axis is 14\,AU.  Therefore mass transfer via 
Roche lobe overflow is effectively ruled out by the ACS observations, presenting something of 
a challenge to the binary origin of its carbon-polluted exterior.  One way to resolve this is to
invoke effective wind capture when the white dwarf progenitor star was an asymptotic giant.  
Prior to the bulk of mass loss during this phase, the binary would have orbited more closely 
by a factor of a few, putting the dwarf carbon star progenitor as close as $5-7$\,AU.  If the 
giant envelope extended to a few AU, its wind could potentially pollute a companion orbiting 
at this semimajor axis.

The dwarf carbon stars were once thought to be a class of halo stars, but these two with 
white dwarf primaries appear to have disk-like kinematics.  Both systems have modest 
proper motions corresponding to tangential speeds of $40-70$\,km\,s$^{-1}$ \citep{zac05}.  
The prototype carbon dwarf and halo star G77-61 has very little {\em GALEX} near-ultraviolet 
flux and no detection in the far-ultraviolet, making it unlikely that any putative white dwarf
companion \citep{dea86} could be warmer than 5000\,K.

\subsection{Wind Capture in Close Binaries}

\citet{zuc03} searched for calcium K absorption lines in the high-resolution spectra of 17 DA 
white dwarfs with known M dwarf companions that were spatially unresolved in ground-based
observations.  Emission from active secondaries prevented a useful search in seven of these 
cases, and five of the remaining ten white dwarfs were found to be metal-polluted (the absorption 
feature in 1210$+$464 is uncertain and this star may not be DAZ; B. Zuckerman 2010, private 
communication).  This result suggests that wind from low mass stellar companions, when present,
can readily pollute the atmosphere of white dwarfs \citep{deb06}.  Eight of the stars observed by 
\citet{zuc03} were imaged with ACS in the current survey.

The binaries 0354$+$463, 0752$-$146, 1026$+$002 are all spatially unresolved and therefore 
likely to be in short period orbits.  In fact, the latter system is a known 14\,hr radial velocity variable
\citep{saf93}, and contains a DAZ white dwarf.  The other two systems should now be considered 
strong candidates for short period systems; the first has a DAZ primary while the second displays 
prominent calcium emission from the M dwarf and was indeterminate \citep{zuc03}.

The doubles 0034$-$211, 0347$-$137, 0933$+$025, 1049$+$103, 1210$+$464 are all spatially 
resolved and hence have wide orbital separations where wind from the M dwarf cannot effectively 
pollute the white dwarf.  Surprisingly then, 1049$+$103 is a DAZ white dwarf.  There is no evidence
of an unseen, closely orbiting third component in this system; this would have to be a very low mass
star or brown dwarf but {\em Spitzer} studies reveal no evidence for these types of companions at 
metal-polluted white dwarfs \citep{far09a}.  The white dwarf primary is rather warm at 20\,600\,K 
\citep{lie05a}, and is currently accreting metals at $1.1\times10^8$\,g\,s$^{-1}$ \citep{koe09}, 
suggesting the likely presence of circumstellar material.
  
\begin{figure}
\epsscale{1.2}
\figurenum{10}
\plotone{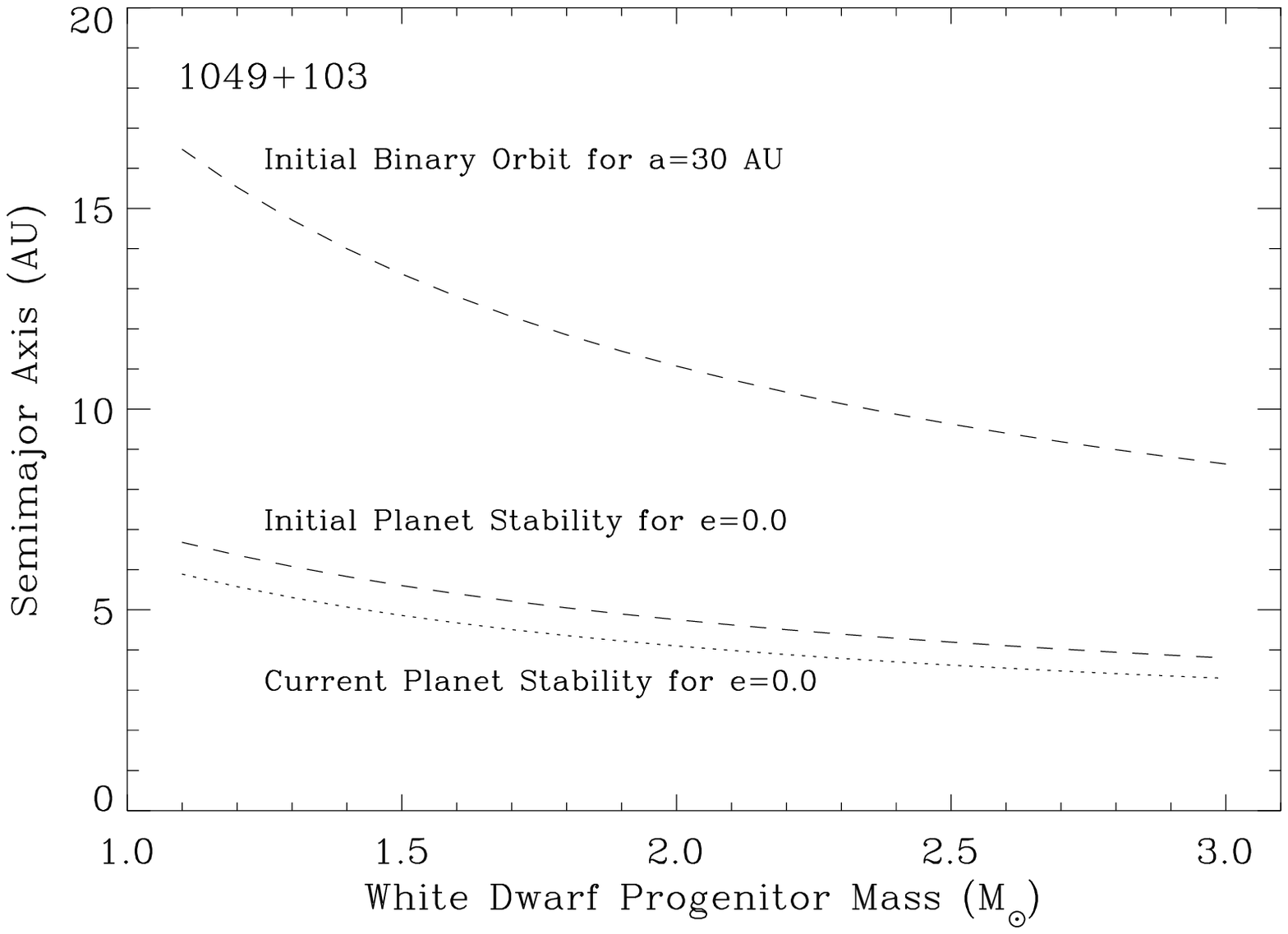}
\caption{The initial orbit of PG\,1049$+$103 plotted as a function of the white dwarf progenitor
mass, using the initial to final mass relationship \citep{dob09}.  Also plotted is the limit for stable 
planetary orbits \citep{hol99} in the initial and current configurations.  Because the orbits of any
planets initially wider than a few AU would expand by a factor of a few, it is unlikely they would 
now persist at the white dwarf.
\label{fig10}}
\epsscale{1.0}
\end{figure}

When both components of 1049$+$103 were on the main sequence, the orbital separation was
smaller by a factor of few, and any surviving planetary system that might pollute the white dwarf 
would have to be contained within a relatively narrow region \citep{hol99}.  The asymptotic giant
predecessor directly engulfed the innermost regions around the white dwarf, while orbits outside
its maximum extent (including the orbit of the companion star) expanded by a factor of a few.  Also
in response to the change in binary mass ratio post-main sequence, the maximum radius for stable 
planetary orbits decreased.  Therefore, the M dwarf companion currently restricts stable planets at 
the white dwarf to lie in a region that should be mostly or totally empty.  Figure \ref{fig10} illustrates 
this; barring changes in eccentricity, there are no stable planetary orbits over the main and 
post-main sequence lifetime of the binary, leaving the nature of the DAZ white dwarf in this 
system somewhat of a puzzle. 

\subsection{Systems with Potential Orbital Motion}

There are 13 systems imaged by ACS that have projected separations within roughly 30\,AU.
If these separations are not significantly different from the actual semimajor axis, then many of 
these binaries should execute one-quarter of an orbit in less than 50\,yr.  While this is still quite 
long for monitoring short-term, near zero inclination systems will already exhibit measurable 
changes in position angle ($\Delta \theta >9\arcdeg$) since they were imaged with ACS in 
2004 and 2005.  Full orbital and stellar parameters for these systems are possible within a 
human lifetime.

\section{IMPLICATIONS AND DISCUSSION}

Under the assumption that all of the spatially-unresolved binaries are in short period orbits, one 
can make a crude estimate of the change in orbital energy of the system post-main sequence.  
For the following, assume a secondary mass of $0.2\,M_{\odot}$, a white dwarf mass of $0.6\,M_
{\odot}$, and a main sequence progenitor mass of $1.8\,M_{\odot}$.  If the system avoids a common
envelope phase, it expands by a factor of 2.5 \citep{jea24} following the main sequence, implying 
initial orbits of 8\,AU for companions which are now near 20\,AU, of which there are several in this 
study.  Applying the same logic to the smallest projected separation of 14\,AU, it may have been 
initially as close as 5.5\,AU yet effectively avoided drag forces due to the giant envelope.

On the other hand, a typical post-common envelope system has a period near 0.5\,d \citep{mor05}.  
The change in specific orbital energy between a main-sequence binary (as described above) in a 
one to few AU orbit, and a post-main sequence binary in a 0.5\,d orbit is essentially independent 
of the initial semimajor axis for $a_0>0.5$\,AU, and is around $3\times10^{14}$\,erg\,g$^{-1}$.  If 
one then deposits all of this energy into $1.2\,M_{\odot}$ of slowly expanding hydrogen atoms during 
the lifetime of the giant envelope, this imparts roughly 250\,km\,s$^{-1}$ of kinetic energy per atom; 
sufficient to escape $2.0\,M_{\odot}$ from an initial distance of 0.1\,AU.  While this calculation is 
simple, it demonstrates that the deposition of orbital energy into the envelope is commensurate 
with the short periods found among white dwarfs with low mass stellar and substellar companions.

Given this likely trade-off, and all else being equal, one would expect the lowest mass companions 
to end up in the shortest period orbits and higher mass companions at longer periods.  There is
likely not enough empirical data to test this in real systems with a continuum of initial orbits, primary
and secondary masses.  For higher mass secondaries (K dwarfs and brighter) in close orbits, the 
white dwarf would likely go unrecognized, but this is a very interesting phase space.  For example,
take the Sirius system: with a semimajor axis of 20\,AU and a likely white dwarf progenitor mass of
$5.0\,M_{\odot}$, the initial semimajor axis was only 4\,AU \citep{lie05b}.  The Sirius binary has an
orbital eccentricity of 0.6 and the white dwarf progenitor should have directly engulfed Sirius A for 
a significant portion of its giant evolution, yet did not cause the binary orbit to shrink at all.

Lastly, there may be one example of a white dwarf with a low mass companion in an orbit that
could be described as intermediate.  G77-61 is the prototype dwarf carbon star, and \citet{dea86}
reported a radial velocity period of 245\,d due to an unseen companion of $M\sin{i}\approx0.55\,
M_{\odot}$.  The mass of this carbon-polluted halo star is uncertain, but its effective temperature
suggests a possible analogy with a late K dwarf or early M dwarf.  If the unseen companion is 
indeed a very cool white dwarf as suggested by the data, then G77-61 is either the longest period 
system known to emerge from a common envelope, or the shortest period system known to have 
avoided one.

In conclusion, the results of an ACS HRC imaging Snapshot survey reveal a bimodal distribution 
of orbital separations for low mass, unevolved companions to white dwarfs.  These data support
the theoretical picture where low mass stellar or substellar companions spiral inward during a 
common envelope phase, or migrate outward if the initial orbital separation is sufficient to avoid 
contact with the asymptotic giant envelope \citep{nor10,sch96,dek93}.  All spatially-unresolved 
binaries in this study are predicted to be in short period orbits amenable to radial velocity and 
photometric variability studies.  The distribution of M dwarf companion spectral types argues 
against any significant mass transfer during the common envelope, indicating the current masses 
of the secondaries are essentially -- to within a few tens of percent or better -- the same as their 
masses at formation.

\acknowledgments

The authors thank the anonymous referee for a valuable and thorough report that improved 
the quality of the manuscript, and D. Koester for the use of his white dwarf spectral models.
This work is based on observations made with the {\em Hubble Space Telescope} which is 
operated by the Association of  Universities for Research in Astronomy under NASA contract 
NAS 5-26555.  Support for Program number 10255 was provided by NASA through grant 
HST-GO-10255 from the Space Telescope Science Institute.  This publication makes use of 
data products from the Two Micron All Sky Survey, which is a joint project of the University of 
Massachusetts and the Infrared Processing and Analysis Center / California Institute of 
Technology, funded by NASA and the National Science Foundation.  This work includes data 
taken with the NASA Galaxy Evolution Explorer, operated for NASA by the California Institute 
of Technology under NASA contract NAS5-98034.  Some data presented herein are part of 
the Sloan Digital Sky Survey, which is managed by the Astrophysical Research Consortium 
for the Participating Institutions (http://www.sdss.org/). 

{\em Facility:}	\facility{{\em HST} (ACS)}

\section{APPENDIX:  NOTES ON INDIVIDUAL OBJECTS}

{\em 0023$+$388}.  The white dwarf parameters derived by \citet{tre07} imply a distance of 
59\,pc, and are consistent with the {\em GALEX} far- and near-ultraviolet fluxes and $V$-band 
photometry \citep{mcc08}.  This is roughly compatible with the estimated 70\,pc distance to the 
spatially-unresolved dM5.5 tertiary, but the dM3 (i.e.\ $V-K\approx5.0$) secondary star has 
a photometric distance of only 43\,pc.  If the $24\farcs6$ distant, common proper motion 
companion is itself a near equal luminosity binary, then all three distance estimates would 
become amenable.  While the widely bound secondary star did not fall within the ACS field 
of view, this system may have a fourth component.

{\em 0034$-$211}.  \citet{zuc03} report H$\beta$ and Ca K line emission in the composite
spectrum.  The ACS images reveal the binary is widely separated, and hence for a system
without additional components, the emission must originate from activity intrinsic to the M 
dwarf.

{\em 0116$-$231}.  There does not exist a reliable effective temperature estimate for the white
dwarf, but it is a soft X-ray source \citep{odw03} and hotter than DA3 as listed in \citet{mcc99}.
\citet{wol96} give a lower limit of 24\,000\,K based on pointed X-ray observations, and it is bright
in both {\em GALEX} ultraviolet bandpasses.

\begin{figure}
\epsscale{1.2}
\figurenum{11}
\plotone{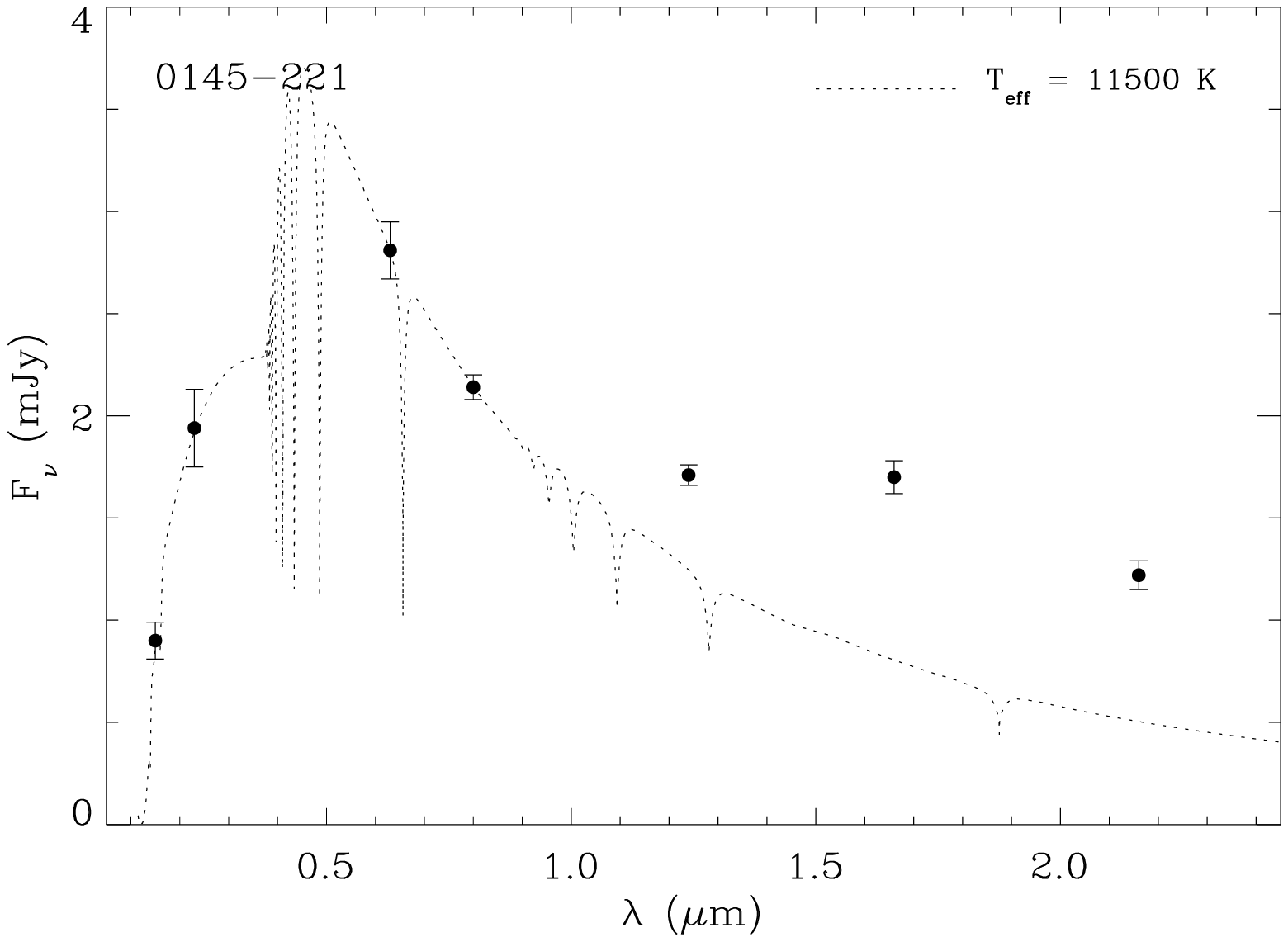}
\caption{Spectral energy distribution (SED) of GD\,1400 overplotted with an 11\,500\,K DA 
model.  The photometric data are {\em GALEX} far- and near-ultraviolet, optical $r'$, ACS 
(transformed) $I$, and 2MASS $JHK_s$.  There is no evidence for the L dwarf companion 
in existing optical data.
\label{fig11}}
\epsscale{1.0}
\end{figure}

{\em 0145$-$221}.  The ACS photometry of GD\,1400 is around 0.25\,mag fainter than expected 
based on a previous analysis \citep{far04b}.  Figure \ref{fig11} plots an 11\,500\,K DA white dwarf
model \citep{koe10} together with {\em GALEX} far- and near-ultraviolet fluxes, $r'$-band and 
ACS (transformed) $I$-band photometry.  Compared to this model and the plotted optical and
ultraviolet data, the previously reported $V=14.85\pm0.12$\,mag photometry \citep{far04b} is 
too bright.  However, GD\,1400 is a large-amplitude ZZ\,Ceti pulsator, varying by up to 0.2\,mag 
with its most significant modes at 462, 728, and 823\,s \citep{fon03}.  The ACS exposures were
1080\,s in total and should represent an average brightness amplitude, while the exposures 
analyzed by \citet{far04b} were three 60\,s images separated by 90\,s.  These latter data may 
have been biased by one or more exposures taken during a bright phase, and combined with
their relatively large photometric error, are likely inferior to the ACS photometry.  The system is
a radial velocity variable with a period of 10\,hr (Burleigh et al.\ 2010, in preparation).

{\em 0208$-$153}.  Paper I reported that the PSF width of the M dwarf was not larger than that 
of the white dwarf, nominally indicating a likely third, cool component broadening the PSF 
of the hotter star.  Re-examination of the older pipeline data (15.4c) confirms that both PSFs 
have equal widths in those images, while in the subsequent pipeline (15.7) dataset the FWHM 
of the M dwarf is larger by 0.12\,pixels, more or less as expected if each star is single.

{\em 0219$+$282}.  This is a triple system; the DAB spectrum of the white dwarf has been
shown to be better explained with a composite DA and DB white dwarf \citep{lim09}.  The 
two spatially-resolved stars in the ACS image have PSFs indicative of a relatively blue and 
red object, consistent with a spatially-unresolved DA+DB and a visual M dwarf.  Radial 
velocity monitoring of the DA and DB components should be able confirm this scenario. 

{\em 0257$-$005}.  An extreme example of a type 1 (see \S3.2) photometric distance mismatch,
with $\Delta (m-M)=+1.6$\,mag.  The disparate distance estimates may be reconciled if the dM 
is an equal luminosity double {\em and} the white dwarf is at the cool, high gravity end allowed 
by its error bars \citep{far09b}.

{\em 0309$-$275, 0949$+$451, 1001$+$203, 1106$+$316, 1412$-$049}.  These stars have 
inaccurate coordinates in the literature, SIMBAD, and \citet{mcc08}.  Accurate ACS positions 
for the white dwarfs are given in Table \ref{tbl2}.

\begin{figure}
\epsscale{1.2}
\figurenum{12}
\plotone{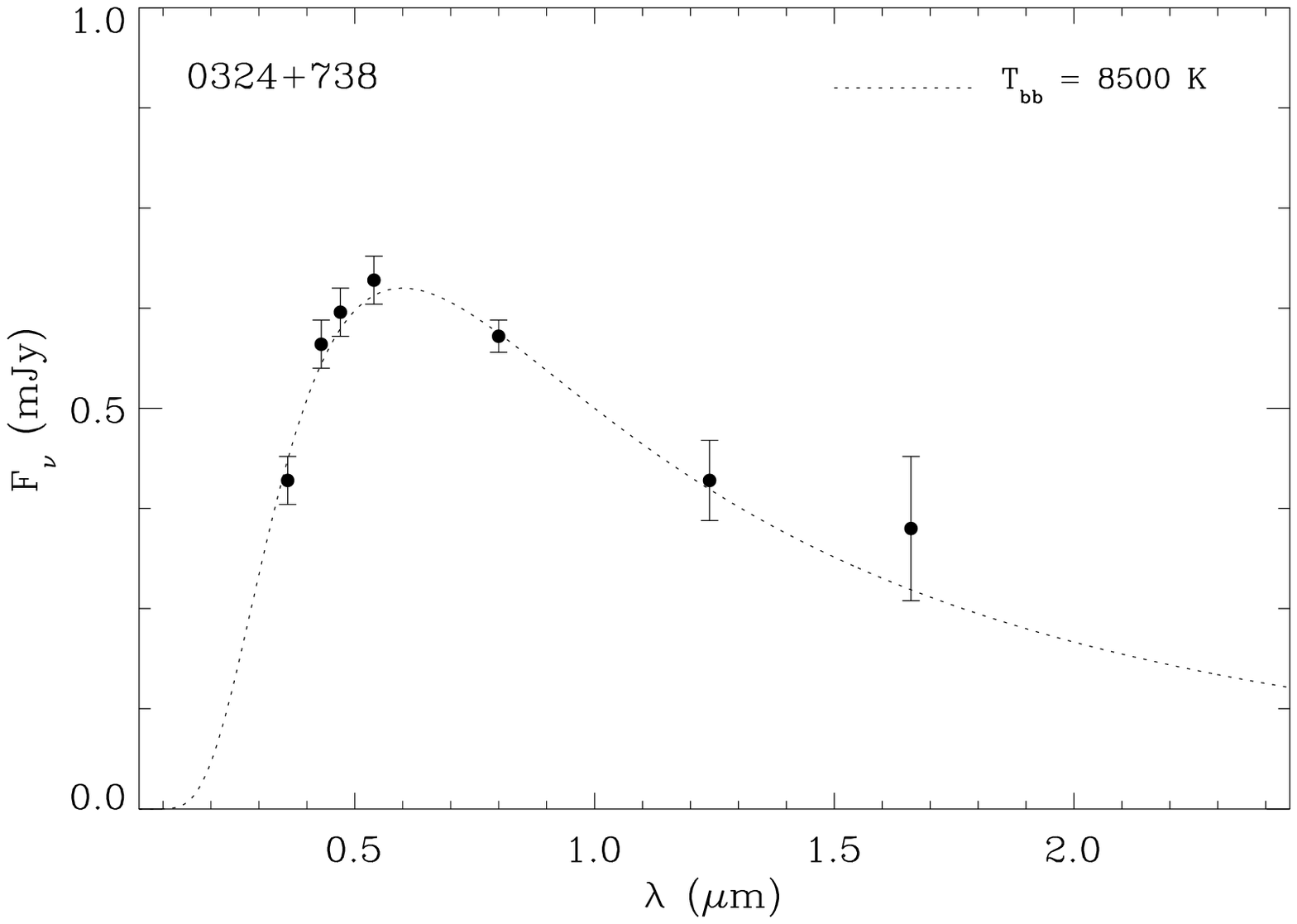}
\caption{SED of G221-10 overplotted with an 8500\,K blackbody model.  The photometric data 
are short wavelength multichannel measurements from \citet{gre76}, ACS (transformed) $I$, and 
2MASS $JH$.  The data are consistent with no near-infrared excess.
\label{fig12}}
\epsscale{1.0}
\end{figure}

{\em 0324$+$738}.  The white dwarf lies within several arcseconds of two background stars 
(Paper I) that are brighter at most or all optical wavelengths, resulting in photometric confusion 
and underestimations of its mass and effective temperature \citep{ber01,gre84}.  Perhaps the
most reliable archival data are the shorter wavelength multichannel spectrophotometry taken
by \citet{gre76}.  Combined with the ACS photometry, Figure \ref{fig12} shows these data are 
well-fitted by an 8500\,K blackbody model, implying a mass of $0.99\,M_{\odot}$ at its 40\,pc 
distance from trigonometric parallax.  A near-infrared excess was suspected on the basis of its 
2MASS $H=16.18\pm0.25$\,mag but from the plot it is apparent that  all the data are consistent 
with a single star.  The common proper motion companion (G221-11) which fell into the ACS 
field of view by chance was itself spatially resolved into two stars, making the system triple.  
The two M dwarf components are not spatially resolved in 2MASS, and their photometric 
deconvolution and subsequent spectral type estimates are somewhat uncertain.

{\em 0347$-$137}.  This white dwarf has three disparate temperature determinations in the 
literature (12\,600\,K; \citealt{gia06}, 15\,500\,K; \citealt{zuc03}, 21\,300\,K; \citealt{koe01}), 
probably owing to spectroscopic contamination by the M dwarf secondary.  Extrapolating from 
$U=14.7$\,mag \citep{mer91} for the white dwarf alone, $\log\,g=8.0$ models predict $I=15.9
$\,mag for a 21\,000\,K star versus $I=15.2$\,mag for a 12\,500\,K star.  The ACS photometry 
yields $I=15.46$\,mag, consistent with $T_{\rm eff}=15\,500$\,K.   The {\em HST} images 
reveal the binary is widely separated, and hence for a system without additional components, 
the Balmer line emission reported by \citet{koe01} must originate from activity intrinsic to the 
dM.

{\em 0354$+$463}.  This DAZ with a late dMe companion is somewhat mysterious as no
radial velocity variability searches have been successful yet the composite spectrum has 
Balmer line emission \citep{zuc03,sch96}.  The velocities of the Ca K absorption in the white 
dwarf, and the Balmer emission lines in the M dwarf all agree reasonably well ($-65\pm5
$\,km\,s$^{-1}$) and collectively represent several measurements over 5 years.  Furthermore, 
the ACS images reveal a single point source, supporting a closely orbiting, post-common 
envelope binary with $a<1.0$\,AU.  Given that the white dwarf atmosphere is polluted with 
several metals, likely due to a stellar wind from the dM7 companion \citep{zuc03}, it is virtually 
certain the system has a short orbital period.  The sum of all the evidence then points to a near 
pole-on binary with small radial velocity changes that may be confirmed through high-resolution 
spectroscopic monitoring.

{\em 0357$-$233}.  This system has a severe photometric distance mismatch between 
its components, $\Delta (m-M)=-1.6$\,mag.  However, the system is not well-studied; 
\citet{gre79} remarks the white dwarf is ``very hot'' but no spectroscopic temperature 
estimate exists.  The $UB$ photometry reported in \citet{far05a} is uncontaminated by light 
from the M dwarf companion, and combined with {\em GALEX} fluxes and the ACS $I$-band 
photometry for the white dwarf, imply a temperature near 35\,000\,K.  In this particular case, it 
seems almost certain the white dwarf luminosity has been significantly underestimated, and 
must have a large effective radius (a binary or low surface gravity), and probably a higher
temperature than estimated here.  A good spectroscopic parameter determination for the 
white dwarf is needed to reconcile the photometric distance discrepancy.

{\em 0430$+$136}.  This system is spatially resolved into two components, both of which are
clearly too bright to be a 36\,000\,K white dwarf \citep{tre07} given the $UBV$ photometry of 
\citet{weg90}.  Specifically, such a white dwarf with $U=15.9$\,mag should have $I=17.6$\,mag 
while both ACS imaged components have $I<17.0$\,mag.  Assuming the brighter component 
is an M dwarf and the dimmer component is an unresolved DA$+$dM produces self-consistent 
results and photometric distances between 100 and 150\,pc for the M dwarf components.  
There is tentative evidence for a spatially-unresolved, tertiary M dwarf; the SPY spectrum of 
the composite system reveals Balmer emission lines which have shifted between two 
observations (D. Koester 2010, private communication).  While the M dwarf distance estimates 
can be brought into agreement if the widely separated component is itself an unresolved double, 
the photometric distance to the white dwarf is at least 320\,pc ($\log\,g=8.11$; P. Bergeron 2009 
private communication), and is by far the most extreme distance mismatch of the entire sample. 
The situation can only be amended if the white dwarf has a significantly lower luminosity via high 
mass, lower temperature or both.

{\em 0458$-$665}.  An error in the coordinates and name of RX\,J0458.9-6628 exists in 
\citet{mcc99}, where it is called 0458$-$662 based on an incorrect position.  Both the position 
and name have been amended in \citet{mcc08}.  The ACS image yields precise coordinates for 
the white dwarf which are listed in Table \ref{tbl2}, and nearly identical to the optical discovery 
coordinates for the binary \citep{hut96}.  

{\em 0518$+$333}.  This star was imaged based on its 2MASS $JHK_s$ photometry, which is 
apparently contaminated by its $7\farcs5$ distant M dwarf secondary.  The ACS photometry and 
the $JHK$ data of \citet{far09b} are consistent with a single white dwarf of 9500\,K.  However, 
the proper motion companion was serendipitously resolved into two components, making the 
system triple.

{\em 0752$-$146}.  The orbital period of this highly probable close binary has never been 
established, having fallen just short of strict variability criterion \citep{max00}.  However, three 
independent radial velocity studies of this system have together seen some variability and large 
velocity separations between emission and absorption components \citep{zuc03,max00,sch96}.
The ACS image reveals an unresolved point source, making the projected separation of the pair
certainly closer than 0.9\,AU, and the orbit shorter than 1\,yr.  Further radial velocity monitoring 
should be able to discern the period and mass ratio of this double-lined spectroscopic binary.

{\em 0807$+$190}.  LHS\,1986 appears to be an erroneous identification by \citet{luy79}, as 
there is no cataloged optical or near-infrared (e.g USNO, 2MASS) point source located within 
$10''$ of the position $08^{\rm h} 10^{\rm m} 02.90^{\rm s} +18\arcdeg 52' 09\farcs7$ (J2000; 
\citealt{bak02}).  The only object with significant proper motion within $1'$ of these coordinates
moves at $0\farcs06$\,yr$^{-1}$ at position $08^{\rm s} 10^{\rm s} 00.29^{\rm s} +18\arcdeg 51' 
46\farcs6$ (J2000; \citealt{zac05}); this is the target observed by ACS.  There is no entry for 
LHS\,1986 in the revised NLTT\,catalog \citep{sal03}, nor the LSPM catalog, confirming the 
likely error (S. Lepine 2009, private communication; \citealt{lep05}).  Furthermore the ACS 
imaged star is not a white dwarf; there is no detection in {\em GALEX} and it is an SDSS and 
2MASS stellar source with $ugrizJHK$ photometry consistent with a mid to late M dwarf at
$d\ga200$\,pc  \citep{boc07,kir94}, consistent with its modest proper motion.  The analysis
by \citet{giz97} and the optical spectrum shown in \citet{rei05} -- assuming this is the same 
object as in the ACS images -- both indicate a K-type star.  \citet{giz97} remark the star may 
have been misidentified, and the original LHS\,catalog gives a color type of 'k' \citep{luy79}.
This star is likely a relatively fast moving but distant, metal-poor sdK star with reddened optical 
colors, whose proper motion was misjudged.
	
{\em 0824$+$288}.  This spectroscopic composite DA$+$dC(e) binary has been resolved into
two closely separated components at $0\farcs077$, plus the wider tertiary M dwarf companion 
at $3\farcs33$.  While it is not certain if the two closest, spatially-resolved components are the 
dC and DA stars (each one could be an unresolved binary, and the other a fourth component), 
this is the most likely explanation and most consistent with all the photometry \citep{far05a},
and the shape and width of the deconvolved ACS PSFs.  At the adopted photometric distance to
the system, the DA$+$dC pair are currently separated by at least 14\,AU, challenging the 
scenario in which a dC star forms via mass transfer \citep{heb93}.

{\em 0908$+$226}.  This star (LP\,369-15) is listed as a part of a common proper motion 
pair with LP\,369-14 at a separation of $200''$ and position angle $277\arcdeg$, both stars 
moving at $0\farcs11$ yr$^{-1}$ along $120\arcdeg$ \citep{sil05,luy97}.  This appears to be
a spurious entry in the LDS\,catalog as there are no such objects within a few arcminutes of 
LP\,369-15, and no entries in the revised NLTT or the LSPM catalogs \citep{lep05,sal03}. Its 
proper motion is given as zero in both the USNO and NOMAD catalogs \citep{zac05,mon03}, 
and blinking the POSS I and II plate scans verifies the star moves little or not at all in 45 years.  
Also, the coordinates of LP\,369-15 in SIMBAD and the Luyten catalogs are inaccurate by 
more than a dozen arcseconds (Table \ref{tbl2} lists its ACS imaged position).  Although its
proper motion was apparently misjudged by \citet{luy97}, its inclusion in a list of white dwarfs 
appears to be correct; SDSS data reveals the star (J091143.09$+$222748.8) to be a DA$+
$dMe spectroscopic composite, with proper motion $0\farcs005$\,yr$^{-1}$.

{\em 0933$+$025, 1033$+$464}.  \citet{sch96} notes H$\alpha$ emission in the composite 
spectra of these binaries, which must be attributed to an active red dwarf since the systems
are spatially resolved and hence widely separated.

{\em 0937$-$095}.  The ACS field of view captured both components of the common 
proper motion binary LDS\,3913, separated by $13\farcs54$ at position angle $354.8
\arcdeg$ and comoving at $(99,-236)$ mas yr$^{-1}$ \citep{sal03,luy97}.  The primary, 
LDS\,3913A (also LP\,668-9, G161-56, NLTT\,22302), is almost certainly a K-type star, 
while the secondary, LDS\,3913B (also, LP\,668-10, NLTT\,22301), also appears to be a 
late K or early M dwarf.  This conclusion is supported by $UBV$ \citep{sil05,sal03,mer91}, 
ACS $I$-band, and 2MASS $JHK_s$ \citep{skr06} photometry, which yield $V-K=2.9$ 
and $V-K=3.4$ -- consistent with spectral types K5 and M0 -- for LDS\,3193A and B
respectively.  Two main-sequence stars of corresponding spectral type would lie at
$d\approx500$\,pc, and with tangential speeds of $v\approx600$ km s$^{-1}$.  Such
high velocities suggests kinematic membership in the galactic thick disk or halo, hence
these cool stars are probably metal-poor subdwarfs.  Assuming the pair lie 1 magnitude 
below the main sequence, their distance would be $d\approx340$\,pc with tangential 
speeds of $v\approx400$ km s$^{-1}$, and $(U,V,W)=(-340,-210,-100)$ km s$^{-1}$.
Therefore, in all likelihood, this pair belongs to the thick disk or halo.

\begin{figure}
\epsscale{1.2}
\figurenum{13}
\plotone{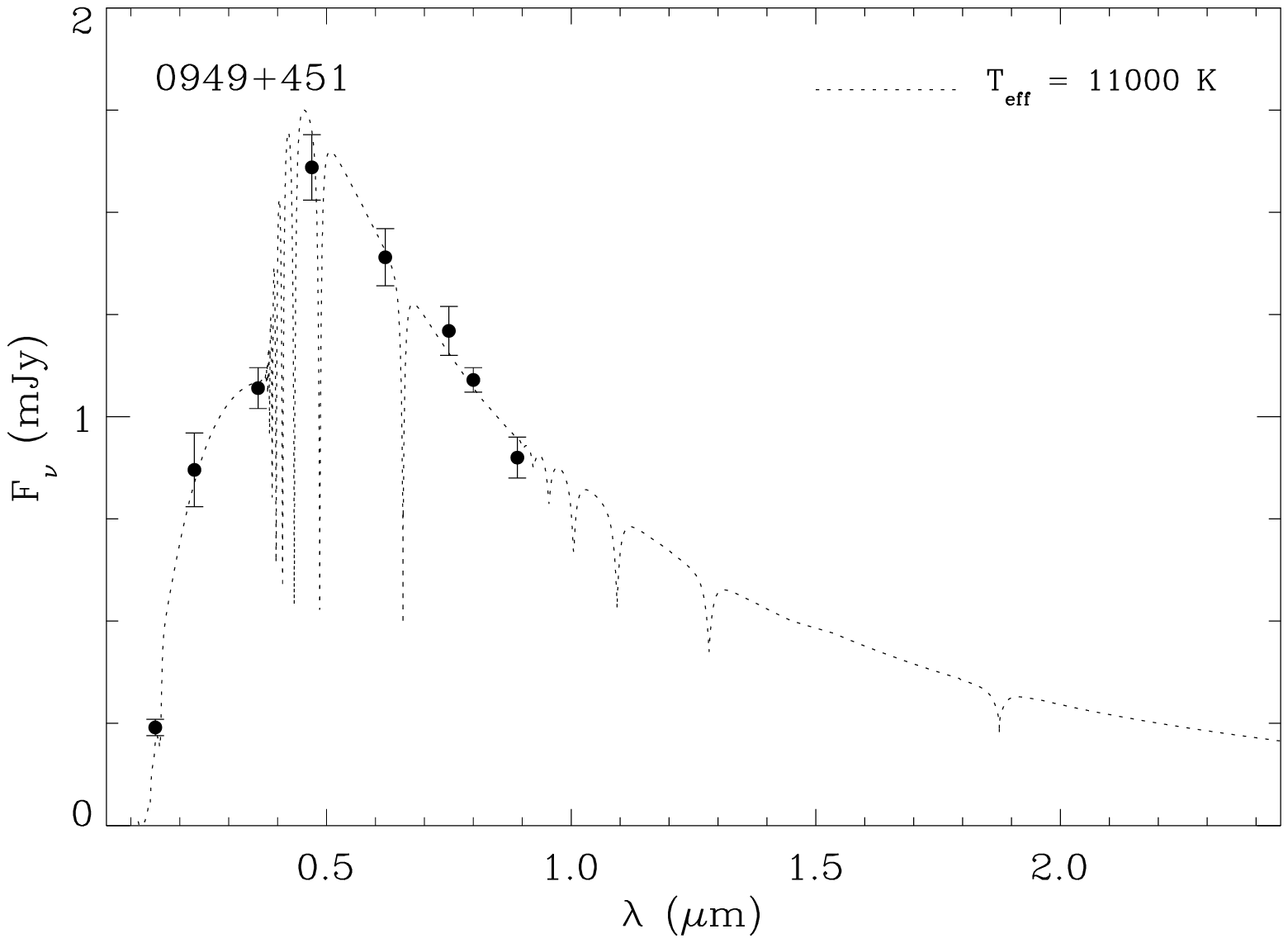}
\caption{SED of HS\,0949$+$4508 overplotted with an 11\,000\,K DA model.  The photometric 
data are {\em GALEX} far- and near-ultraviolet, SDSS $ugriz$, and ACS (transformed) $I$.
\label{fig13}}
\epsscale{1.0}
\end{figure}

{\em 0949$+$451}.  This system is not well-studied and Figure \ref{fig13} plots {\em GALEX}, 
SDSS, and the ACS $I$-band photometry for the white dwarf, fitted with an 11\,000\,K DA 
white dwarf model.  Although the $ugriz$ photometry is somewhat uncertain due to the close 
visual M dwarf companions, the data are altogether well modeled.  At a photometric distance of 
65\,pc, the double M dwarf has a projected separation of only 2.5\,AU.

{\em 1015$-$173}.  This pair has an extreme photometric distance mismatch between 
its components.  The system is not well studied, but {\em GALEX} and $UBV$ photometry 
\citep{kil97} support a relatively hot white dwarf.  Still, the 25\,000\,K estimate found here 
produces a distance far too close to match the M dwarf distance estimate; even an equally 
luminous double white dwarf is insufficient to bridge the gap.  Therefore, the white dwarf
is likely to be intrinsically brighter via temperature or radius.

{\em 1036$-$204}.  This white dwarf has a peculiar SED described by very different 
temperatures in the ultraviolet-blue versus the optical red and near-infrared \citep{far09b}  
The broad and deep optical ${\rm C}_2$ absorption bands are likely responsible for a 
significant increase in its near-infrared fluxes via the redistribution of emergent energy; 
there is no evidence for a companion in this system.

{\em 1051$+$516}.  \citet{sil06} derive $T_{\rm eff}=29\,600$\,K, $\log\,g=8.1$ for this star
based on the SDSS data, but these values are likely to be uncertain owing to contamination
from the bright M dwarf companion  The {\em GALEX} fluxes for the white dwarf favor the lower 
effective temperature estimate used here, which leads to better agreement in the component
distances.

{\em 1106$-$211}.  There is no spectroscopic, photometric, nor kinematical evidence that this 
star is a white dwarf \citep{mcc08}.  As \citet{hoa07} notes correctly, there is no star with high 
proper motion within several arcminutes of the coordinates given in \citet{eva92}.  The object 
located at those coordinates has a proper motion of $0\farcs03$\,yr$^{-1}$ \citep{zac05}, and 
optical through near-infrared colors consistent with a late K dwarf (e.g. $V-K=3.3$).

{\em 1108$+$325}.  Component identification in this system was not clear-cut.  The 
deconvolved FWHMs are nearly identical, as are the instrumental magnitudes prior to color 
correction, although the former were certainly affected by close binarity.  The northwest and 
southeast objects were identified as the cool companion and white dwarf respectively.  This 
choice fits the expected brightness of the white dwarf at $I$-band, based on its {\em GALEX} 
and SDSS $ug$ photometry, whereas if the roles are reversed, the photometric distances 
become more discrepant.  Additionally, the best {\sf daophot} PSF fits for the northwest and 
southeast source were found using M dwarf and white dwarf templates, respectively.  The 
$JHK$ photometry for this system was taken from \citet{far09b} as the 2MASS data have 
low S/N at the two longer wavelengths.

{\em 1133$+$358}.  This rare DC$+$dM binary is unresolved in the ACS images, implying a
projected separation under 1.3\,AU at its trigonometric parallax distance of 50\,pc \citep{dah88}.
The spectrum in \citet{put97} shows several Balmer emission lines from an active and optically
dominant M dwarf.  Without such emission potentially masking metal absorption in the white 
dwarf, one might expect the cool helium atmosphere star to be readily polluted by the wind 
from the secondary \citep{zuc03}.  The star is detected in the {\em GALEX} near-ultraviolet 
bandpass, and combined with ground-based ultraviolet photometry, confirms the presence 
of a 6500\,K white dwarf \citep{gre76}.  For this temperature, the parallax distance yields 
$M_V=13.37$\,mag and a white dwarf mass of 0.41\,$M_{\odot}$, suggesting the M dwarf 
companion is in a close orbit and ejected sufficient mass during the first ascent giant phase 
to prevent helium fusion.

{\em 1156$+$132}.  Unfortunately, inaccurate coordinates for this star in \citet{mcc08,mcc99}
caused the white dwarf to fall outside of the $25''\times 25''$ field of view of the ACS / HRC; 
hence no data were obtained for this star.  Table \ref{tbl2} lists both the incorrect and correct 
position for the star, and the error has been corrected in SIMBAD and the current online white 
dwarf catalog.

{\em 1156$+$129}.  This white dwarf is poorly studied, and the effective temperature estimate 
made here relies on the SDSS $u$-band and {\em GALEX} fluxes, but is still relatively uncertain.

{\em 1236$-$004}.  The white dwarf and M dwarf components of this binary were misidentified
in Paper I.  The ACS stellar profiles suggest the western component is the M dwarf (with a larger 
FWHM), and the SDSS color composite finding chart reveals a clear change in the halo of the 
stellar image from blue in the east, to white-orange in the west.  The newer analysis presented
here should be considered a correction to Paper I.  

{\em 1247$+$550}.  This cool white dwarf has published $JHK$ photometry \citep{ber01}
which differs significantly from the relatively low S/N photometry of 2MASS; these superior
quality data suggest the star does not have a near-infrared excess.

{\em 1333$+$487}.  Surprisingly, the {\em HST} \ ACS observations are the first published 
data to establish the spatially-resolved, visual binary nature of GD\,325, at angular separation 
$2\farcs95$ and position angle $72.1\arcdeg$ (Paper I; \citealt{gre74}).  The USNO-B1 optical 
and 2MASS near-infrared catalog positions for the source are offset by $2\farcs3$ at $76\arcdeg$.
Based on $UBV$ data \citep{mcc99}, the ACS photometry of the spatially-resolved white dwarf
is too bright for a single star of 16\,000\,K \citep{cas06}.  Furthermore, the FWHM of the white
dwarf is wider than that of the spatially-resolved M dwarf, a singular example among the 29
well-resolved pairs (GD\,325 is a clear outlier in Figure \ref{fig2}).  Assigning the photometric
excess and wide PSF to a spatially-unresolved tertiary produces self-consistent results and 
distance estimates of 46 and 51\,pc to the M dwarf components.  However, the white dwarf has 
a trigonometric parallax distance of 34\,pc \citep{dah82,mcc99}, but it is possible the astrometry
was biased due to binarity; there is no mention of the visual companion in the literature.  

\begin{figure}
\epsscale{1.2}
\figurenum{14}
\plotone{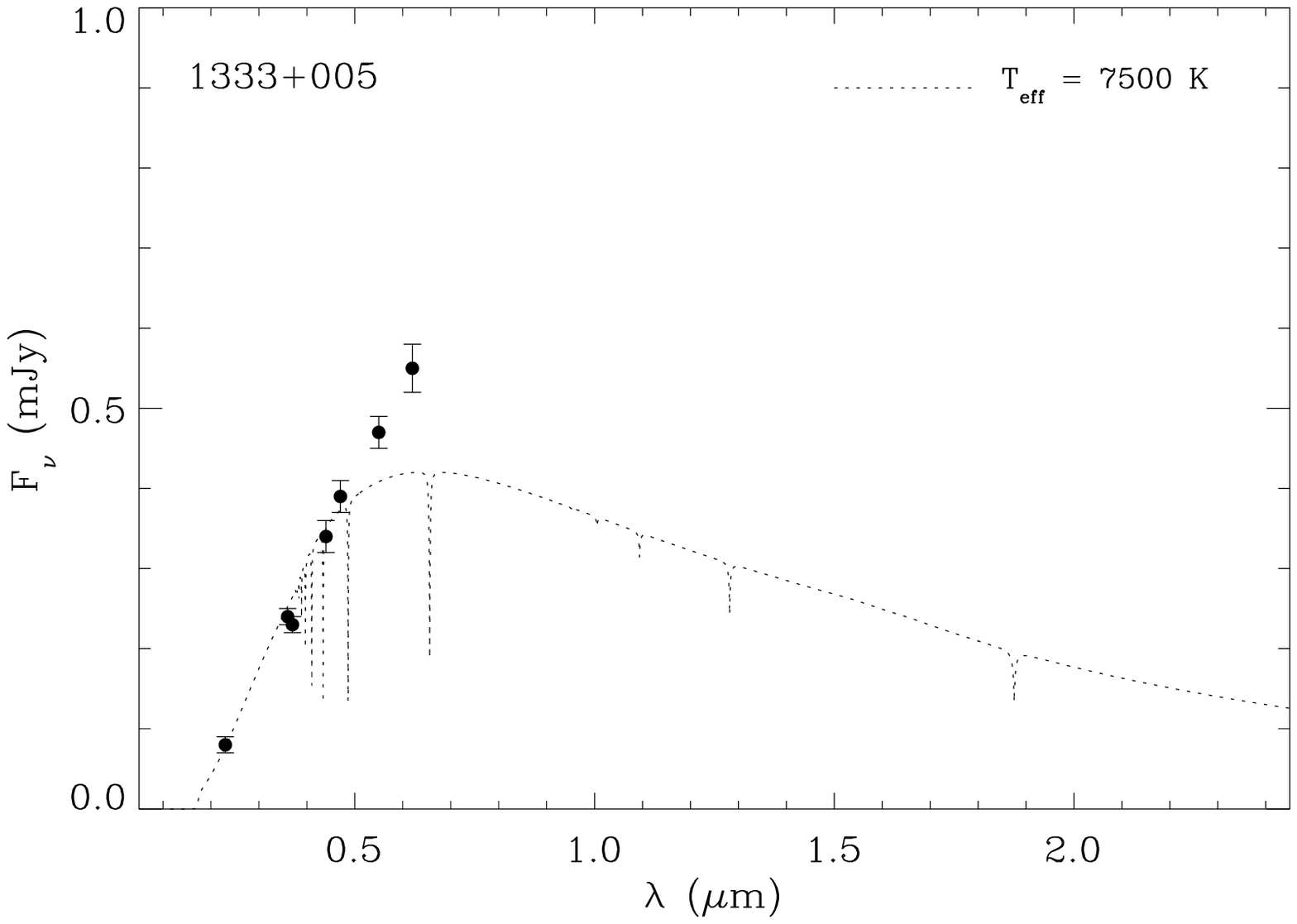}
\caption{SED of LP618-14 overplotted with an 7500\,K DA model.  The photometric data are 
{\em GALEX} near-ultraviolet, SDSS $ugr$, and optical $UBV$ \citep{far05a}.  The photometry
redward of 5000\,\AA \ was ignored in the model fit due to the influence of the M dwarf; these
longer wavelength fluxes are all above 1.0\,mJy.
\label{fig14}}
\epsscale{1.0}
\end{figure}

{\em 1333$+$005}.  There is an extreme distance discrepancy between the components in 
this spatially-unresolved binary, strongly suggesting the white dwarf is a low mass, helium core 
degenerate.  {\em GALEX} near-ultraviolet and short wavelength ground-based photometry 
suggest the white dwarf is warmer than previous spectroscopic estimates \citep{kil06}, likely 
influenced by the M dwarf continuum flux.  Figure \ref{fig14} shows a 7500\,K DA model yields
a good fit to the ultraviolet through blue photometry.  At the  photometric distance to the M dwarf, 
the white dwarf would have $\log\,g<7.0$ and a mass below 0.2\,$M_{\odot}$.  Such an extreme 
solution indicates the secondary is likely to be subluminous relative to the M dwarf templates 
used here.  In any event, the white dwarf is likely to have a low mass and a commensurately
large radius.

{\em 1339$+$346}.  The optical companion in the ACS images appears to be a background
early K-type star; a result of chance alignment and not physical.  The cool star imaged with ACS 
is part of a wide, common proper motion pair of K stars with $0\farcs05$\,yr$^{-1}$ \citep{zac05}; 
the K star nearest to the white dwarf is $2\farcs5$ distant at $213\arcdeg$, while the second K 
star is offset by $21\farcs2$ at $298\arcdeg$.  The two cool stars have $ugrizJHK$ photometry 
indicating approximate spectral types of K2 and K5 at around 720\,pc \citep{far09b,aba09}, 
while the white dwarf distance is near 100\,pc \citep{lie05a}.  Examination of photographic plate
scans taken more than 55 years ago reveals the white dwarf has moved relative to the nearest 
K star, confirming they are not physically associated \citep{far09b}.

{\em 1412$-$049}.  There are virtually no constraints on the properties of this white dwarf; 
Table \ref{tbl3} lists a best guess based on {\em GALEX} fluxes, and constraints from the red 
dwarf spectral type estimate.

{\em 1419$+$576}.  The white dwarf does not have any published parameters, and the 
SDSS spectrum shows the M dwarfs dominating the emergent flux beginning at 5000\AA.  
The {\em GALEX}, $u$-band, and ACS (transformed) $I$-band photometry support a rough
effective temperature estimate of 25\,000\,K.  For a reasonable range of photometric distances,
the visual double M dwarf has a projected separation less than 11\,AU.

{\em 1433$+$538}.  This spatially-unresolved binary is somewhat mysterious.  Its spectrum
contains no obvious emission lines expected in a close binary; the H$\alpha$ region was flat 
during four temporally separate observations \citep{sch96}, while the SDSS spectrum shows 
well-defined H$\alpha$ absorption \citep{sil06}.  Close inspection of the unbinned SDSS data
reveal possible weak H$\alpha$ and Na 8190\AA \ emission features, but these may be due to 
noise.  There are two published trigonometric parallaxes for this point source, both of which
are smaller than the quoted measurement error \citep{mcc99,dah88}, and hence neither of
these distances are used.

{\em 1434$+$289}.  A near-infrared spectrum of this white dwarf reveals no excess emission;
the 2MASS data have low S/N at $HK_s$ \citep{far09b}.

{\em 1436$-$216}.  This system is a spatially-unresolved DA$+$dMe recently discovered by 
\citet{zuc03}.  There are few constraints on the white dwarf parameters, and those used here 
are a best guess based on optical $UBV$ and {\em GALEX} fluxes.

{\em 1458$+$171}.  \citet{lie05a} derive parameters that indicate this is a low mass, helium core
white dwarf, consistent with a short period binary.

{\em 1558$+$616}.  The parameters of this white dwarf are unconstrained in the literature
\citep{mcc08}.  It has no {\em GALEX} data, but the SDSS $u-g$ color suggests an effective 
temperature near 15\,000\,K, assuming little or no contamination in the $g$ band by the M 
dwarf.

{\em 1603$+$125}.  This star is not a white dwarf; the original designation \citep{mcc99} was 
likely intended to be 1603$+$175 which has a corresponding source in the KUV catalog (P. 
Bergeron 2008, private communication); and has been changed accordingly in \citet{mcc08}.
The source imaged by ACS is likely a K star based on its $ugrizJHK$ photometry and small
proper motion \citep{aba09,zac05}.

{\em 1619$+$525}.   This possible triple system is discussed in detail by \citet{far09b} and in 
Paper I.  A bound triple is possible at a shared distance near 290\,pc;  the photometric distance 
to both spatially-resolved M dwarf components.  However, this is significantly further than the 
100\,pc distance estimate to the white dwarf from Balmer line spectroscopy \citep{lie05a}.  A 
chance alignment is highly improbable, and no relative motion is seen between any of the 
three stars.  While the physical association of these three stars remains tentative, the white 
dwarf parameters may be skewed due to spectroscopic contamination.

{\em 1631$+$781}.  This triple system is discussed in detail in Paper I.  Here, owing to 
recently published white dwarf parameters \citep{tre07} and the newer ACS analysis which 
photometrically deconvolves the two M dwarfs, excellent agreement is now found between 
the component photometric distances.   The M dwarfs are widely separated from the white 
dwarf yet show clear signs of activity via Balmer line emission variability in strength but not
velocity \citep{sio95}.  Hence, it is possible the observed EUV and variable X-ray emission 
may originate from a pair of coronally active M dwarfs, rather than the white dwarf as
previously reported \citep{coo92}.

{\em 1658$+$440}.  The apparent 2MASS $K_s$-band excess of this massive, magnetic
white dwarf is spurious; {\em Spitzer} IRAC data reveal no infrared excess \citep{far08b}.

{\em 1717$-$345}.  This system has uncertain white dwarf parameters in the literature, as 
the optical spectrum is dominated by the red dwarf \citep{kaw04,rei88}.  Additionally, there 
is a lack of {\em GALEX} or ground-based photoelectric ultraviolet or blue photometry which 
might provide better constraints.

{\em 1833$+$644}.  This is a likely triple system, although the candidate tertiary detected at
$1\farcs8$ has yet to be confirmed with additional observations.  The evidence in favor of a 
bound system is 1) the modest galactic latitude $b=+26\arcdeg$ at $\ell=94\arcdeg$ (normal 
to the direction towards the galactic center), 2) the angular separation, and 3) the very wide 
FWHM expected for a late M dwarf.  The probability of a late M dwarf being found within a 
cylinder centered at the white dwarf, of radius $2"$ and length 100\,pc is $5\times10^{-6}$ 
\citep{cru07}.

{\em 1845$+$683, 2211$+$372, 2323$+$256, 2349-283}.  Not binary suspects (\citealt{far09b}; 
Paper I).

{\em 2151$-$015}.  Paper I suggested the white dwarf PSF appeared too red for a single star, 
possibly indicating a third component.  The position of the white dwarf in Figure \ref{fig2} firmly 
discounts this possibility.  The ACS observations confirm the suspicions of both \citet{zuc03} and
\citet{max99} that this system is not a close binary.

{\em 2237$-$365}.  This white dwarf -- M dwarf system is correctly designated HE\,2237$-$3630 
but is not the same star as LHS\,3841.  This unfortunate co-identification was made in \citet{fri00} 
who took a spectrum of the DC$+$dM binary and who list the correct position for the white dwarf
system.  SuperCOSMOS images reveal a close visual white dwarf -- M dwarf pair, solidified by 
{\em GALEX} and 2MASS photometry.  Believing the two designations corresponded to the same
source, the ACS observations targeted LHS\,3841, a metal-poor halo star \citep{rei05}, and hence 
no data were obtained for the white dwarf binary system.  Table \ref{tbl2} lists the positions of both
LHS\,3841 and HE\,2237$-$3630.

{\em 2257$+$162}.  The white dwarf is most likely a low mass, helium core degenerate 
\citep{lie05a} and the system is almost certainly a short period binary.  Neither \citet{lie05a} nor
\citet{tre07} reports the detection of the red dwarf in their optical spectra.

{\em 2336$-$187}.  This appears to be a single white dwarf without any $K$-band excess;
the 2MASS flux is $2\sigma$ brighter than that measured independently by \citep{far09b}.
The resulting $I-K$ color is consistent with a 8100\,K DA white dwarf photosphere.

\clearpage 

\LongTables 

\tabletypesize{\footnotesize}

\begin{deluxetable*}{cccccccccc}
\tablecaption{Target ACS Photometric and Astrometric Data\label{tbl1}}
\tablewidth{0pt}
\tablehead{
\colhead{WD}                			&
\colhead{Name}  				&
\colhead{Spatially}				&
\colhead{FWHM}				&
\colhead{$a/b$}				&
\colhead{$\alpha$}				&
\colhead{P.A.}					&
\colhead{F814W\tablenotemark{a}}	&
\colhead{$I_c$}				&
\colhead{Source}				\\
&
&Resolved?
&(arcsec)
&
&(arcsec)
&(deg)
&(mag)
&(mag)
&Notes}

\startdata

0023$+$388	&G171-B10A			&No		&0.07367		&1.0294		&\nodata				&\nodata			&$15.22\pm0.02$	&$15.22\pm0.03$	&1\\
0034$-$211	&LTT\,0329			&Yes	&0.07470		&1.0350		&\nodata				&\nodata 			&$15.11\pm0.02$	&$15.07\pm0.03$	&2,5\\
			&					&		&0.07412		&1.0222		&$0.327\pm0.001$		&$105.25\pm0.09$	&$12.76\pm0.02$	&$12.81\pm0.03$	&5\\
0116$-$231	&GD\,695				&Yes	&0.07393		&1.0071		&\nodata				&\nodata			&$16.59\pm0.02$	&$16.54\pm0.03$	&2\\
			&					&		&0.07570		&1.0211		&$1.106\pm0.001$		&$9.95\pm0.05$	&$16.15\pm0.02$	&$16.23\pm0.03$	&\\
0131$-$163	&GD\,984				&Yes	&0.07412		&1.0147		&\nodata				&\nodata			&$14.29\pm0.02$	&$14.23\pm0.03$	&2,5\\
			&					&		&0.07435		&1.0186		&$0.190\pm0.001$		&$145.34\pm0.17$	&$14.50\pm0.02$	&$14.57\pm0.03$	&5\\
0145$-$221	&GD\,1400			&No		&0.07358		&1.0025		&\nodata				&\nodata			&$15.16\pm0.02$	&$15.13\pm0.03$	&1\\
0145$-$257	&GD\,1401			&Yes	&0.07338		&1.0169		&\nodata				&\nodata			&$14.97\pm0.02$	&$14.93\pm0.03$	&2\\
			&					&		&0.07568		&1.0103		&$2.294\pm0.001$		&$154.06\pm0.01$	&$13.83\pm0.02$	&$13.90\pm0.03$	&\\
0205$+$133	&PG					&Yes	&0.07362		&1.0433		&\nodata				&\nodata			&$15.45\pm0.02$	&$15.40\pm0.03$	&2\\
			&					&		&0.07573		&1.0360		&$1.254\pm0.001$		&$10.60\pm0.02$	&$13.94\pm0.02$	&$13.97\pm0.03$	&\\
0208$-$153	&MCT\,0208$-$1520	&Yes	&0.07362		&1.0211		&\nodata				&\nodata			&$15.94\pm0.02$	&$15.90\pm0.03$	&2\\
			&					&		&0.07563		&1.0194		&$2.648\pm0.001$		&$217.43\pm0.02$	&$13.80\pm0.02$	&$13.85\pm0.03$	&\\
0219$+$282	&KUV\,02196$+$2816	&Yes	&0.07473		&1.0089		&\nodata				&\nodata			&$17.33\pm0.03$	&$17.28\pm0.04$	&2,5\\
			&					&		&0.07868		&1.0310		&$0.116\pm0.002$		&$193.37\pm0.19$	&$18.14\pm0.04$	&$18.26\pm0.05$	&5\\
0237$+$115	&PG					&Yes	&0.07338		&1.0240		&\nodata				&\nodata			&$16.38\pm0.02$	&$16.32\pm0.03$	&2,5\\
			&					&		&0.07455		&1.0098		&$0.124\pm0.002$		&$268.52\pm0.21$	&$15.00\pm0.02$	&$15.05\pm0.03$	&5\\
0257$-$005	&KUV\,02579$-$0036	&Yes	&0.07230		&1.0233		&\nodata				&\nodata			&$17.37\pm0.03$	&$17.32\pm0.04$	&2\\
			&					&		&0.07525		&1.0501		&$0.978\pm0.002$		&$112.99\pm0.15$	&$18.61\pm0.05$	&$18.70\pm0.06$	&\\
0303$-$007	&KUV\,03036$-$0043	&No		&0.07535		&1.0232		&\nodata				&\nodata			&$14.54\pm0.02$	&$14.59\pm0.03$	&1\\
0309$-$275	&					&Yes	&0.07498		&1.0359		&\nodata				&\nodata			&$16.84\pm0.04$	&$16.78\pm0.05$	&2,5\\
			&					&		&0.07538		&1.0076		&$0.099\pm0.002$		&$18.32\pm0.43$	&$14.59\pm0.02$	&$14.61\pm0.03$	&5\\
0324$+$738	&G221-10				&No		&0.07398		&1.0435		&\nodata				&\nodata			&$16.55\pm0.02$	&$16.53\pm0.03$	&1,7\\
			&G221-11				&Yes	&0.07545		&1.0322		&\nodata				&\nodata			&$13.69\pm0.02$	&$13.82\pm0.03$	&2,5,6\\
			&					&		&0.07757		&1.0322		&$0.297\pm0.001$		&$31.84\pm0.05$	&$15.11\pm0.02$	&$15.30\pm0.03$	&5\\
0331$-$356	&HE\,0331$-$3541		&No		&0.07590		&1.0271		&\nodata				&\nodata			&$12.72\pm0.02$	&$12.77\pm0.03$	&1\\
0347$-$137	&GD\,51				&Yes	&0.07378		&1.0168		&\nodata				&\nodata			&$15.50\pm0.02$	&$15.46\pm0.03$	&2\\
			&					&		&0.07690		&1.0247		&$1.052\pm0.001$		&$265.10\pm0.03$	&$13.57\pm0.02$	&$13.65\pm0.03$	&\\
0354$+$463	&Rubin\,80			&No		&0.07495		&1.0181		&\nodata				&\nodata			&$14.89\pm0.02$	&$14.90\pm0.03$	&1\\
0357$-$233	&Ton\,S\,392			&Yes	&0.07322		&1.0176		&\nodata				&\nodata			&$16.24\pm0.02$	&$16.18\pm0.03$	&2\\
			&					&		&0.07533		&1.0094		&$1.191\pm0.001$		&$350.86\pm0.05$	&$16.47\pm0.02$	&$16.53\pm0.03$	&\\
0430$+$136	&KUV\,04304$+$1339	&Yes	&0.07428		&1.0408		&\nodata				&\nodata			&$16.75\pm0.03$	&$16.76\pm0.04$	&2,5\\
			&					&		&0.07683		&1.0370		&$0.260\pm0.001$		&$47.37\pm0.04$	&$15.11\pm0.02$	&$15.19\pm0.03$	&5\\
0458$-$665	&RX\,J0458.9$-$6628	&No		&0.07470		&1.0219		&\nodata				&\nodata			&$14.56\pm0.02$	&$14.61\pm0.03$	&1\\
0518$+$333	&G86-1B				&No		&0.07383		&1.0229		&\nodata				&\nodata			&$15.85\pm0.02$	&$15.83\pm0.03$	&1,7\\
			&G86-1A				&Yes	&0.07500		&1.0381		&\nodata				&\nodata			&$12.41\pm0.02$	&$12.46\pm0.03$	&2,5,6\\
			&					&		&0.07555		&1.0341		&$0.158\pm0.001$		&$266.31\pm0.13$	&$12.77\pm0.02$	&$12.83\pm0.03$	&5\\
0752$-$146	&LTT\,2980			&No		&0.07378		&1.0395		&\nodata				&\nodata			&$13.47\pm0.02$	&$13.44\pm0.03$	&1\\
0807$+$190	&LHS\,1986			&No		&0.07595		&1.0308		&\nodata				&\nodata			&$16.74\pm0.02$	&$16.84\pm0.03$	&1,7\\
0812$+$478	&PG					&No		&0.07490		&1.0239		&\nodata				&\nodata			&$15.15\pm0.02$	&$15.13\pm0.03$	&1\\
0824$+$288	&PG					&Yes	&0.07445		&1.0180		&\nodata				&\nodata			&$15.12\pm0.03$	&$15.07\pm0.04$	&3,5\\
			&					&		&0.07390		&1.0324		&$0.077\pm0.002$		&$228.28\pm0.45$	&$13.92\pm0.02$	&$13.92\pm0.03$	&5\\
			&					&		&0.07513		&1.0418		&$3.330\pm0.001$		&$120.56\pm0.01$	&$14.47\pm0.02$	&$14.51\pm0.03$	&\\			
0908$+$226	&LP\,369-15			&No		&0.07542		&1.0386		&\nodata				&\nodata			&$16.46\pm0.02$	&$16.52\pm0.03$	&1\\
0915$+$201	&LB\,3016			&Yes	&0.07187		&1.0380		&\nodata				&\nodata			&$16.92\pm0.02$	&$16.86\pm0.03$	&2\\
			&					&		&0.07470		&1.0224		&$2.312\pm0.002$		&$251.74\pm0.04$	&$17.11\pm0.03$	&$17.16\pm0.04$	&\\
0933$+$025	&PG					&Yes	&0.07343		&1.0529		&\nodata				&\nodata			&$16.52\pm0.02$	&$16.48\pm0.03$	&2\\
			&					&		&0.07437		&1.0499		&$1.232\pm0.001$		&$346.60\pm0.05$	&$14.74\pm0.02$	&$14.82\pm0.03$	&\\
0937$-$095	&LDS\,3913A			&No		&0.07388		&1.0455		&\nodata				&\nodata			&$14.60\pm0.02$	&$14.60\pm0.03$	&1,7\\
			&LDS\,3913B			&No		&0.07410		&1.0359		&\nodata				&\nodata			&$15.63\pm0.02$	&$15.63\pm0.03$	&1\\
0949$+$451	&HS\,0949$+$4508		&Yes	&0.07455		&1.0177		&\nodata				&\nodata			&$15.89\pm0.02$	&$15.86\pm0.03$	&2\\
			&					&		&0.07890		&1.1256		&$2.892\pm0.001$		&$119.63\pm0.01$	&$13.50\pm0.02$	&$13.60\pm0.03$	&4\\	
1001$+$203	&Ton\,1150			&No		&0.07498		&1.0199		&\nodata				&\nodata			&$13.75\pm0.02$	&$13.78\pm0.03$	&1\\
1015$-$173	&EC\,10150$-$1722		&Yes	&0.07258		&1.0530		&\nodata				&\nodata			&$17.30\pm0.06$	&$17.25\pm0.07$	&2,5\\
			&					&		&0.07510		&1.0049		&$0.060\pm0.002$		&$131.26\pm0.67$	&$16.69\pm0.04$	&$16.74\pm0.05$	&5\\
1026$+$002	&PG					&No		&0.07485		&1.0520		&\nodata				&\nodata			&$12.99\pm0.02$	&$13.00\pm0.03$	&1\\
1033$+$464	&GD\,123				&Yes	&0.07325		&1.0083		&\nodata				&\nodata			&$14.72\pm0.02$	&$14.67\pm0.03$	&2\\
			&					&		&0.07542		&1.0048		&$0.731\pm0.001$		&$311.77\pm0.04$	&$14.34\pm0.02$	&$14.47\pm0.03$	&\\
1036$-$204	&LHS\,2293			&No		&0.07547		&1.0150		&\nodata				&\nodata			&$15.32\pm0.02$	&$15.31\pm0.03$	&1,7\\
1037$+$512	&PG					&No		&0.07693		&1.0375		&\nodata				&\nodata			&$15.03\pm0.02$	&$15.06\pm0.03$	&1\\
1049$+$103	&PG					&Yes	&0.07402		&1.0349		&\nodata				&\nodata			&$16.05\pm0.02$	&$16.01\pm0.03$	&2,5\\
			&					&		&0.07570		&1.0311		&$0.262\pm0.001$		&$130.41\pm0.04$	&$14.87\pm0.02$	&$14.97\pm0.03$	&,5\\
1051$+$516	&SBS				&No		&0.07483		&1.0086		&\nodata				&\nodata			&$14.63\pm0.02$	&$14.68\pm0.03$	&1\\
1106$+$316	&Ton\,28				&Yes	&0.07423		&1.0118		&\nodata				&\nodata			&$17.57\pm0.03$	&$17.52\pm0.04$	&2\\
			&					&		&0.07560		&1.0069		&$0.478\pm0.002$		&$104.94\pm0.20$	&$16.44\pm0.02$	&$16.49\pm0.03$	&\\
1106$-$211	&					&No		&0.07395		&1.0109		&\nodata				&\nodata			&$15.60\pm0.02$	&$15.60\pm0.03$	&1,7\\
1108$+$325\tablenotemark{b}
			&Ton\,60				&Yes	&0.07355		&1.0403		&\nodata				&\nodata			&$17.23\pm0.03$	&$17.17\pm0.04$	&2,5\\
			&					&		&0.07325		&1.0278		&$0.167\pm0.001$		&$298.36\pm0.12$	&$17.29\pm0.03$	&$17.34\pm0.04$	&5\\
1133$+$489	&PG					&Yes	&0.07483		&1.0446		&\nodata				&\nodata			&$17.51\pm0.03$	&$17.45\pm0.04$	&2,5\\
			&					&		&0.07738		&1.0243		&$0.090\pm0.002$		&$10.31\pm0.32$	&$18.16\pm0.04$	&$18.29\pm0.05$	&5\\
1133$+$358	&G147-65				&No		&0.07545		&1.0109		&\nodata				&\nodata			&$13.21\pm0.02$	&$13.29\pm0.03$	&1\\
1140$+$004	&					&No		&0.07448		&1.0070		&\nodata				&\nodata			&$16.93\pm0.02$	&$16.96\pm0.03$	&1\\
1156$+$132	&LP\,494-12			&\nodata	&\nodata		&\nodata		&\nodata				&\nodata			&\nodata			&\nodata			&\\
1156$+$129	&					&Yes	&0.07385		&1.0305		&\nodata				&\nodata			&$18.02\pm0.04$	&$18.00\pm0.05$	&2\\
			&					&		&0.07615		&1.0346		&$0.564\pm0.002$		&$274.78\pm0.18$	&$16.20\pm0.02$	&$16.28\pm0.03$	&\\
1210$+$464	&PG					&Yes	&0.07270		&1.0485		&\nodata				&\nodata			&$16.36\pm0.02$	&$16.31\pm0.03$	&2\\
			&					&		&0.07343		&1.0433		&$1.043\pm0.001$		&$324.32\pm0.05$	&$13.12\pm0.02$	&$13.16\pm0.03$	&\\
1218$+$497	&PG					&Yes	&0.07388		&1.0301		&\nodata				&\nodata			&$16.73\pm0.02$	&$16.68\pm0.03$	&2,5\\
			&					&		&0.07595		&1.0247		&$0.303\pm0.001$		&$335.14\pm0.06$	&$16.14\pm0.02$	&$16.22\pm0.03$	&5\\
1236$-$004	&					&Yes	&0.07400		&1.0155		&\nodata				&\nodata			&$18.10\pm0.04$	&$18.05\pm0.05$	&2\\
			&					&		&0.07440		&1.0312		&$0.663\pm0.002$		&$267.12\pm0.20$	&$17.83\pm0.04$	&$17.89\pm0.05$	&\\
1247$+$550	&LHS\,342			&No		&0.07520		&1.0132		&\nodata				&\nodata			&$16.37\pm0.02$	&$16.37\pm0.03$	&1,7\\
1247$-$176	&EC\,12477$-$1738		&No		&0.07468		&1.0336		&\nodata				&\nodata			&$14.82\pm0.02$	&$14.86\pm0.03$	&1\\
1307$-$141	&EC\,13077$-$1411		&Yes	&0.07350		&1.0111		&\nodata				&\nodata			&$16.90\pm0.02$	&$16.86\pm0.03$	&2\\
			&					&		&0.07670		&1.0218		&$2.133\pm0.001$		&$165.51\pm0.03$	&$15.32\pm0.02$	&$15.40\pm0.03$	&\\
1333$+$487	&GD\,325				&Yes	&0.07542		&1.0197		&\nodata				&\nodata			&$14.19\pm0.02$	&$14.17\pm0.03$	&2\\
			&					&		&0.07500		&1.0174		&$2.950\pm0.001$		&$72.08\pm0.01$	&$13.35\pm0.02$	&$13.43\pm0.03$	&\\
1333$+$005	&LP\,618-14			&No		&0.07728		&1.0302		&\nodata				&\nodata			&$15.59\pm0.02$	&$15.62\pm0.03$	&1\\
1334$-$326	&EC\,13349$-$3237 	&No		&0.07345		&1.0648		&\nodata				&\nodata			&$15.03\pm0.02$	&$15.04\pm0.03$	&1\\
1339$+$606	&SBS				&No		&0.07388		&1.0424		&\nodata				&\nodata			&$16.49\pm0.02$	&$16.49\pm0.03$	&1\\
1339$+$346	&PG					&Yes	&0.07335		&1.0288		&\nodata				&\nodata			&$16.06\pm0.02$	&$16.02\pm0.03$	&2,7\\
			&					&		&0.07527		&1.0320		&$2.521\pm0.001$		&$212.69\pm0.02$	&$14.71\pm0.02$	&$14.70\pm0.03$	&8\\
1412$-$049	&PG					&Yes	&0.07208		&1.0433		&\nodata				&\nodata			&$17.12\pm0.03$	&$17.06\pm0.04$	&2\\
			&					&		&0.07513		&1.0340		&$3.507\pm0.001$		&$255.50\pm0.02$	&$14.79\pm0.02$	&$14.81\pm0.03$	&\\
1419$+$576	&SBS				&Yes	&0.07475		&1.0167		&\nodata				&\nodata			&$17.52\pm0.03$	&$17.48\pm0.04$	&2\\
			&					&		&0.07672		&1.1125		&$0.655\pm0.002$		&$304.04\pm0.12$	&$15.06\pm0.02$	&$15.10\pm0.03$	&4\\
1433$+$538	&GD\,337				&No		&0.07440		&1.0056		&\nodata				&\nodata			&$15.67\pm0.02$	&$15.67\pm0.03$	&1\\
1434$+$289	&Ton\,210				&No		&0.07425		&1.0381		&\nodata				&\nodata			&$16.08\pm0.02$	&$16.02\pm0.03$	&1,7\\
1435$+$370	&CBS\,194			&Yes	&0.07253		&1.0629		&\nodata				&\nodata			&$16.94\pm0.02$	&$16.90\pm0.03$	&2\\
			&					&		&0.07513		&1.0268		&$1.253\pm0.001$		&$318.95\pm0.05$	&$14.81\pm0.02$	&$14.86\pm0.03$	&\\
1436$-$216	&EC\,14363$-$2137		&No		&0.07530		&1.0267		&\nodata				&\nodata			&$14.48\pm0.02$	&$14.51\pm0.03$	&1\\
1443$+$336	&PG					&Yes	&0.07190		&1.0299		&\nodata				&\nodata			&$16.91\pm0.02$	&$16.86\pm0.03$	&2\\
			&					&		&0.07603		&1.0177		&$0.679\pm0.001$		&$286.25\pm0.09$	&$15.53\pm0.02$	&$15.57\pm0.03$	&\\
1458$+$171	&PG					&No		&0.07473		&1.0231		&\nodata				&\nodata			&$15.79\pm0.02$	&$15.79\pm0.03$	&1\\
1502$+$349	&CBS\,223			&Yes	&0.07478		&1.0238		&\nodata				&\nodata			&$16.85\pm0.02$	&$16.81\pm0.03$	&2\\
			&					&		&0.07628		&1.0310		&$1.912\pm0.002$		&$30.13\pm0.04$	&$17.04\pm0.03$	&$17.16\pm0.04$	&\\
1504$+$546	&CBS\,301			&No		&0.07555		&1.0352		&\nodata				&\nodata			&$15.01\pm0.02$	&$15.04\pm0.03$	&1\\
1517$+$502	&CBS\,311			&No		&0.07483		&1.0398		&\nodata				&\nodata			&$16.66\pm0.02$	&$16.68\pm0.03$	&1\\
1522$+$508	&CBS\,318			&No		&0.07560		&1.0307		&\nodata				&\nodata			&$15.96\pm0.02$	&$16.00\pm0.03$	&1\\
1558$+$616	&HS\,1558$+$6140		&Yes	&0.07465		&1.0234		&\nodata				&\nodata			&$17.27\pm0.03$	&$17.23\pm0.04$	&2\\
			&					&		&0.07480		&1.0224		&$0.719\pm0.001$		&$336.00\pm0.10$	&$15.71\pm0.02$	&$15.80\pm0.03$	&\\
1603$+$125	&					&No		&0.07248		&1.0349		&\nodata				&\nodata			&$14.18\pm0.02$	&$14.17\pm0.03$	&1,7\\
1619$+$525	&PG					&Yes	&0.07340		&1.0187		&\nodata				&\nodata			&$15.81\pm0.02$	&$15.77\pm0.03$	&3\\
			&					&		&0.07445		&1.0112		&$2.593\pm0.001$		&$282.58\pm0.02$	&$15.20\pm0.02$	&$15.24\pm0.03$	&\\
			&					&		&0.07545		&1.0378		&$0.465\pm0.002$		&$24.02\pm0.22$	&$17.84\pm0.04$	&$17.96\pm0.05$	&\\
1619$+$414	&KUV\,16195$+$4125	&Yes	&0.07443		&1.0105		&\nodata				&\nodata			&$17.38\pm0.03$	&$17.35\pm0.04$	&2,5\\
			&					&		&0.07630		&1.0074		&$0.232\pm0.001$		&$188.48\pm0.09$	&$15.68\pm0.02$	&$15.80\pm0.03$	&5\\
1622$+$323	&PG					&Yes	&0.07478		&1.0414		&\nodata				&\nodata			&$16.92\pm0.03$	&$16.86\pm0.04$	&2,5\\
			&					&		&0.07420		&1.0408		&$0.093\pm0.002$		&$298.18\pm0.37$	&$15.77\pm0.02$	&$15.81\pm0.03$	&5\\
1631$+$781	&RX\,J1629.0$+$7804	&Yes	&0.07435		&1.0082		&\nodata				&\nodata			&$13.61\pm0.02$	&$13.56\pm0.03$	&2,5\\
			&					&		&0.07570		&1.1008		&$0.302\pm0.001$		&$175.96\pm0.10$	&$12.28\pm0.02$	&$12.33\pm0.03$	&4,5\\
1643$+$143	&PG					&Yes	&0.07428		&1.0093		&\nodata				&\nodata			&$16.27\pm0.02$	&$16.22\pm0.03$	&2,5\\
			&					&		&0.07510		&1.0129		&$0.312\pm0.001$		&$20.87\pm0.06$	&$13.91\pm0.02$	&$13.95\pm0.03$	&5\\
1646$+$062	&PG					&Yes	&0.07308		&1.0284		&\nodata				&\nodata			&$16.45\pm0.03$	&$16.40\pm0.04$	&2,5\\
			&					&		&0.07433		&1.0327		&$0.161\pm0.001$		&$269.03\pm0.14$	&$15.42\pm0.02$	&$15.49\pm0.03$	&5\\
1658$+$440	&PG					&No		&0.07345		&1.0213		&\nodata				&\nodata			&$15.09\pm0.02$	&$15.04\pm0.03$	&1\\
1717$-$345	&					&No		&0.07617		&1.0086		&\nodata				&\nodata			&$14.17\pm0.02$	&$14.22\pm0.03$	&1\\
1833$+$644	&KUV\,18332$+$6429	&Yes	&0.07360		&1.0672		&\nodata				&\nodata			&$17.31\pm0.04$	&$17.26\pm0.05$	&3\\
			&					&		&0.07588		&1.0400		&$0.079\pm0.002$		&$221.91\pm0.48$	&$15.37\pm0.02$	&$15.43\pm0.03$	&\\
			&					&		&0.07700		&1.0489		&$1.820\pm0.002$		&$129.47\pm0.03$	&$19.60\pm0.08$	&$19.83\pm0.09$	&\\
1845$+$683	&KUV18453$+$6819	&No		&0.07343		&1.0098		&\nodata				&\nodata			&$15.67\pm0.02$	&$15.61\pm0.03$	&1,7\\
2009$+$622	&GD\,543				&No		&0.07468		&1.0291		&\nodata				&\nodata			&$15.03\pm0.02$	&$15.02\pm0.03$	&1\\
2151$-$015	&LTT\,8747			&Yes	&0.07443		&1.0252		&\nodata				&\nodata			&$14.32\pm0.02$	&$14.29\pm0.03$	&2\\
			&					&		&0.07790		&1.0227		&$1.082\pm0.001$		&$193.78\pm0.03$	&$15.06\pm0.02$	&$15.31\pm0.03$	&\\
2211$+$372	&LHS\,3779			&No		&0.07228		&1.0119		&\nodata				&\nodata			&$16.40\pm0.02$	&$16.38\pm0.03$	&1,7\\
2237$-$365	&HE\,2237$-$3630		&\nodata	&\nodata		&\nodata		&\nodata				&\nodata			&\nodata			&\nodata			&\\
			&LHS\,3841			&No		&0.07558		&1.0107		&\nodata				&\nodata			&$14.86\pm0.02$	&$14.90\pm0.03$	&1\\
2257$+$162	&PG					&No		&0.07500		&1.0081		&\nodata				&\nodata			&$16.06\pm0.02$	&$16.04\pm0.03$	&1\\
2317$+$268	&KUV\,23176$+$2650	&No		&0.07492		&1.0257		&\nodata				&\nodata			&$15.64\pm0.02$	&$15.65\pm0.03$	&1\\
2323$+$256	&G128-62				&No		&0.07428		&1.0224		&\nodata				&\nodata			&$16.30\pm0.02$	&$16.28\pm0.03$	&1,7\\
2336$-$187	&G273-97				&No		&0.07483		&1.0049		&\nodata				&\nodata			&$15.18\pm0.02$	&$15.16\pm0.03$	&1,7\\
2349$-$283	&GD\,1617			&No		&0.07358		&1.0107		&\nodata				&\nodata			&$15.65\pm0.02$	&$15.61\pm0.03$	&1,7

\enddata

\tablecomments{
(1) Single point source;
(2) Double point source, two Airy disks;
(3) Triple point source, three Airy disks
(4) Single but elongated Airy disk;
(5) Measurements affected by $\alpha<0\farcs4$;
(6) Common proper-motion companion;
(7) Not a white dwarf binary (Appendix, Table \ref{tbl8});
(8) Background star, optical but not physical companion.}

\tablenotetext{a}{All photometry is in Vega magnitudes.}

\tablenotetext{b}{The component identification in this system was not unambiguous.  See Appendix for details.}

\end{deluxetable*}

\tabletypesize{\normalsize}

\begin{deluxetable}{cccc}
\tablecaption{Primary Target Coordinates\label{tbl2}}
\tablewidth{0pt}
\tablehead{
\colhead{WD}			&
\colhead{R.A.}			&
\colhead{Dec.}			&
\colhead{Epoch}		\\
&($\rm ^h\,^m\,^s$)
&($\arcdeg\,\arcmin\,\arcsec$)
&(yr)}

\startdata

0023$+$388		&00 26 33.2		&$+$39 09 03		&2004.56\\
0034$-$211		&00 37 24.9		&$-$20 53 43 		&2004.56\\
0116$-$231		&01 18 37.2		&$-$22 54 59		&2004.67\\
0131$-$163		&01 34 24.1		&$-$16 07 08		&2004.67\\
0145$-$221		&01 47 21.8		&$-$21 56 51		&2005.43\\
0145$+$257		&01 48 08.1		&$-$25 32 44		&2004.71\\
0205$+$133		&02 08 03.5		&$+$13 36 25		&2004.66\\
0208$-$153		&02 10 43.0		&$-$15 06 33		&2004.55\\
0219$+$282		&02 22 28.4		&$+$28 30 08		&2004.62\\
0237$+$115		&02 40 06.7		&$+$11 48 30		&2004.60\\
0257$-$005		&03 00 24.5		&$-$00 23 42		&2005.50\\
0303$-$007		&03 06 07.1		&$-$00 31 14		&2004.56\\
0309$-$275		&03 11 33.1		&$-$27 19 25		&2005.43\\
0324$+$738		&03 30 13.9		&$+$74 01 57		&2004.70\\
0331$-$356		&03 33 52.4		&$-$35 31 18		&2005.37\\
0347$-$137		&03 50 14.6		&$-$13 35 14		&2004.57\\
0354$+$463		&03 58 17.1		&$+$46 28 41		&2004.70\\
0357$-$233		&03 59 04.9		&$-$23 12 25		&2004.67\\
0430$+$136		&04 33 10.3		&$+$13 45 17		&2005.16\\
0458$-$665		&04 58 54.0		&$-$66 28 13		&2004.62\\
0518$+$333		&05 21 43.5		&$+$33 21 59		&2005.16\\
0752$-$146		&07 55 08.9		&$-$14 45 53		&2005.22\\
0807$+$190\tablenotemark{a}		
				&08 10 00.3		&$+$18 51 47		&2005.33\\
0812$+$478		&08 15 48.9		&$+$47 40 39		&2005.18\\
0824$+$288		&08 27 04.9		&$+$28 44 03		&2005.17\\
0908$+$226		&09 11 43.1		&$+$22 27 49		&2005.17\\
0915$+$201		&09 18 33.0		&$+$19 53 08		&2005.18\\
0933$+$025		&09 35 40.7		&$+$02 21 58		&2005.17\\
0937$-$095\tablenotemark{a}		
				&09 39 49.8		&$-$09 46 11		&2005.20\\
0949$+$451		&09 52 21.9		&$+$44 54 31 		&2004.72\\
1001$+$203		&10 04 04.2		&$+$20 09 23		&2005.29\\
1015$-$173		&10 17 28.8		&$-$17 37 07		&2005.23\\
1026$+$002		&10 28 34.9		&$-$00 00 30		&2005.29\\
1033$+$464		&10 36 25.2		&$+$46 08 31		&2005.20\\
1036$-$204		&10 38 55.3		&$-$20 40 55		&2005.25\\
1037$+$512		&10 40 16.8		&$+$51 56 48		&2005.19\\
1049$+$103		&10 52 27.7		&$+$10 03 38		&2005.23\\
1051$+$516		&10 54 21.9		&$+$51 22 54		&2004.73\\
1106$+$316		&11 08 43.0		&$+$31 23 56		&2005.17\\
1106$-$211\tablenotemark{a}		
				&11 09 10.9		&$-$21 23 33		&2005.24\\
1108$+$325		&11 10 45.9		&$+$32 14 47		&2005.32\\
1133$+$489		&11 36 09.5		&$+$48 43 19		&2004.56\\
1133$+$358		&11 35 42.7		&$+$35 34 24		&2005.22\\
1140$+$004		&11 43 12.5		&$+$00 09 26		&2005.33\\
1156$+$132\tablenotemark{b}	
				&11 59 32.6		&$+$13 00 13		&2005.19\\
				&11 59 33.1		&$+$13 00 32		&\nodata\\
1156$+$129		&11 59 15.6		&$+$12 39 29		&2005.41\\
1210$+$464		&12 12 59.8		&$+$46 09 47		&2005.18\\
1218$+$497		&12 21 05.3		&$+$49 27 20		&2004.56\\
1236$-$004		&12 38 36.4		&$-$00 40 43		&2004.55\\
1247$+$550		&12 50 07.5		&$+$54 47 00		&2004.66\\
1247$-$176		&12 50 22.1		&$-$17 54 48		&2005.18\\
1307$-$141		&13 10 22.5		&$-$14 27 08		&2005.26\\
1333$+$487		&13 36 01.7		&$+$48 28 45		&2004.64\\
1333$+$005		&13 36 16.0		&$+$00 17 31		&2004.56\\
1334$-$326		&13 37 50.6		&$-$32 52 22		&2005.17\\
1339$+$606		&13 41 00.0		&$+$60 26 10		&2004.58\\
1339$+$346		&13 41 18.1		&$+$34 21 55		&2005.24\\
1412$-$049		&14 15 02.5		&$-$05 11 03		&2004.56\\
1419$+$576		&14 21 05.2		&$+$57 24 56		&2004.70\\
1433$+$538		&14 34 43.1		&$+$53 35 25		&2004.56\\
1434$+$289		&14 36 41.6		&$+$28 44 52		&2005.25\\
1435$+$370		&14 37 36.7		&$+$36 51 35 		&2004.59\\
1436$-$216		&14 39 12.8		&$-$21 50 12		&2005.18\\
1443$+$336		&14 46 00.6		&$+$33 28 50		&2004.67\\
1458$+$171		&15 00 19.4		&$+$16 59 16		&2004.57\\
1502$+$349		&15 04 31.8		&$+$34 47 01		&2004.70\\
1504$+$546		&15 06 05.3		&$+$54 28 19		&2004.56\\
1517$+$502		&15 19 05.9		&$+$50 07 03		&2004.56\\
1522$+$508		&15 24 25.2		&$+$50 40 11		&2005.22\\
1558$+$616		&15 58 55.4		&$+$61 32 03		&2004.66\\
1603$+$125\tablenotemark{a}		
				&16 05 32.1		&$+$12 25 42		&2004.61\\
1619$+$525		&16 20 24.5		&$+$52 23 22		&2004.62\\
1619$+$414		&16 21 12.6		&$+$41 18 10		&2004.64\\
1622$+$323		&16 24 49.0		&$+$32 17 02		&2004.62\\
1631$+$781		&16 29 09.9		&$+$78 04 40		&2004.64\\
1643$+$143		&16 45 39.2		&$+$14 17 45		&2005.23\\
1646$+$062		&16 49 07.8		&$+$06 08 46		&2004.64\\
1658$+$440		&16 59 48.4		&$+$44 01 05		&2005.26\\
1717$-$345		&17 21 10.3		&$-$34 33 28		&2005.20\\
1833$+$644		&18 33 29.2		&$+$64 31 52		&2005.19\\
1845$+$683		&18 45 09.2		&$+$68 22 34		&2004.66\\
2009$+$622		&20 10 42.6		&$+$62 25 31		&2004.73\\
2151$-$015		&21 54 06.5		&$-$01 17 11		&2004.59\\
2211$+$372		&22 14 08.9		&$+$37 27 11		&2004.63\\
2237$-$365\tablenotemark{b}	
				&22 39 59.4		&$-$36 16 01		&2004.64\\
				&22 40 05.3		&$-$36 15 19		&\nodata\\
2257$+$162		&22 59 46.9		&$+$16 29 17		&2005.45\\
2317$+$268		&23 20 04.1		&$+$26 06 23		&2004.58\\
2323$+$256		&23 25 57.9		&$+$25 52 21		&2004.66\\
2336$-$187		&23 38 52.8		&$-$18 26 14		&2005.32\\
2349$-$283		&23 52 23.2		&$-$28 03 16		&2004.65

\enddata

\tablenotetext{a}{Not a white dwarf (Appendix, Table \ref{tbl8}).}

\tablenotetext{b}{These white dwarfs were not imaged by ACS; the first row of coordinates gives
the position of the imaged (incorrect) star, while the second set of coordinates gives the correct
position for the white dwarf.}

\end{deluxetable}

\clearpage

\begin{deluxetable}{lccccc}
\tablecaption{White Dwarf Parameters\label{tbl3}}
\tablewidth{0pt}
\tablehead{
\colhead{WD}			&
\colhead{$T_{\rm eff}$}	&
\colhead{$\log\,g$}		&
\colhead{$V$}			&
\colhead{$d_{\rm wd}$}	&
\colhead{References}	\\
&(K)
&(cm\,s$^{-2}$)
&(mag)
&(pc)
&}

\startdata

0023$+$388	&10\,800	&8.14	&15.97	&59		&1,2\\		
0034$-$211	&17\,200	&8.04	&14.89	&61		&1,2\\	
0116$-$231	&25\,000	&7.9		&16.29	&164	&3\\
0131$-$163	&49\,000	&7.81	&13.94	&105	&2\\		
0145$-$221	&11\,600	&8.10	&14.85	&39		&4,5\\		
0145$-$257	&26\,200	&7.93	&14.66	&80		&6,7\\		
0205$+$133	&57\,400	&7.63	&15.06	&223	&8\\		
0208$-$153	&20\,000	&7.9		&15.69	&101	&1\\		
0219$+$282	&36\,300	&8.09	&16.97	&310	&1,9\\		
			&27\,200	&8.09	&\nodata	&\nodata	&9\\	
0237$+$115	&70\,000	&8.0		&15.97	&273	&1,10\\		
0257$-$005	&80\,900	&7.13	&16.97	&1040	&1,11\\	
0303$-$007	&18\,700	&7.97	&16.59	&144	&2\\		
0309$-$275	&40\,000	&7.8		&16.46	&292	&1\\		
0324$+$738	&8500	&8.65	&16.85	&40		&1\\		
0331$-$356	&31\,400	&7.70	&15.11	&141	&4\\		
0347$-$137	&15\,500	&8.0		&15.32	&66		&12\\		
0354$+$463	&8000	&8.0		&15.55	&32		&1,12\\		
0357$-$233	&35\,000	&7.9		&15.87	&199	&1,13\\
0430$+$136	&36000	&7.90	&17.33	&376	&,1,2\\
0458$-$665	&20\,000	&7.9		&17.72	&258	&14\\		
0518$+$333	&9500	&7.77	&16.10	&65		&1,11\\	
0752$-$146	&18\,500	&8.0		&13.53	&35		&1,15\\		
0812$+$478	&60\,900	&7.58	&15.22	&259	&8\\
0824$+$288	&50\,700	&7.74	&14.80	&168	&1,8\\		
0908$+$226	&15\,000	&8.0		&19.22	&402	&1\\
0915$+$201	&70\,000	&7.33	&16.52	&643	&8\\		
0933$+$025	&22\,400	&8.04	&16.24	&135	&8\\
0949$+$451	&11\,000	&8.0		&15.89	&65		&1\\
1001$+$203	&21\,500	&7.97	&15.91	&117	&8\\		
1015$-$173	&25\,000	&7.9		&16.99	&226	&1\\		
1026$+$002	&17\,200	&7.97	&13.95	&40		&8\\		
1033$+$464	&29\,400	&7.88	&14.37	&83		&8\\		
1036$-$204	&7500	&8.0		&16.28	&29		&16,11\\	
1037$+$512	&20\,100	&8.03	&16.38	&133	&8\\		
1049$+$103	&20\,600	&7.91	&15.79	&112	&1,8\\		
1051$+$516	&20\,000	&7.9		&17.42	&225	&1,17\\		
1106$+$316	&25\,000	&7.9		&17.26	&257	&1,18\\
1108$+$325	&63\,000	&7.59	&16.83	&550	&1,8\\	
1133$+$358	&6500	&7.67	&16.87	&50		&1,19,20\\		
1133$+$489	&48\,000	&8.00	&17.12	&384	&1,21\\		
1140$+$004	&16\,000	&8.0		&18.30	&278	&1,17\\		
1156$+$129	&15\,000	&8.0		&17.85	&214	&1\\		
1210$+$464	&27\,700	&7.85	&16.03	&171	&8\\	
1218$+$497	&35\,700	&7.87	&16.36	&250	&1,8\\		
1236$-$004	&44\,700	&7.60	&17.72	&673	&17\\		
1247$-$176	&20\,900	&8.06	&16.50	&141	&22\\		
1307$-$141	&20\,000	&7.9		&16.65	&158	&1\\		
1333$+$487	&16\,000	&8.20	&14.15	&35		&23,24\\		
1333$+$005	&7500	&8.0		&17.47	&69		&1,25\\		
1334$-$326	&35\,000	&7.52	&17.29	&619	&4,26\\		
1339$+$606	&46\,300	&7.80	&17.10	&439	&17\\		
1339$+$346	&16\,000	&7.82	&15.93	&103	&8\\	
1412$-$049	&50\,000	&7.8		&16.73	&398	&1\\		
1419$+$576	&25\,000	&7.9		&17.27	&258	&1\\		
1433$+358$	&22\,400	&7.80	&16.13	&151	&8\\		
1434$+$289	&32\,800	&8.00 	&15.71	&156	&8\\		
1435$+$370	&15\,300	&7.99	&16.75	&129	&2\\		
1436$-$216	&25\,000	&7.9		&16.64	&193	&1\\
1443$+$336	&29\,800	&7.83	&16.56	&239	&8\\		
1458$+$171	&22\,000	&7.43	&16.39	&226	&8\\		
1502$+$349	&21\,300	&7.96	&16.58	&161	&2\\		
1504$+$546	&24\,700	&7.86	&17.07	&243	&2\\		
1517$+$502	&31\,100	&7.84	&17.95	&406	&27\\
1522$+$508	&21\,600	&8.10	&17.65	&238	&17\\
1558$+$616	&15\,000	&8.0		&16.95	&141	&1\\
1619$+$525	&18\,000	&7.90	&15.62	&94		&1,8\\
1619$+$414	&14\,100	&7.93	&17.22	&156	&1,2\\		
1622$+$323	&68\,300	&7.56	&16.52	&513	&1,8\\	
1631$+$781	&44\,900	&7.76	&13.23	&75		&1,2\\
1643$+$143	&26\,800	&7.91	&15.95	&153	&1,8\\		
1646$+$062	&29\,900	&7.98	&16.10	&174	&1,8\\
1717$-$345	&12\,700	&8.0		&17.20	&137	&1,28\\		
1833$+$644	&25\,000	&7.9		&16.99	&227	&1\\
2009$+$622	&25\,900	&7.70	&15.29	&125	&1,29\\		
2151$-$015	&9100	&8.21	&14.54	&22		&1,2\\
2257$+$162	&24\,600	&7.49	&16.14	&213	&8\\		
2317$+$268	&31\,500	&7.70	&16.63	&284	&1,2,30\\		
2336$-$187	&8100	&8.05	&15.51	&31		&4		

\enddata

\tablecomments{A single digit following the decimal place for $\log\,g$ indicates an assumption 
of $M=0.60\,M_{\odot}$.  White dwarf $V$-band magnitudes are disentangled from the light of 
their companions, by virtue of being derived via:  1) spatially-resolved photometry; 2) effective 
temperature, model colors, and uncontaminated photometry in other bandpasses (e.g. F814W, 
$I$); or 3) best estimates based on photographic photometry when no other data are available.
Distances were calculated from absolute magnitudes using white dwarf models \citep{ber95a,
ber95b} and specified references.}

\tablerefs{
(1) This work and Paper I;
(2) \citealt{tre07};
(3) \citealt{wol96};
(4) \citealt{koe01};
(5) \citealt{fon03};
(6) \citealt{fin97};
(7) \citealt{ven97};
(8) \citealt{lie05a};
(9) \citealt{lim09};
(10) \citealt{dre96};
(11) \citealt{far09b};
(12) \citealt{zuc03};
(13) \citealt{far05a};
(14) \citealt{hut96};
(15) \citealt{egg65};
(16) \citealt{lie03};
(17) \citealt{sil06};
(18) \citealt{wag86};
(19) \citealt{dah88};
(20) \citealt{gre76};
(21) \citealt{hug06};
(22) \citealt{kil97};
(23) \citealt{cas06};
(24) \citealt{dah82};
(25) \citealt{kil06};
(26) \citealt{tap09};
(27) \citealt{lie94};
(28) \citealt{kaw04};
(29) \citealt{ber92};
(30) \citealt{osw84}}

\end{deluxetable}

\clearpage

\tabletypesize{\small}

\begin{deluxetable}{lccccc}
\tablecaption{Parameters of Resolved Secondary \& Tertiary Stars\label{tbl4}}
\tablewidth{0pt}
\tablehead{
\colhead{Primary}		&
\colhead{Companion}	&
\colhead{SpT}			&
\colhead{$I-K$}			&
\colhead{$d_{\rm rd}$}	&
\colhead{References}	\\
&
&
&
&(pc)
&}

\startdata

0034$-$211	&LTT\,0329B			&dM3	&2.15	&75		&1,2\\
0116$-$231	&GD\,695B			&dM4	&2.35	&166	&1,3,4\\
0131$-$163	&GD\,984B			&dM3.5	&2.26	&109	&1,5\\
0145$-$257 	&GD\,1401B			&dM3.5	&2.28	&80		&1,6,7\\
0205$+$133	&PG\,0205$+$133B		&dM2	&1.99	&157	&1,3,8\\
0208$-$153	&MCT\,0208$-$1520B	&dM2.5	&2.06	&121	&1,3\\
0219$+$282	&KUV\,02196$+$2816B	&dM5	&2.85	&324	&1,3\\
0237$+$115	&PG\,0237$+$115B		&dM2.5	&2.13	&207	&1,8,9\\
0257$-$005	&KUV\,02579$-$0036B	&dM4.5	&2.56	&494	&1,3,10\\
0309$-$275	&WD\,0309$-$275B		&dM0	&1.87	&339	&1,3\\
0324$+$738\tablenotemark{a}
			&G221-11A			&dM5	&2.68	&40		&1\\
			&G221-10B			&dM6	&3.12	&40		&1\\
0347$-$137	&GD\,51B				&dM4	&2.33	&51		&1,5,11\\
0357$-$233	&Ton\,S 392B			&dM3	&2.17	&414	&1,3,11\\
0430$+$136	&KUV\,04304$+$1339B	&dM4	&2.35	&103	&1,3\\
0518$+$333\tablenotemark{a}
			&G86-B1A			&dM3	&2.14	&65		&1\\
			&G86-B1C			&dM3	&2.17	&65		&1\\
0824$+$288	&PG\,0824$+$288B		&dC		&2.05	&\nodata	&1,12\\
			&PG\,0824$+$288C		&dM2	&2.06	&195	&1,11\\
0915$+$201	&LB\,3016B			&dM3	&2.20	&550	&1,3\\
0933$+$025	&PG\,0933$+$025B		&dM4	&2.32	&87		&1,5,11\\
0949$+$451	&HS\,0949$+$4508B	&dM4.5	&2.46	&58		&1,9\\
			&HS\,0949$+$4508C	&dM5	&2.85	&72		&1\\
1015$-$173	&EC\,10150$-$1722B	&dM3	&2.14	&459	&1,3\\
1033$+$464	&GD\,123B			&dM5	&2.68	&59		&1,5,11\\
1049$+$103	&PG\,1049$+$103B		&dM4.5	&2.46	&91		&1,11\\
1106$+$316	&Ton\,58B			&dM2.5	&2.12	&404	&1,3\\
1108$+$325	&Ton\,60B			&dM2.5	&2.12	&597	&1,3\\
1133$+$489	&PG\,1133$+$489B		&dM5	&2.64	&344	&1\\
1156$+$129	&WD\,1156$+$129B		&dM4	&2.36	&169	&1,3\\
1210$+464$	&PG\,1210$+$464B		&dM2	&1.99	&108	&1,5,11\\
1218$+$497	&PG\,1218$+$497B		&dM4	&2.34	&165	&1,3\\
1236$-$004	&WD\,1236$-$004B		&dM2.5	&2.13	&773	&1,3\\
1307$-$141	&EC\,13077$-$1411B	&dM4	&2.37	&112	&1,3\\
1333$+$487	&GD\,325B			&dM4	&2.30	&34		&1,2,13\\
1412$-$049	&PG\,1412$-$049B		&dM0	&1.81	&377	&1,3\\
1419$+$576	&SBS\,1419$+$576B	&dM2	&2.00	&374	&1,14\\
			&SBS\,1419$+$576C	&dM2	&2.02	&378	&1\\
1435$+$370	&CBS\,194B			&dM2.5	&2.09	&192	&1,3\\
1443$+$346	&PG\,1443$+$346B		&dM2	&2.02	&326	&1,3\\
1502$+$349	&CBS\,223B			&dM5	&2.78	&198	&1,3\\
1558$+$616	&HS\,1558$+$6140B	&dM4.5	&2.40	&135	&1,3\\
1619$+$525	&PG\,1619$+$525B		&dM2	&1.85	&291	&1,3\\
			&PG\,1619$+$525C		&dM5	&2.62	&297	&1,3\\
1619$+$414	&KUV\,16195$+$4125B	&dM5	&2.76	&106	&1,3\\
1622$+$323	&PG\,1622$+$323B		&dM2	&2.01	&365	&1,3\\
1631$+$781	&RX\,J1629.0$+$7804B	&dM3	&2.14	&73		&1,5,15\\
			&RX\,J1629.0$+$7804C	&dM3.5	&2.27	&72		&1\\
1643$+$143	&PG\,1643$+$143B		&dM2	&1.98	&156	&1,5,11\\
1646$+$062	&PG\,1646$+$062B		&dM3.5	&2.22	&168	&1\\
1833$+$644	&KUV\,18332$+$6429B	&dM3	&2.17	&249	&1,9\\
			&KUV\,18332$+$6429C	&dM6.5	&3.2		&249	&1\\
2151$-$015	&LTT\,8747B			&dM7.5	&3.82	&22		&1,5,11

\enddata
\tablenotetext{a}{These stars are common proper motion companions which were 
serendipitously resolved into two components, the entries are secondary and tertiary 
companions to their respective white dwarfs.}

\tablerefs{
(1) This work and Paper I;
(2) \citealt{pro83};
(3) \citealt{wac03};
(4) \citealt{lam00};
(5) \citealt{sch96};
(6) \citealt{gre00};
(7) \citealt{fin97};
(8) \citealt{gre86};
(9) \citealt{hoa07};
(10) \citealt{far09b};
(11) \citealt{far05a};
(12) \citealt{heb93};
(13) \citealt{gre74};
(14) \citealt{ste01};
(15) \citealt{coo92}}

\end{deluxetable}

\clearpage

\tabletypesize{\small}

\begin{deluxetable}{lccccc}
\tablecaption{Parameters of Unresolved Secondary Stars\label{tbl5}}
\tablewidth{0pt}
\tablehead{
\colhead{Primary}		&
\colhead{Companion}	&
\colhead{SpT}			&
\colhead{$I-K$}			&
\colhead{$d_{\rm rd}$}	&
\colhead{References}	\\
&
&
&
&(pc)
&}

\startdata

0023$+$388	&G171-B10C			&dM5.5	&3.02	&70		&1,2,3\\
0145$-$221	&GD\,1400B			&dL6.5	&5.8		&35		&1,2,4,5,6\\
0303$-$007	&KUV\,03036$-$0043B	&dM4	&2.32	&84		&1,2,7\\
0331$-$356	&HE\,0331$-$3541B	&dM2.5	&2.08	&77		&1,8\\
0354$+$463	&Rubin\,80B			&dM7	&3.63	&44		&1,9,10\\
0430$+$136\tablenotemark{a}
			&KUV\,04304$+$1339C	&dM5.5	&2.82	&151	&1\\
0458$-$665	&RX\,J0458.9$-$6628B	&dM3	&2.15	&176	&1,2,11\\
0752$-$146	&LTT\,2980B			&dM6	&3.07	&33		&1,9,10\\
0812$+$478	&PG\,0812$+$478B		&dM4	&2.33	&174	&1,2\\
0908$+$226	&LP\,369-15B			&dM3	&2.20	&425	&1,2\\
1001$+$203	&Ton\,1150B			&dM2.5	&2.13	&123	&1,9,10\\
1026$+$002	&PG\,1026$+$002B		&dM4.5	&2.50	&45		&1,9,10,12\\
1037$+$512	&PG\,1037$+$512B		&dM4	&2.37	&110	&1,2\\
1051$+$516	&SBS\,1051$+$516B	&dM3	&2.16	&183	&1,13,14\\
1133$+$358	&G147-65B			&dM4.5	&2.55	&50		&1,15,16\\
1140$+$004	&WD\,1140$+$004B		&dM4.5	&2.42	&265	&1,17\\
1247$-$176	&EC\,12477$-$1738B	&dM4.5	&2.46	&95		&1,2,8\\
1333$+$487\tablenotemark{a}	
			&GD\,325C			&dM6.5	&3.28	&34		&1\\
1333$+$005	&LP\,618-14B			&dM4.5	&2.39	&148	&1,9\\
1334$-$326	&EC\,13349$-$3237B	&dM0	&1.88	&432	&1,18\\
1339$+$606	&SBS\,1339$+$606B	&dM3.5	&2.25	&347	&1,19,20\\
1433$+$358	&GD\,337B			&dM4.5	&2.49	&181	&1,15\\
1436$-$216	&EC\,14363$-$2137B	&dM2.5	&2.11	&173	&1,2,21\\
1458$+$171	&PG\,1458$+$171B		&dM4.5	&2.57	&177	&1,2\\
1504$+$546	&CBS\,301B			&dM3	&2.16	&223	&1,2,13\\
1517$+$502	&CBS\,311B			&dC		&2.80	&\nodata	&1,2,22\\
1522$+$508	&CBS\,318B			&dM3.5	&2.23	&294	&1,2,14\\
1717$-$345	&WD\,1717$-$345B		&dM4	&2.34	&68		&1,23\\
2009$+$622	&GD\,543B			&dM4	&2.34	&151	&1,10\\
2257$+$162	&PG\,2257$+$162B		&dM4.5	&2.52	&275	&1,2\\
2317$+$268	&KUV\,23176$+$2650B	&dM3.5	&2.24	&216	&1,2

\enddata

\tablecomments{All of the binary systems represented by the table entries should be radial
velocity variables, some of which have been confirmed since the ACS targets were selected.
Additionally, the white dwarf primaries may be polluted by wind from the companions, as in
0354$+$463, and 1026$+$002.}

\tablenotetext{a}{These tentatively identified tertiary companions are identified on the basis 
of clear photometric excess in the F814W bandpass, and overly wide PSFs, but require 
spectroscopic confirmation.}

\tablerefs{
(1) This work and Paper I;
(2) \citealt{wac03};
(3) \citealt{rei96};
(4) \citealt{far04b};
(5) \citealt{far05b};
(6) \citealt{dob05};
(7) \citealt{weg87};
(8) \citealt{koe01};
(9) \citealt{far05a};
(10) \citealt{sch96};
(11) \citealt{hut96};
(12) \citealt{saf93};
(13) \citealt{ste01};
(14) \citealt{sil06};
(15) \citealt{pro83};
(16) \citealt{gre76};
(17) \citealt{ray03};
(18) \citealt{kil97};
(19) \citealt{fle96};
(20) \citealt{hoa07};
(21) \citealt{zuc03};
(22) \citealt{lie94};
(23) \citealt{rei88}}

\end{deluxetable}

\bigskip
\bigskip

\tabletypesize{\small}

\begin{deluxetable}{ccccc}
\tablecaption{Three Very Close (Visual) Double M Dwarfs\label{tbl6}}
\tablewidth{0pt}
\tablehead{
\colhead{Triple System}		&
\colhead{$\alpha$}			&
\colhead{P.A.}				&
\colhead{$I_B$}			&
\colhead{$I_C$}			\\
&(arcsec)
&(deg)
&(mag)
&(mag)}

\startdata

0949$+$451	&$0.038\pm0.004$		&$170.1\pm1.4$	&$14.00\pm0.08$	&$14.99\pm0.19$\\
1419$+$576	&$0.030\pm0.005$		&$22.0\pm3.1$		&$15.86\pm0.13$	&$15.89\pm0.13$\\
1631$+$781	&$0.029\pm0.005$		&$15.4\pm3.3$		&$12.76\pm0.09$	&$13.67\pm0.13$

\enddata

\end{deluxetable}

\clearpage

\tabletypesize{\footnotesize}

\begin{deluxetable}{ccccccccccc}
\tablecaption{Parameters for All Double Stars\label{tbl7}}
\tablewidth{0pt}
\tablehead{
\colhead{White Dwarf}			&
\colhead{Binary}				&
\colhead{}						&
\colhead{}						&
\colhead{}						&
\colhead{}						&
\colhead{}						&
\colhead{Minimum}				&
\colhead{Maximum}				&
\colhead{Binary}				&
\colhead{System}				\\
System
&Components		
&$\alpha$\tablenotemark{a}
&$d_{\rm wd}$	
&$d_{\rm rd}$			
&$\Delta (m-M)$
&$d_{\rm adopt}$
&Separation
&Separation
&Type
&Notes\\
&
&(arcsec)
&(pc)				
&(pc)	
&(mag)				
&(pc)
&(AU)
&(AU)
&						
&}

\startdata

0023$+$388	&AB		&24.6	&59		&43		&$+$0.67		&65		&1600		&\nodata		&DA$+$dM	&1\\
			&AC		&\nodata	&59		&70		&$-$0.36		&65		&\nodata		&1.6			&DA$+$dMe	&6\\
0034$-$211	&AB		&0.327	&61		&75		&$-$0.44		&68		&22			&\nodata		&DA$+$dMe	&5\\
0116$-$231	&AB		&1.106	&164	&166	&$-$0.02		&165	&182		&\nodata		&DA$+$dM	&\\
0131$-$163	&AB		&0.190	&105	&109	&$-$0.09		&107	&20			&\nodata		&DA$+$dM	&\\
0145$-$221	&AB		&\nodata	&43		&...		&...			&43		&\nodata		&1.1			&DA$+$dL	&7\\
0145$+$257	&AB		&2.294	&80		&80		&$+$0.01		&80		&184		&\nodata		&DA$+$dM	&\\
0205$+$133	&AB		&1.254	&223	&157	&$+$0.76		&190	&238		&\nodata		&DA$+$dM 	&1\\
0208$-$153	&AB		&2.648	&101	&121	&$-$0.39		&111	&294		&\nodata		&DA$+$dM	&\\
0219$+$282	&AB		&\nodata	&310	&...		&$-$0.09		&317	&\nodata		&7.9			&DA$+$DB 	&6\\
			&AC		&0.116	&310	&324	&$-$0.09		&317	&37			&\nodata		&DA$+$dM	&\\
0237$+$115	&AB		&0.124	&273	&207	&$+$0.59		&240	&30			&\nodata		&DO$+$dM 	&1\\ 
0257$-$005	&AB		&0.978	&1040	&490	&$+$1.61		&765	&748		&\nodata		&DAO$+$dM	&1\\
0303$-$007	&AB		&\nodata	&144	&84		&$+$1.17		&114	&\nodata		&2.9			&DA$+$dMe	&2,6\\	
0309$-$275	&AB		&0.099	&292	&339	&$-$0.32		&316	&31			&\nodata		&DA$+$dM	&\\
0324$+$738	&AB		&12.75	&40		&40		&...			&40		&510		&\nodata		&DC$+$dM	&8\\
			&BC		&0.297	&40		&40		&...			&40		&12			&\nodata		&dM$+$dM	&8\\
0331$-$356	&AB		&\nodata	&141	&77		&$+$1.31		&109	&\nodata		&2.7			&DA$+$dMe	&2,6\\
0347$-$137	&AB		&1.052	&66		&51		&$+$0.57		&59		&62			&\nodata		&DA$+$dMe	&1,5\\
0354$+$463	&AB		&\nodata	&32		&44		&$-$0.67		&38		&\nodata		&1.0			&DAZ$+$dMe	&4,6\\
0357$-$233	&AB		&1.191	&199	&414	&$-$1.60		&307	&365		&\nodata		&DA$+$dM	&3\\
0430$+$136	&AB		&0.260	&376	&103	&$+$2.82		&236	&61			&\nodata		&DA$+$dMe	&1\\
			&AC:		&\nodata	&376	&151	&$+$1.99		&236	&\nodata		&5.9			&DA$+$dMe	&2,6\\
0458$-$665	&AB		&\nodata	&258	&176	&$+$0.83		&217	&\nodata		&5.4			&DA$+$dMe	&2,6\\
0518$+$333	&AB		&7.475	&65		&65		&...			&65		&486		&\nodata		&DA$+$dM	&8\\
			&BC		&0.158	&65		&65		&...			&65		&10			&\nodata		&dM$+$dM	&8\\
0752$-$146	&AB		&\nodata	&35		&33		&$+$0.10		&34		&\nodata		&0.9			&DA$+$dMe	&6\\
0812$+$478	&AB		&\nodata	&259	&174	&$+$0.86		&217	&\nodata		&5.4			&DA$+$dM	&2,6\\
0824$+$288	&AB		&3.330	&168	&195	&$-$0.32		&182	&604		&\nodata		&DA$+$dM	&\\
			&AC		&0.077	&168	&...		&...			&182	&14			&\nodata		&DA$+$dCe	&5\\
0908$+$226	&AB		&\nodata	&402	&425	&$-$0.12		&414	&\nodata		&10			&DA$+$dMe	&6\\
0915$+$201	&AB		&2.312	&643	&550	&$+$0.34		&597	&1380		&\nodata		&DA$+$dM	&\\
0933$+$025	&AB		&1.232	&135	&87		&$+$0.95		&111	&137		&\nodata		&DA$+$dMe	&1,5\\
0949$+$451	&AB		&2.892	&65		&65		&$+$0.01		&65		&188		&\nodata		&DA$+$dM	&\\
			&BC		&0.038	&...		&65		&$+$0.01		&65		&2.5			&\nodata		&dM$+$dM	&\\	
1001$+$203	&AB		&\nodata	&117	&123	&$-$0.10		&120	&\nodata		&3.0			&DA$+$dMe	&6\\
1015$-$173	&AB		&0.060	&226	&459	&$-$1.53		&343	&21			&\nodata		&DA$+$dM	&3\\
1026$+$002	&AB		&\nodata	&40		&45		&$-$0.28		&42		&\nodata		&1.1			&DAZ$+$dMe	&7\\		
1033$+$464	&AB		&0.731	&83		&59		&$+$0.75		&71		&52			&\nodata		&DA$+$dMe	&1,5\\
1037$+$512	&AB		&\nodata	&133	&110	&$+$0.40		&121	&\nodata		&3.0			&DA$+$dM	&6\\
1049$+$103	&AB		&0.262	&112	&91		&$+$0.45		&102	&27			&\nodata		&DAZ$+$dM	&\\	
1051$+$516	&AB		&\nodata	&225	&183	&$+$0.45		&204	&\nodata		&5.1			&DA$+$dMe	&6\\
1106$+$316	&AB		&0.478	&257	&404	&$-$0.98		&331	&158		&\nodata		&DA$+$dM	&3\\
1108$+$325	&AB		&0.167	&550	&597	&$-$0.18		&574	&96			&\nodata		&DA$+$dM	&\\	
1133$+$489	&AB		&0.090	&384	&344	&$+$0.23		&364	&33			&\nodata		&DO$+$dM	&\\
1133$+$358	&AB		&\nodata	&50		&50		&...			&50		&\nodata		&1.3			&DC$+$dMe	&6,8\\
1140$+$004	&AB		&\nodata	&278	&265	&$+$0.10		&272	&\nodata		&6.8			&DA$+$dMe	&6\\
1156$+$129	&AB		&0.564	&214	&169	&$+$0.51		&192	&108		&\nodata		&DA$+$dM	&1\\
1210$+$464	&AB		&1.043	&171	&108	&$+$0.99		&140	&145		&\nodata		&DA$+$dMe	&1,5\\
1218$+$497	&AB		&0.303	&250	&165	&$+$0.90		&208	&63			&\nodata		&DA$+$dM	&1\\
1236$-$004	&AB		&0.663	&575	&773	&$-$0.64		&674	&447		&\nodata		&DA$+$dMe	&3,5\\
1247$-$176	&AB		&\nodata	&141	&95		&$+$0.85		&118	&\nodata		&3.0			&DA$+$dM	&2,7\\
1307$-$141	&AB		&2.133	&158	&112	&$+$0.76		&135	&288		&\nodata		&DA$+$dM	&1\\
1333$+$487	&AB		&2.950	&34		&34		&...			&34		&100		&\nodata		&DB$+$dM	&8\\
			&AC:		&\nodata	&34		&34		&...			&34		&\nodata		&0.9			&DB$+$dM	&6,8\\
1333$+$005	&AB		&\nodata	&69		&148	&$-$1.66		&109	&\nodata		&2.7			&DA$+$dM	&4,6\\
1334$-$326	&AB		&\nodata	&619	&432	&$+$0.78		&526	&\nodata		&13			&DA$+$dMe	&2,7\\
1339$+$606	&AB		&\nodata	&439	&347	&$+$0.51		&393	&\nodata		&9.8			&DA$+$dMe	&2,6\\
1412$-$049	&AB		&3.507	&398	&377	&$+$0.12		&388	&1360		&\nodata		&DA$+$dM	&\\
1419$+$576	&AB		&0.655	&258	&376	&$-$0.82		&317	&208		&\nodata		&DB$+$dM	&2\\
			&BC		&0.030	&...		&376	&...			&317	&9.5			&\nodata		&dM$+$dM	&\\
1433$+$538	&AB		&\nodata	&151	&181	&$-$0.39		&166	&\nodata		&4.2			&DA$+$dM	&6\\
1435$+$370	&AB		&1.253	&129	&192	&$-$0.87		&161	&201		&\nodata		&DA$+$dM	&3\\
1436$-$216	&AB		&\nodata	&193	&173	&$+$0.24		&183	&\nodata		&4.6			&DA$+$dMe	&6\\
1443$+$336	&AB		&0.679	&239	&326	&$-$0.68		&283	&192		&\nodata		&DA$+$dM	&3\\
1458$+$171	&AB		&\nodata	&226	&177	&$+$0.53		&202	&\nodata		&5.0			&DA$+$dMe	&2,6\\	
1502$+$349	&AB		&1.912	&161	&198	&$-$0.46		&180	&343		&\nodata		&DA$+$dM	&\\	
1504$+$546	&AB		&\nodata	&243	&223	&$+$0.19		&233	&\nodata		&5.8			&DA$+$dMe	&6\\
1517$+$502	&AB		&\nodata	&406	&...		&...			&406	&\nodata		&10			&DA$+$dCe	&6\\
1522$+$508	&AB		&\nodata	&238	&294	&$-$0.46		&266	&\nodata		&6.7			&DA$+$dMe	&6\\
1558$+$616	&AB		&0.719	&141	&135	&$+$0.10		&138	&99			&\nodata		&DA$+$dM	&\\	
1619$+$525	&AB		&2.593	&94		&291	&$-$2.46		&194	&503		&\nodata		&DA$+$dM	&1\\
			&AC		&0.465	&94		&297	&$-$2.51		&194	&90			&\nodata		&DA$+$dM	&1\\
1619$+$414	&AB		&0.232	&156	&106	&$+$0.83		&131	&30			&\nodata		&DA$+$dM	&1\\
1622$+$323	&AB		&0.093	&513	&365	&$+$0.74		&439	&41			&\nodata		&DA$+$dM	&1\\
1631$+$781	&AB		&0.302	&75		&73		&$+$0.07		&74		&22			&\nodata		&DA$+$dMe	&5\\
			&BC		&0.029	&...		&73		&...			&74		&2.1			&\nodata		&dM$+$dM	&\\	
1643$+$143	&AB		&0.312	&153	&156	&$-$0.05		&155	&48			&\nodata		&DA$+$dM	&\\
1646$+$062	&AB		&0.161	&174	&168	&$+$0.07		&171	&28			&\nodata		&DA$+$dM	&\\
1717$-$345	&AB		&\nodata	&137	&68		&$+$1.53		&103	&\nodata		&2.6			&DA$+$dMe	&2,6\\
1833$+$644	&AB		&0.079	&227	&249	&$-$0.21		&238	&19			&\nodata		&DA$+$dM	&\\
			&AC		&1.820	&227	&...		&...			&238	&429		&\nodata		&DA$+$dM	&\\
2009$+$622	&AB		&\nodata	&125	&151	&$-$0.41		&138	&\nodata		&3.5			&DA$+$dM	&7\\
2151$-$015	&AB		&1.082	&22		&22		&$+$0.04		&22		&24			&\nodata		&DA$+$dMe	&5\\
2257$+$162	&AB		&\nodata	&213	&275	&$-$0.56		&244	&\nodata		&6.1			&DA$+$dM	&4,6\\
2317$+$268	&AB		&\nodata	&284	&216	&$+$0.59		&250	&\nodata		&6.3			&DA$+$dM 	&3,6

\enddata

\tablecomments{
(1) Wide Binary with $\Delta (m-M)>+0.5$\,mag;
(2) Close Binary with $\Delta (m-M)>+0.5$\,mag;
(3) Wide Binary with $\Delta (m-M)<-0.5$\,mag;
(4) Close Binary with $\Delta (m-M)<-0.5$\,mag;
(5) Wide Binary with active dMe secondary; 
(6) Radial velocity variable candidate;
(7) Radial velocity variable with known period (\S3.6);
(8) Distance from trigonometric parallax.}

\tablenotetext{a}{For spatially-unresolved binaries, an upper limit of $0\farcs025$ was assumed
based on the separations of the closest spatially-resolved pairs in Table \ref{tbl6}.}

\end{deluxetable}

\bigskip
\bigskip

\tabletypesize{\small}

\begin{deluxetable}{lcccc}
\tabletypesize{\small}
\tablecaption{Suspected or Confirmed Non-Binaries\label{tbl8}}
\tablewidth{0pt}
\tablehead{
\colhead{WD}			&
\colhead{$I-K$}			&
\colhead{SpT}			&
\colhead{References}}

\startdata

0324$+$738	&$+0.25$		&DC6	&1\\
0518$+$333	&$-0.07$		&DA5	&1,2\\
0807$+$190	&$+2.49$		&sdK	&1,3,4\\
0937$-$025	&$+1.57$		&sdK5	&1\\
1036$-$204	&$+1.22$		&DQp7	&1,2,5\\
1106$-$211	&$+1.78$		&dK7	&1\\
1247$+$550	&$+0.53$		&DC12	&1,6\\
1339$+$346	&$-0.13$		&DA3	&1,2\\
1434$+$289	&$-0.68$		&DA2	&1,2\\
1603$+$125	&$+1.19$		&dK2	&1\\
1658$+$440	&$-0.49$		&DA2	&1,7\\
1845$+$683	&$-0.69$		&DA1	&1,2\\
2211$+$372	&$+0.49$		&DC8	&1,3\\
2323$+$256	&$+0.82$		&DA9	&1,8\\
2336$-$187	&$+0.27$		&DA6	&1,2,9\\
2349$-$283	&$+0.02$		&DA3	&1,10

\enddata

\tablerefs{
(1) This work and Paper I;
(2) \citealt{far09b};
(3) \citealt{rei05};
(4) \citealt{giz97};
(5) \citealt{lie03};
(6) \citealt{ber01};
(7) \citealt{far08b};
(8) \citealt{hin86};
(9) \citealt{tre07};
(10) \citealt{koe01}}

\tablecomments{These stars are no longer suspected of having a near-infrared excess 
consistent with a spatially-unresolved low mass companion within several square arcsec of 
their position.  The stars 0324$+$738 and 0518$+$333 are part of wide binary systems, but 
were previously suspected of harboring additional components based on their 2MASS 
photometry.}

\end{deluxetable}
		

\begin{thebibliography}{}

\bibitem[Abazajian et al.(2009)]{aba09} Abazajian, K. N., et al.\ 2009, \apjs, 182, 543
		
\bibitem[Bakos et al.(2002)]{bak02} Bakos, G. A., Sahu, K. c., \& Nemeth, P. 2002, \apjs, 141, 187
				
\bibitem[Bergeron et al.(2001)]{ber01} Bergeron, P., Leggett, S. K., \& Ruiz, M. T. 2001, \apjs, 133, 413
		
\bibitem[Bergeron et al.(1992)]{ber92} Bergeron, P., Saffer, R. A., \& Liebert, J. 1992, \apj, 394, 228
		
\bibitem[Bergeron et al.(1995a)]{ber95a} Bergeron, P., Saumon, D., \& Wesemael, F. 1995a, \apj, 443, 764

\bibitem[Bergeron et al.(1995b)]{ber95b} Bergeron, P., Wesemael, F., \& Beauchamp, A. 1995b, \pasp, 107, 1047

\bibitem[Bochanski et al.(2007)]{boc07} Bochanski, J. J., West, A. A., Hawley, S. L., \& Covey, K. R. 2007, \aj,133, 531
	
\bibitem[Bohlin(2007)]{boh07} Bohlin, R. C. 2007, in ASP Conference Series 364, The Future of Photometric, Spectrophotometric and Polarimetric Standardization, ed. C. Sterken (San Francisco: ASP), 315

\bibitem[Burleigh et al.(2002)]{bur02} Burleigh, M. R., Clarke, F. J., \& Hodgkin, S. T. 2002, \mnras, 331, L41         
	
\bibitem[Castanheira et al.(2006)]{cas06} Castanheira, B. G., Kepler, S. O., Handler, G., \& Koester, D. 2006, \aap, 450, 331
	
\bibitem[Catal\'an et al.(2008)]{cat08} Catal\'an, S., Isern, J., Garc\'ia-Berro, E., Ribas, I., Allende Prieto, C., \& Bonanos, A. Z. 2008, \aap, 477, 213
				
\bibitem[Cooke et al.(1992)]{coo92} Cooke, B. A., et al.\ 1992, \nat, 335, 61
		
\bibitem[Copenhagen University Observatory(2006)]{cmc06} Copenhagen University Observatory 2006, The Carlsberg Meridian Catalog 14 (Strasbourg: CDS)

\bibitem[Cruz et al.(2003)]{cru03} Cruz, K. L., Reid, I. N., Liebert, J., Kirkpatrick, J. D., \& Lowrance, P. J. 2003, \aj, 126, 2421
	
\bibitem[Cruz et al.(2007)]{cru07} Cruz, K. L., et al.\ 2007, \aj, 133, 439
	
\bibitem[Dahn et al.(1977)]{dah77} Dahn, C. C., Liebert, J., Kron, R. G., Spinrad, H., \& Hintzen, P. M. 1977, \apj, 216, 757

\bibitem[Dahn et al.(1982)]{dah82} Dahn, C. C., et al.\ 1982, \aj, 87, 419

\bibitem[Dahn et al.(1988)]{dah88} Dahn, C. C., et al.\ 1988, \aj, 95, 237

\bibitem[Dahn et al.(2002)]{dah02} Dahn, C. C., et al.\ 2002, \aj, 124, 1170

\bibitem[Dearborn et al.(1986)]{dea86} Dearborn, D. S. P., Liebert, J., Aaronson, M., Dahn, C. C., Harrington, R., Mould, J., Greenstein, J. L. 1986, \apj, 300, 314
		
\bibitem[Debes(2006)]{deb06} Debes, J. H. 2006, \apj, 652, 636

\bibitem[de Kool \& Ritter(1993)]{dek93} de Kool, M., \& Ritter, H., 1993, \aap, 267, 397

\bibitem[De Marco(2009)]{dem09} De Marco, O. 2009, \pasp,121, 316
			
\bibitem[DENIS Consortium(2005)]{den05} DENIS Consortium 2005, The DENIS Database, $3^{\rm rd}$ Release (Strasbourg: CDS)
				
\bibitem[Dobbie et al.(2005)]{dob05} Dobbie, P. D., Burleigh, M. R., Levan, A. J., Barstow, M. A., Napiwotzki, R., Holberg, J. B., Hubeny, I., \& Howell, S. B. 2005, \mnras, 357, 1049
		
\bibitem[Dobbie et al.(2009)]{dob09} Dobbie, P. D., Napiwotzki, R., Burleigh, M. R., Williams, K. A,; Sharp, R., Barstow, M. A., Casewell, S. L., \& Hubeny, I.  2009, \mnras, 395, 2248	
	
\bibitem[Dreizler \& Werner(1996)]{dre96} Dreizler, S., \& Werner K. 1996, \aap, 314, 217
			
\bibitem[Eggen \& Greenstein(1965)]{egg65} Eggen, O. J., \& Greenstein, J. L. 1965, \apj, 142, 925 

\bibitem[Evans(1992)]{eva92}  Evans, D. W. 1992, \mnras, 255, 521

\bibitem[Farihi(2004)]{far04a} Farihi, J. 2004, Ph.D. Thesis, UCLA
	
\bibitem[Farihi et al.(2005a)]{far05a} Farihi, J., Becklin, E. E., \& Zuckerman, B. 2005a, \apjs, 161, 394
		
\bibitem[Farihi et al.(2008a)]{far08a} Farihi, J., Becklin, E. E., \& Zuckerman, B. 2008a, \apj, 681, 1470

\bibitem[Farihi \& Christopher(2004)]{far04b} Farihi, J., \& Christopher, M. 2004, \aj, 128, 1868

\bibitem[Farihi et al.(2006)]{far06} Farihi, J., Hoard, D. W., \& Wachter S. 2006, \apj, 646, 480

\bibitem[Farihi (2009)]{far09b} Farihi, J. 2009, \mnras, 398, 2091

\bibitem[Farihi et al.(2009)]{far09a} Farihi, J., Jura, M., \& Zuckerman, B. 2009, \apj, 694, 805
			
\bibitem[Farihi et al.(2005b)]{far05b} Farihi, J., Zuckerman, B., \& Becklin, E. E. 2005b, \aj, 130, 2237
				
\bibitem[Farihi et al.(2008b)]{far08b} Farihi, J., Zuckerman, B., \& Becklin, E. E. 2008b, \apj, 674, 431

\bibitem[Finley et al.(1997)]{fin97} Finley, D. S., Koester, D., \& Basri, G. 1997, \apj, 488, 375

\bibitem[Fleming et al.(1996)]{fle96} Fleming, T. A., Snowden, S. L., Pfefferman, E., Briel, U., \& Greiner, J. 1996, \aap, 316, 147

\bibitem[Fontaine et al.(2003)]{fon03} Fontaine, G., Bergeron, P., Bill\`eres, M., \& Charpinet, S. 2003, \apj, 591, 1184 

\bibitem[Fontaine et al.(2001)]{fon01} Fontaine, G., Brassard, P., \& Bergeron, P. 2001, \pasp, 113, 409 

\bibitem[Ford et al.(1998)]{for98} Ford, H. C., et al.\ 1998, SPIE, 3356, 234
	
\bibitem[Friedrich et al.(2000)]{fri00} Friedrich, S., Koester, D., Christlieb, N., Reimers, D., \& Wisotzki, L. 2000, \aap, 363, 1040
	
\bibitem[Gianninas et al.(2006)]{gia06} Gianninas, A., Bergeron, P., \& Fontaine, G. 2006, \aj, 132, 831
	
\bibitem[Gizis \& Reid(1997)]{giz97} Gizis, J. E., \& Reid, I. N. 1997, \pasp, 109, 849

\bibitem[Green et al.(2000)]{gre00} Green, P. J., Ali, B., \& Napiwotzki, R. 2000, \apj, 540, 992 

\bibitem[Green et al.(1986)]{gre86} Green, R. F., Schmidt, M., \& Liebert, J. 1986, \apjs, 61, 305

\bibitem[Greenstein(1974)]{gre74} Greenstein, J. L. 1974, \aj, 79, 964

\bibitem[Greenstein(1976)]{gre76} Greenstein, J. L. 1976, \apj, 207, L119

\bibitem[Greenstein(1979)]{gre79} Greenstein, J. L. 1979, \apj, 227, 224
				
\bibitem[Greenstein(1984)]{gre84} Greenstein, J. L. 1984, \apj, 276, 602
	
\bibitem[Heber et al.(1993)]{heb93} Heber, U., Bade, N., Jordan, S., \& Voges, W. 1993, \aap, 267, L31	
	
\bibitem[Hintzen(1986)]{hin86} Hintzen, P. 1986, \aj, 92, 431
	
\bibitem[Hoard et al.(2007)]{hoa07} Hoard, D. W., Wachter, S., Sturch, L. K., Widhalm, A. M., Weiler, K. P., Pretorius, M. L., Wellhouse, J. W., \& Gibiansky, M. 2007, \aj, 134, 26

\bibitem[Hogan et al.(2009)]{hog09} Hogan, E., Burleigh, M. R., \& Clarke, F. J. 2009, \mnras, 396, 2074
	
\bibitem[Holberg \& Bergeron(2006)]{hol06} Holberg, J. B., \& Bergeron, P. 2006, \aj,132, 1221
		
\bibitem[Holman \& Wiegiert(1999)]{hol99} Holman, M. J., \& Wiegiert, P. A. 1999, \apj, 117, 621

\bibitem[H\"ugelmeyer et al.(2006)]{hug06} H\"ugelmeyer, S. D., Dreizler, S., Homeier, D., Krzesi\'nski, J., Werner, K., Nitta, A., \& Kleinman, S. J. 2006, \aap, 454, 617
	
\bibitem[Hutchings et al.(1996)]{hut96} Hutchings, J. B., Crampton, D., Cowley, A. P., Schmidtke, P. C., McGrath, T. K., \& Chu, Y. H. 1995, \pasp, 107, 931

\bibitem[Jeans(1924)]{jea24} Jeans, J. H. 1924, \mnras, 85, 2
												
\bibitem[Kawka et al.(2004)]{kaw04} Kawka, A., Vennes, S., \& Thorstensen, J. R. 2004, \aj, 127, 1702
	
\bibitem[Kilic et al.(2006a)]{kil06} Kilic, M., et al.\ 2006, \aj, 131, 582
	
\bibitem[Kilkenny et al.(1997)]{kil97} Kilkenny, D., O'Donoghue, D., Koen, C., Stobie, R. S., \& Chen A. 1997, \mnras, 287, 867

\bibitem[Kirkpatrick \& McCarthy(1994)]{kir94} Kirkpatrick, J. D., \& McCarthy, D. W. 1994, \aj, 107, 333  
	
\bibitem[Koester et al.(2001)]{koe01} Koester, D., et al.\ 2001, \aap, 378, 556

\bibitem[Koester(2009)]{koe09} Koester, D. 2009, A\&A, 498, 517

\bibitem[Koester(2010)]{koe10} Koester, D. 2010, to appear in Memorie della Societ\`a Astronomica Italiana, based on lectures given at the School of  Astrophysics  ``F. Lucchin'', Tarquinia, June 2008 (arXiv:0812.0482)	
			
\bibitem[Lamontagne et al.(2000)]{lam00} Lamontagne, R., Demers, S., Wesemael, F., Fontaine, G., \& Irwin, M. J. 2000, \aj, 119, 241
		
\bibitem[Leggett(1992)]{leg92} Leggett, S. K. 1992, \apjs, 82, 351
		
\bibitem[L\'epine \& Shara(2005)]{lep05} L\'epine, S., \& Shara, M. 2005, \aj, 129, 1483
			
\bibitem[Liebert et al.(1994)]{lie94} Liebert, J., Schmidt, G. D., Lesser, M., Stepanian, J. A., Lipovetsky, V. A., Chaffe, F. H., Foltz, C. B., \& Bergeron, P. 1994, \apj, 421, 733	
			
\bibitem[Liebert et al.(2003)]{lie03} Liebert, J., Bergeron, P., \& Holberg, J. B. 2003, \aj, 125, 348

\bibitem[Liebert et al.(2005a)]{lie05a} Liebert, J., Bergeron, P., \& Holberg, J. B. 2005, \apjs, 156, 47

\bibitem[Liebert et al.(2005b)]{lie05b} Liebert, J., Young, P. A., Arnett, D., Holberg, J. B., \& Williams, K. A. 2005, \apj, 630, L69	

\bibitem[Limoges et al.(2009)]{lim09} Limoges, M. M., Bergeron, P., \& Dufour, P. 2009, \apj, 696, 1461

\bibitem[Luyten(1979)]{luy79} Luyten, W. J. 1979, The LHS\,Catalog (Minneapolis: University of Minnesota)

\bibitem[Luyten(1997)]{luy97} Luyten, W. J. 1997, The LDS\,Catalogue (Minneapolis: University of Minnesota)
			 	
\bibitem[Martin et al.(2005)]{mar05} Martin, D. C., et al.\ 2005, \apj, 619, L1
			
\bibitem[Maxted \& Marsh (1999)]{max99} Maxted, P. F. L., \& Marsh, T. R. 1999, \mnras, 307, 122
	
\bibitem[Maxted et al.(2000)]{max00} Maxted, P. F. L., Marsh, T. R., \& Moran, C. K. J., 2000, \mnras, 319, 305
				
\bibitem[McCook \& Sion(2008)]{mcc08} McCook, G. P., \& Sion, E. M. 2008, Catalog of Spectroscopically Identified White Dwarfs (Strasbourg: CDS)

\bibitem[McCook \& Sion(1999)]{mcc99} McCook, G. P., \& Sion, E. M. 1999, \apjs, 121, 1

\bibitem[Mermilliod(1991)]{mer91} Mermilliod, J. C. 1991, Homogeneous Means in the UBV System (Strasbourg: CDS)	
		
\bibitem[Monet et al.(2003)]{mon03} Monet, D., et al.\ 2003, \aj, 125, 984

\bibitem[Morales-Rueda et al.(2005)]{mor05} Morales-Rueda, L., Marsh, T. R., Maxted, P. F. L., Nelemans, G., Karl, C., Napiwotzki, R., \& Moran, C. K. J. 2005, \mnras, 359, 648
					
\bibitem[Mullally et al.(2007)]{mul07} Mullally, F., Kilic, M., Reach, W. T., Kuchner, M. J., von Hippel, T., Burrows, A., \& Winget, D. E. 2007, \apjs, 171, 206
	
\bibitem[Nelemans et al.(2005)]{nel05} Nelemans, G., et al.\ 2005, \aap, 440, 1087

\bibitem[Nordhaus et al.(2010)]{nor10} Nordhaus, J., Spiegel, D. S., Ibgui, L., Goodman, J., \& Burrows, A. 2010, \mnras, in press(arXiv:1002.2216)
	
\bibitem[O'Dwyer et al.(2003)]{odw03} O'Dwyer, I. J., Chu, Y. H., Gruendel, R. A., Guerrero, M. A., \& Webbink, R. F. 2003, \aj, 125, 2239

\bibitem[Oswalt et al.(1984)]{osw84} Oswalt, T. D., Foltz, C. B., \& Peterson, B. M. 1984, \aj, 89, 421	
	
\bibitem[Paczynski(1976)]{pac76} Paczynski, B. 1976, Proceedings of IAU Symposium 73, eds. P. Eggleton, S. Mitton, \& J. Whelan (Dordrecht: D. Reidel), 75
				
\bibitem[Probst(1983)]{pro83} Probst, R. 1983, \apjs, 53, 335

\bibitem[Putney(1997)]{put97} Putney, A. 1997, \apjs, 112, 527

\bibitem[Raymond et al.(2003)]{ray03} Raymond, S. N., et al.\ 2003, \aj, 125, 2621

\bibitem[Rebassa-Mansergas et al.(2007)]{reb07} Rebassa-Mansergas, A., G\"ansicke, B. T., Rodr\'iguez-Gil, P., Schreiber, M. R., \& Koester, D. 2007, \mnras, 382, 1377

\bibitem[Rebassa-Mansergas et al.(2008)]{reb08} Rebassa-Mansergas, A., et al.\ 2008, \mnras, 390,1635	
		
\bibitem[Reid(1996)]{rei96} Reid, I. N. 1996, \aj, 111, 2000

\bibitem[Reid \& Gizis(2005)]{rei05} Reid, I. N., \& Gizis, J. E. 2005, \pasp, 117, 676

\bibitem[Reid et al.(1988)]{rei88} Reid, I. N., Wegner, G., Wickramasinghe, D. T., \& Bessell, M. S. 1988, \aj, 96, 275

\bibitem[Saffer et al.(1993)]{saf93} Saffer, R. A., Wade, R. A., Liebert, J., Green, R. F., Sion, E. M., Bechtold, J., Foss, D., \& Kidder, K. 1993, \aj, 105, 1945
		
\bibitem[Salim \& Gould(2003)]{sal03} Salim, S., \& Gould, A. 2003, \apj, 582, 1011
	
\bibitem[Schreiber et al.(2010)]{sch10} Schreiber, M. R., et al.\ 2010, \aap, 513, L7

\bibitem[Schultz et al.(1996)]{sch96} Schultz, G., Zuckerman, B., \& Becklin E. E. 1996, \apj, 460, 402

\bibitem[Silvestri et al.(2005)]{sil05} Silvestri, N. M., Hawley, S. L., \& Oswalt, T. D. 2005, \aj, 129, 2428

\bibitem[Silvestri et al.(2006)]{sil06} Silvestri, N. M., et. al. 2006, \aj, 131, 1674

\bibitem[Sion et al.(1995)]{sio95} Sion, E. M., Holberg, J. B., Barstow, M. A., \& Kidder, K. M. 1995, \pasp, 107, 232

\bibitem[Sirianni et al.(2005)]{sir05} Sirianni, M. et al.\ 2005, \pasp, 117, 1049

\bibitem[Skrutskie et al.(2006)]{skr06} Skrutskie, M. F., et al.\ 2006, \aj, 131, 1163

\bibitem[Space Telescope Science Institute(2006)]{sts06} Space Telescope Science Institute 2006, The Guide Star Catalog Version 2.3, (Baltimore: STScI)	

\bibitem[Stepanian et al.(2001)]{ste01} Stepanian, J. A., Green, R. F., Foltz, C. B., Chaffee, F., 	Chavushyan, V. H., Lipovetsky, V. A., \& Erastova, L. K. 2001, \aj, 122, 3361

\bibitem[Tappert et al.(2009)]{tap09} Tappert, C., G\"ansicke, B. T.,  Zorotovic, M., Toledo, I., Southworth, J., Papadaki, C., \& Mennickent, R. E. 2009, \aap, 504, 491

\bibitem[Trembley \& Bergeron(2007)]{tre07} Tremblay, P. E., \& Bergeron, P. 2007, \apj, 657, 1013	
		
\bibitem[Vennes et al.(1997)]{ven97} Vennes, S., Thejll, P. A., Galvan, R. G., \& Dupuis, J. 1997, \apj, 480, 714
					
\bibitem[Wachter et al.(2003)]{wac03} Wachter, S., Hoard, D. W., Hansen, K. H., Wilcox, R. E., Taylor, H. M., \& Finkelstein, S. L. 2003, \apj, 586, 1356

\bibitem[Wagner et al.(1986)]{wag86} Wagner, R. M., Sion, E. M., Liebert, J., Starrfield, S. G., \& Zotov, N. 1986, \pasp, 98, 552

\bibitem[Wegner et al.(1990)]{weg90} Wegner, G., Africano, J. L., \& Boodrich,	B. 1990, \aj, 99, 1907
	 
\bibitem[Wegner et al.(1987)]{weg87} Wegner, G., McMahan, R. K., \& Boley, F. I. 1987, \aj, 94, 1271

\bibitem[Wolff et al.(1996)]{wol96} Wolff, B., Jordan, S., \& Koester, D. 1996, \aap, 307, 149	
		
\bibitem[Zacharias et al.(2005)]{zac05} Zacharias, N., Monet, D. G., Levine, S. E., Urban, S. E., Gaume, R., \& Wycoff, G. L. 2005, The NOMAD Catalog (Strasbourg: CDS)
					
\bibitem[Zuckerman \& Becklin(1992)]{zuc92} Zuckerman, B., \& Becklin,	E. E. 1992, \apj, 386, 260
							
\bibitem[Zuckerman et al.(2003)]{zuc03} Zuckerman, B., Koester, D., Reid, I. N., \& H\"unsch, M. 2003, \apj, 596, 477
		
\end{thebibliography}
\end{document}